\documentclass[10pt,amsmath,amssymb,twocolumn,superscriptaddress,nofootinbib,aps,prd,twocolumn,floatfix]{revtex4-2}

\usepackage{graphicx}  
\graphicspath{{./plots/}}
\usepackage{bm}        
\usepackage{enumitem}
\usepackage{etoolbox}

\usepackage[colorlinks=true]{hyperref}

\newcommand{\lr}[1]{ \left( #1 \right) }
\newcommand{\lrs}[1]{ \left[ #1 \right] }
\newcommand{\lrc}[1]{ \left\{ #1 \right\} }
\newcommand{\vev}[1]{ \langle \, #1 \, \rangle }

\newcommand{\ket}[1]{ \, | #1 \rangle }
\newcommand{\bra}[1]{ \langle #1 | \, }
\newcommand{\braket}[2]{ \langle #1 | #2 \rangle }
\newcommand{\tcv}[2] {\left(\begin{array}{c} #1 \\ #2 \\ \end{array}\right)}

\newcommand{\tr}{ {\rm Tr} \, }
\newcommand{\re}{ {\rm Re} \, }

\newcommand{\expa}[1]{ \exp{\left( #1 \right)} }

\newtoggle{twocolumn}
\toggletrue{twocolumn}

\begin{document}
\sloppy

\title{Quantum chaos in supersymmetric quantum mechanics: an exact diagonalization study}

\author{P.~V.~Buividovich}
\email{pavel.buividovich@liverpool.ac.uk}
\affiliation{Department of Mathematical Sciences, University of Liverpool, Liverpool, L69 7ZL, UK}

\date{July 18th, 2022}

\begin{abstract}
We use exact diagonalization to study energy level statistics and out-of-time-order correlators (OTOCs) for the simplest supersymmetric extension $\hat{H}_S = \hat{H}_B \otimes I + \hat{x}_1 \otimes \sigma_1 + \hat{x}_2 \otimes \sigma_3$ of the bosonic Hamiltonian $\hat{H}_B = \hat{p}_1^2 + \hat{p}_2^2 + \hat{x}_1^2 \, \hat{x}_2^2$. For a long time, this bosonic Hamiltonian was considered one of the simplest systems which exhibit dynamical chaos both classically and quantum-mechanically. Its structure closely resembles that of spatially compactified pure Yang-Mills theory. Correspondingly, the structure of our supersymmetric Hamiltonian is similar to that of spatially compactified supersymmetric Yang-Mills theory, also known as the Banks-Fischler-Shenker-Susskind (BFSS) model. We present numerical evidence that a continuous energy spectrum of the supersymmetric model leads to monotonous growth of OTOCs down to the lowest temperatures, a property that is also expected for the BFSS model from holographic duality. We find that this growth is saturated by low-energy eigenstates with effectively one-dimensional wave functions and a completely non-chaotic energy level distribution. We observe a sharp boundary separating these low-energy states from the bulk of chaotic high-energy states. Our data suggests, although with a limited confidence, that at low temperatures the OTOC growth might be exponential over a finite range of time, with the corresponding Lyapunov exponent scaling linearly with temperature. In contrast, the gapped low-energy spectrum of the bosonic Hamiltonian leads to oscillating OTOCs at low temperatures without any signatures of exponential growth. We also find that the OTOCs for the bosonic Hamiltonian are never sufficiently close to the classical Lyapunov distance. On the other hand, the OTOCs for the supersymmetric system agree with the classical limit reasonably well over a finite range of temperatures and evolution times.
\end{abstract}

\maketitle

\section{Introduction}
\label{sec:intro}

Out-of-time-order correlators (OTOCs), first introduced in \cite{Larkin:JETP1969}, have recently attracted a lot of attention as probes of quantum chaos in strongly correlated many-body quantum systems. For a system with a Hamiltonian operator $\hat{H}$ at temperature $T$, the OTOC of two operators $\hat{A}$ and $\hat{B}$ is defined as
\begin{eqnarray}
\label{OTOCs_def}
 C\lr{t} = - \tr\lr{\hat{\rho} \lrs{\hat{A}\lr{t}, \hat{B}\lr{0}}^2 } ,
\end{eqnarray}
where $\hat{A}\lr{t} = e^{i \hat{H} t} \hat{A} e^{-i \hat{H} t}$ is the time dependent operator $\hat{A}$ in the Heisenberg representation, and $\hat{\rho} = \mathcal{Z}^{-1} \, e^{-\hat{H}/T}$ is the thermal density matrix.

OTOCs measure the sensitivity of a time evolution of a physical observable $\hat{A}\lr{t}$ to small perturbations of initial quantum state by an operator $\hat{B}\lr{0}$. If the operator $\hat{A}\lr{t}$ corresponds to some canonical coordinate $\hat{x}\lr{t}$ and $\hat{B}\lr{0}$ - to its conjugate momentum $\hat{p}\lr{0}$, the commutator $\lrs{\hat{x}\lr{t}, \hat{p}}$ in the OTOC (\ref{OTOCs_def}) corresponds to the classical Poisson bracket
\begin{eqnarray}
\label{class_Poisson}
 \lrc{x\lr{t}, p\lr{0}}
 =
 \frac{\partial x\lr{t}}{\partial x\lr{0}} \, \frac{\partial p\lr{0}}{\partial p\lr{0}}
 =
 \frac{\partial x\lr{t}}{\partial x\lr{0}}
\end{eqnarray}
that measures the sensitivity of time evolution of a dynamical system to its initial conditions. For a chaotic system, $\lrc{x\lr{t}, p\lr{0}}$ is expected to grow exponentially as $e^{\lambda \, t}$, where $\lambda$ is the leading (largest) Lyapunov exponent. In what follows we refer to $\lambda$ as simply the Lyapunov exponent. Generalizing the thermal average of the squared partial derivative $\frac{\partial x\lr{t}}{\partial x\lr{0}}$ in (\ref{class_Poisson}), OTOCs (\ref{OTOCs_def}) provide us with a quantum definition of the Lyapunov exponent $\lambda$. Namely, for a quantum chaotic system the OTOCs (\ref{OTOCs_def}) are expected to grow as $e^{2 \lambda \, t}$ for some period of time.

Many studies of OTOCs were to a large extent motivated by the derivation of a rigorous bound
\begin{eqnarray}
\label{MSS_bound}
 \lambda \leq 2 \pi \, T
\end{eqnarray}
on the growth of OTOCs in thermal systems at temperature $T$ by Maldacena, Shenker and Stanford \cite{Maldacena:1503.01409}, and the demonstration that for sufficiently low temperatures this bound is saturated by the Sachdev-Ye-Kitaev (SYK) model \cite{Stanford:1604.07818,Polchinski:1601.06768}. More generally, it was found that the MSS bound (\ref{MSS_bound}) is saturated for systems with a holographic dual description. Interestingly, the bound might be also saturated by simple experimentally accessible systems such as trapped ions in an external potential \cite{Tian:2007.05949}.

However, the demonstration of OTOC growth in generic chaotic quantum many-body systems and in quantum field theory turned out to be a very challenging endeavour. The progress was mostly limited to conformal field theory, quantum circuits and simple spin chain models \cite{Shen:1608.02438,Motrunich:1801.01636,Bohrdt:1612.02434}, often with quenched disorder \cite{Riddell:1810.00038,Hanada:1809.01671}. For example, so far it was not possible to explicitly demonstrate the saturation of the MSS bound (\ref{MSS_bound}) in the BFSS matrix model \cite{Susskind:hep-th/9610043}, which has a well-established holographic description in terms of black $D0$-branes in type IIA superstring theory or $M$-theory \cite{Hanada:2110.01312,Costa:1411.5541}. Even in the SYK model, the equality $\lambda = 2 \pi T$ can only be demonstrated using a model-specific diagrammatic technique \cite{Stanford:1604.07818,Polchinski:1601.06768,Sachdev:1904.02174}, while generally applicable techniques such as exact diagonalization so far could not reach the relevant low-temperature, large-system regime \cite{Sachdev:1603.05246,Sonner:1707.08013,Hanada:1611.04650}. Needless to say, the analysis of OTOCs and quantum Lyapunov exponents in higher-dimensional quantum field theories such as QCD is an even more formidable task.

In this situation, a lot of attention was attracted to simple quantum-mechanical systems in which the OTOCs can be calculated exactly using either analytic or numerical techniques, for example, by explicitly solving the Schrödinger equation. At the level of pure states, exponential OTOC growth was demonstrated in a kicked rotor system \cite{Rozenbaum:1609.01707} and in quantum stadia \cite{Rozenbaum:1801.10591}. Thermal OTOCs for simple quantum systems and for quantum billiards were extensively considered in \cite{Hashimoto:1703.09435}. In \cite{Romatschke:2008.06056}, OTOCs were calculated semi-analytically for the quartic anharmonic oscillator. In these cases, no exponential growth was clearly observed. Interestingly, an exponential OTOC growth can be observed in a double-well potential for temperatures or energies that are close the height of the saddle point separating the two wells \cite{Morita:2105.09603,Hashimoto:2007.04746}. In this case, the saddle point resembles an inverted harmonic oscillator, which also exhibits an exponential OTOC growth \cite{Murugan:2007.01232}.

Since real-time dynamics of many-body systems remains to a large extent inaccessible for simulations on classical computers, an important methodological application of simple chaotic quantum-mechanical systems might be the development and testing of numerical methods for diagnosing real-time quantum chaos.

One of the most popular and simple quantum-mechanical models for studying various aspects of quantum chaos is a system with two bosonic degrees of freedom $\hat{x}_1$ and $\hat{x}_2$ and the Hamiltonian of the form (up to the choice of pre-factors for the kinetic and the potential terms)
\begin{eqnarray}
\label{HB}
 \hat{H}_B = \hat{p}_1^2 + \hat{p}_2^2 + \hat{x}_1^2 \, \hat{x}_2^2 .
\end{eqnarray}
This Hamiltonian can be obtained by projecting the Hamiltonian of $SU\lr{2}$ bosonic matrix model to the sector with zero angular momentum \cite{Sekino:1403.1392} (see also \cite{Berenstein:1608.08972,Kares:hep-th/0401179}). In turn, $SU\lr{2}$ bosonic matrix model is a dimensional reduction of $SU\lr{2}$ Yang-Mills theory. It is also a bosonic part of the Hamiltonian of $N=2$ BFSS matrix model.

Classical dynamics of the Hamiltonian (\ref{HB}) is known to be chaotic at all temperatures or energies \cite{Savvidy:NPB84,Arefeva:hep-th/9710032}, with Lyapunov exponent scaling as
\begin{eqnarray}
\label{classical_Lyapunov_scaling}
 \, \lambda = c \, T^{1/4} ,
\end{eqnarray}
where $c \approx 1.32$ for our definitions of the kinetic and potential terms in (\ref{HB}). The emergence of fractal structures in the classical configuration space of this Hamiltonian was demonstrated recently in \cite{Yoshida:2204.06391}. A generalization of the Hamiltonian (\ref{HB}) with $N$ bosonic degrees of freedom was considered recently in \cite{Kolganov:2205.05663}, and an exponential growth of OTOCs was demonstrated analytically in the next-to-leading order of expansion in $1/N$.

The flat directions $x_1 = 0$ and $x_2 = 0$ of the classical Hamiltonian are lifted due to quantum effects, so that the energy levels of the quantum Hamiltonian (\ref{HB}) are discrete, and all the corresponding wave functions are localized. At sufficiently high energies, the bosonic Hamiltonian (\ref{HB}) exhibits a random-matrix-type level statistics \cite{Haller:PRL1984} and an exponential OTOC growth \cite{Akutagawa:2004.04381} at sufficiently high temperatures or energies. In this regime, the energy spectrum can be considered as continuous, and manifestations of quantum chaos are closely related to classical chaotic dynamics. However, the discrete energy spectrum of the system (\ref{HB}) implies that quantum chaos cannot be observed at low temperatures of the order of the gap between the two lowest energy levels. The crossover\footnote{We are using the term ``crossover'' because phase transitions cannot exist for systems with a finite number of degrees of freedom.} between the regular oscillatory dynamics at low temperatures and chaotic behavior at large temperatures/energies is a purely quantum phenomenon and is similar to the confinement-deconfinement transition in pure Yang-Mills theory or bosonic matrix models \cite{Nishimura:0706.3517,Hanada:1802.02985}.

In this paper we consider a minimal supersymmetric extension of the simple bosonic Hamiltonian (\ref{HB}), which exhibits continuous energy spectrum and OTOC growth all the way down to zero temperatures. In this respect, it is qualitatively similar to supersymmetric matrix models and the SYK model, and is very different from bounded quantum mechanical systems with discrete energy spectrum like (\ref{HB}). The Hamiltonian of our model can be written as
\begin{eqnarray}
\label{HS}
 \hat{H}_S = \hat{H}_B \otimes I + \hat{x}_1 \otimes \sigma_1 + \hat{x}_2 \otimes \sigma_3
 = \nonumber \\ =
 \left(
   \begin{array}{cc}
     \hat{H}_B + \hat{x}_2 & \hat{x}_1 \\
     \hat{x}_1             & \hat{H}_B - \hat{x}_2 \\
   \end{array}
 \right) ,
\end{eqnarray}
where $\sigma_1$, $\sigma_2$ and $\sigma_3$ are the $2 \times 2$ Pauli matrices. The Hilbert space of this model is therefore a direct product of the Hilbert space of the bosonic model (\ref{HB}) and a two-dimensional ``fermionic'' Hilbert space on which the Pauli matrices act. The Hamiltonian (\ref{HS}) can also be represented as a square $\hat{H}_S = \hat{Q}^2$ of the supersymmetry generator
\begin{eqnarray}
\label{QS}
 \hat{Q} = \hat{x}_1 \hat{x}_2 \otimes \sigma_2 + \hat{p}_1 \otimes \sigma_1 - \hat{p}_2 \otimes \sigma_3 .
\end{eqnarray}
This representation makes it obvious that the energy spectrum of the supersymmetric Hamiltonian is bounded from below, even though the fermionic terms $\hat{x}_1 \otimes \sigma_1 + \hat{x}_2 \otimes \sigma_3$ in (\ref{HS}) are unbounded. The supersymmetric model (\ref{HS}) was first introduced in \cite{Nicolai:NPB1989} as a toy model of supersymmetric membrane. In this work it was demonstrated that quantum corrections that lift the classical flat directions $x_1 = 0$ and $x_2 = 0$ are cancelled out due to supersymmetry, so that the directions $x_1 = 0$ and $x_2 = 0$ remain flat for the supersymmetric Hamiltonian (\ref{HS}). As a result, the energy spectrum of $\hat{H}_S$ is continuous, and wave functions extend to infinity along the lines $x_1 = 0$ and $x_2 = 0$. This demonstration was further extended to the BFSS model to argue that supersymmetric membranes are intrinsically unstable, in contract to bosonic membranes \footnote{In \cite{Nicolai:NPB1989}, the supersymmetric Hamiltonian was written as $\hat{H}_S = \hat{H}_B \otimes I + \hat{x}_1 \otimes \sigma_1 + \hat{x}_2 \otimes \sigma_2$. In this paper, we choose a unitary equivalent Hamiltonian (\ref{HS}) that is manifestly real. A unitary transformation relating both Hamiltonians is $U = \hat{I}\otimes\frac{I - i \sigma_1}{\sqrt{2}}$}. If we consider the bosonic Hamiltonian (\ref{HB}) as a minimal model that is similar to the bosonic matrix model of bosonic membranes, the model (\ref{HS}) is a minimal model that is similar to supersymmetric matrix/membrane models such as the BFSS \cite{Susskind:hep-th/9610043} matrix model.

We will see that much like in the case of the BFSS model, additional ``fermionic'' terms in the model (\ref{HS}) eliminate the transition (more precisely, the crossover in our case) to the non-chaotic ``confinement'' regime, so that the model exhibits monotonously growing OTOCs for all temperatures \footnote{Note that our SUSY Hamiltonian is very different from fully integrable supersymmetric Hamiltonians considered in \cite{Dutta:2010.07089,Bhagat:2008.03280}.}. Somewhat counterintuitively, we find that the energy level statistics still exhibits a rather sharp change between chaotic and regular behavior at large and low energies. Furthermore, we will demonstrate that at high temperatures the supersymmetric model (\ref{HS}) has much better agreement with classical dynamics than the purely bosonic one (\ref{HB}).

\section{Numerical method}
\label{sec:method}

In this work, we perform numerical diagonalization of the supersymmetric Hamiltonian (\ref{HS}). To highlight the difference between the bosonic and the supersymmetric model, we also diagonalize the bosonic Hamiltonian (\ref{HB}). Since both Hamiltonians act on infinite-dimensional Hilbert spaces, the first step is to truncate the Hilbert space to a finite number of states that can be treated numerically. To this end we consider the matrices of both Hamiltonians in the basis of two-dimensional harmonic oscillator states of the form $\ket{k_1} \otimes \ket{k_2}$, where the states $\ket{k_1}$ and $\ket{k_2}$ belong to the Hilbert space of functions of $x_1$ and $x_2$, respectively. The corresponding wave functions are of the form
\begin{eqnarray}
\label{basis_wavefuncs}
 \Psi_{k_1, k_2}\lr{x_1, x_2} = \psi_{k_1}\lr{x_1} \, \psi_{k_2}\lr{x_2} ,
\end{eqnarray}
where
\begin{eqnarray}
\label{basis_wavefunc_1D}
\psi_k\lr{x} = \frac{1}{\sqrt{2^k \, k! \, \sqrt{\pi} L}}
    \expa{-\frac{x^2}{2 L^2}} H_k\lr{k, x/L} ,
\end{eqnarray}
\vspace{1em}

are the wave functions that correspond to eigenstates of a one-dimensional harmonic oscillator Hamiltonian $\hat{H}_0 = \frac{L^2 \, \hat{p^2}}{2} + \frac{\hat{x}^2}{2 \, L^2}$, $H_k\lr{k, z} = \lr{-1}^k \, e^{z^2} \frac{d^k}{d z^k} e^{-z^2}$ are the Hermite polynomials, and $L$ is the length parameter. We discuss the tuning of $L$ a bit later.

Matrix elements of the bosonic Hamiltonian in this basis take the form
\begin{eqnarray}
\label{HB_matrix}
 \lr{H_B}_{k_1,k_2; l_1,l_2} = \bra{k_1} \otimes \bra{k_2} \hat{H}_B \ket{l_1} \otimes \ket{l_2}
 = \nonumber \\ =
 \bra{k_1} \hat{p}^2 \ket{l_1} \delta_{k_2 \, l_2}
 +
 \bra{k_2} \hat{p}^2 \ket{l_2} \delta_{k_1 \, l_1}
 + \nonumber \\ +
 \bra{k_1} \hat{x}^2 \ket{l_1} \,
 \bra{k_2} \hat{x}^2 \ket{l_2} ,
\end{eqnarray}
where
\begin{eqnarray}
\label{matrix_elements_1D}
 \bra{k} \hat{p}^2 \ket{l} = \frac{1}{2 L^{2}} \left(\delta_{k,l} \, \lr{2 \, k + 1}
 - \right. \nonumber \\ \left. -
 \delta_{k+2,l} \, \sqrt{\lr{k+1} \lr{k+2}}
 -
 \delta_{k-2,l} \, \sqrt{k \lr{k-1} } \right) ,
 \nonumber \\
 \bra{k} \hat{x}^2 \ket{l} = \frac{L^2}{2} \left(\delta_{k,l} \, \lr{2 \, k + 1}
 + \right. \nonumber \\ \left. -
 \delta_{k+2,l} \, \sqrt{\lr{k+1} \lr{k+2}}
 +
 \delta_{k-2,l} \, \sqrt{k \lr{k-1} } \right)
\end{eqnarray}

are the matrix elements of the operators $\hat{x}^2$ and $\hat{p}^2$ in the basis of one-dimensional wave functions (\ref{basis_wavefunc_1D}). Correspondingly, the matrix of the supersymmetric Hamiltonian (\ref{HS}) can be represented in the following block form:
\begin{widetext}
\begin{eqnarray}
\label{HS_matrix}
 \bra{k_1} \otimes \bra{k_2} \hat{H}_S \ket{l_1} \otimes \ket{l_2}
 =
 \left(
 \begin{array}{cc}
   \lr{H_B}_{k_1,k_2; l_1,l_2} + \delta_{k_1, l_1} \, \bra{k_2} \hat{x} \ket{l_2} & \delta_{k_2, l_2} \, \bra{k_1} \hat{x} \ket{l_1} \\
   \delta_{k_2, l_2} \, \bra{k_1} \hat{x} \ket{l_1} & \lr{H_B}_{k_1,k_2; l_1,l_2}  - \delta_{k_1, l_1} \, \bra{k_2} \hat{x} \ket{l_2}
 \end{array}
 \right) ,
\end{eqnarray}
\end{widetext}
where $\bra{k} \hat{x} \ket{l} = \frac{L}{\sqrt{2}} \left(\delta_{k+1,l} \sqrt{k+1} + \delta_{k-1,l} \sqrt{k} \right)$ are matrix elements of the coordinate operator $\hat{x}$ in the basis (\ref{basis_wavefunc_1D}).

To make the matrix representations (\ref{HB_matrix}) and (\ref{HS_matrix}) finite and thus numerically tractable, we truncate the infinite-dimensional discrete Hilbert space spanned on the basis vectors $\ket{k_1} \otimes \ket{k_2}$ to a finite number of states with
\begin{eqnarray}
\label{truncation_prescription}
 k_1 + k_2 < 2 \, M ,
\end{eqnarray}
where $M$ is the truncation parameter. To reduce the computational cost of exact diagonalization, we further use the discrete parity symmetry of the Hamiltonians (\ref{HB}) and (\ref{HS}) to represent the matrices (\ref{HB_matrix}) and (\ref{HS_matrix}) in block diagonal form. We then perform numerical diagonalization of the matrices (\ref{HB_matrix}) and (\ref{HS_matrix}). For not very large $M \lesssim 100$ we use the QR algorithm as implemented in the \texttt{LAPACK} routine \texttt{dsyev}, finding all eigenvectors and eigenvalues of the matrices (\ref{HB_matrix}) and (\ref{HS_matrix}). For larger values of $M$ we use the Arnoldi algorithm as implemented in the \texttt{ARPACKPP} library to find $n \ll M^2$ lowest eigenvalues, with $n$ taking values between $10^2$ and $10^3$. Our production code that performs exact diagonalization and calculates the OTOCs is publicly available on GitHub \cite{SUSYQMGitHubRep}. Numerical data for OTOCs and energy levels are included as ancillary files in the ArXiv submission.

To analyze the bosonic Hamiltonian (\ref{HB}), the length parameter $L$ is adjusted to the value $L = 2^{1/6} \approx 1.12246$ that minimizes the expectation value ${\bra{0} \otimes \bra{0} \hat{H}_B \ket{0} \otimes \ket{0}} = L^{-2} + L^4/4$ of $\hat{H}_B$ in the ``perturbative'' vacuum state $\ket{0} \otimes \ket{0}$. Another strategy would be to minimize the ground state energy upon exact diagonalization. However, we have found that our results depend very weakly on the choice of $L$ once the number of basis vectors is sufficiently large.

\begin{figure}[h!tpb]
  \centering
  \includegraphics[width=0.48\textwidth]{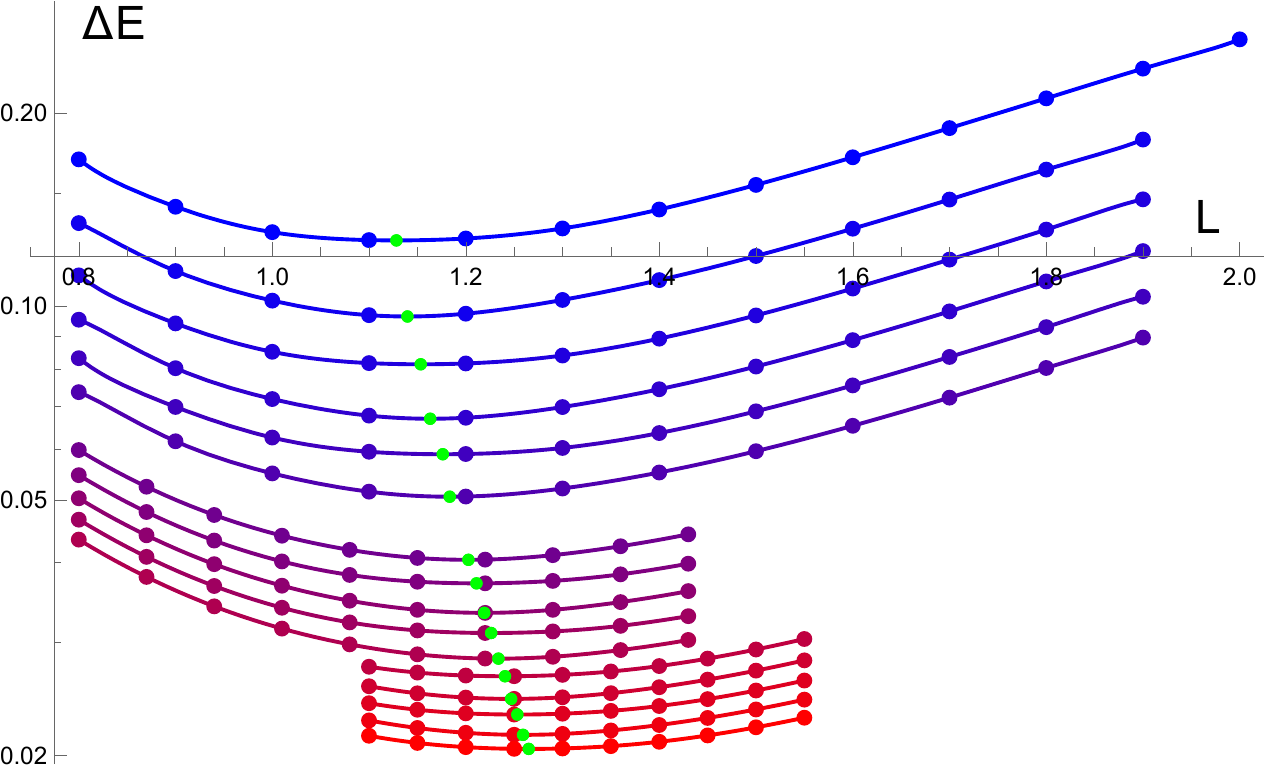}
  \caption{The gap $\Delta E = E_3 - E_1$ between the lowest and the next-to-lowest energy levels $E_1$ and $E_3$ of the matrix (\ref{HS_matrix}) of the supersymmetric Hamiltonian (\ref{HS}) as a function of the length parameter $L$ in the basis wave-functions (\ref{basis_wavefunc_1D}). The value of the truncation parameter $M$ changes between $M=40$ (top curve plotted in blue) to $M=200$ (lowest curve plotted in red) in steps of $10$ (the only missing value is $M=100$). Green points indicate the position of the minima, estimated using spline interpolation. }
  \label{fig:gap_supersymmetric_vs_M_and_L}
\end{figure} 

\begin{figure}[h!tpb]
  \centering
  \includegraphics[width=0.48\textwidth]{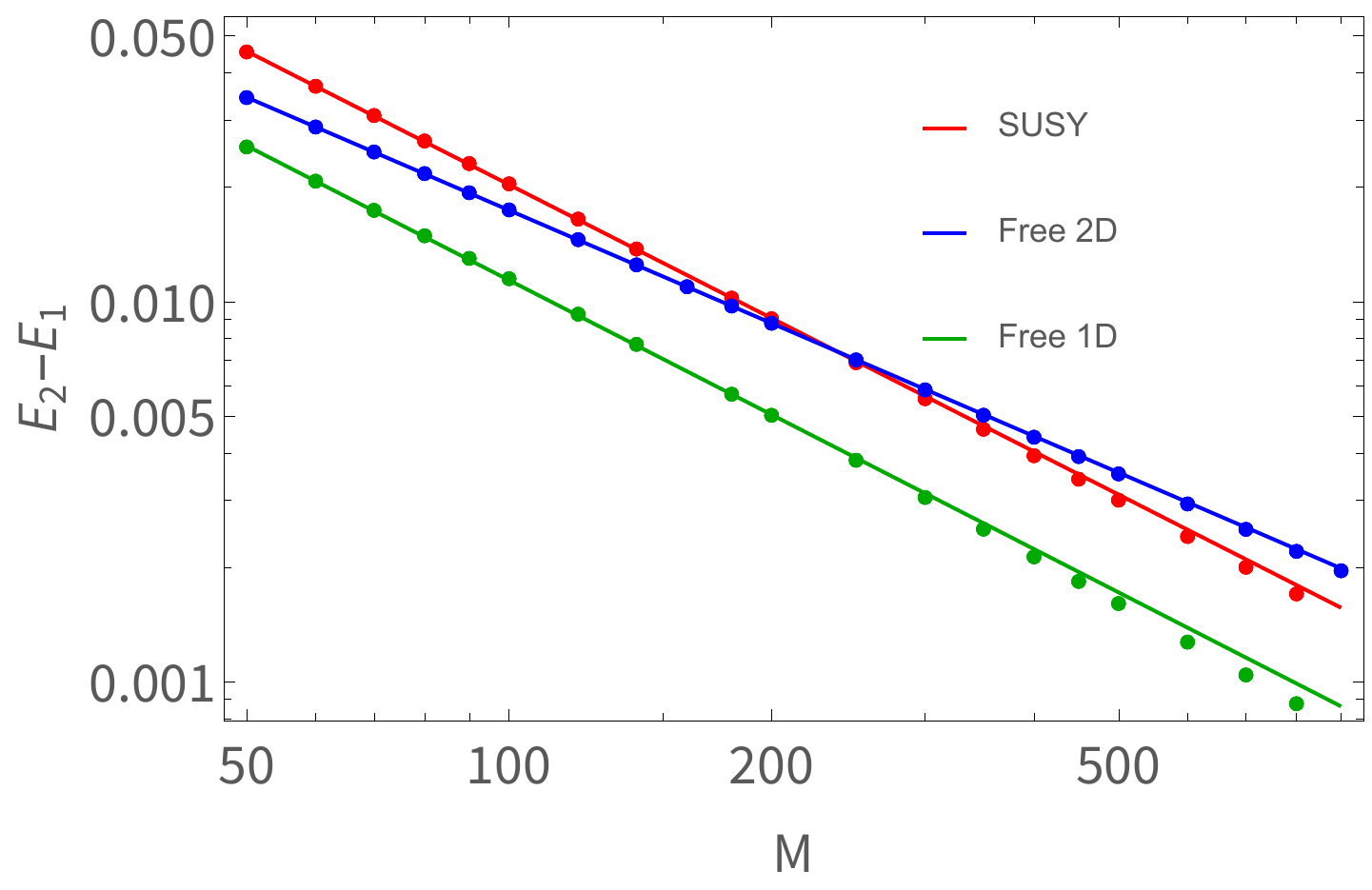}
  \caption{The gap $\Delta E$ between the lowest and the next-to-lowest energy levels of the supersymmetric Hamiltonian (\ref{HS}) as a function of the truncation parameter $M$. For each value of $M$, the optimal value of $L$ is chosen according to (\ref{L_vs_M_supersymmetric}). For comparison, we also show the gap of free one- and two-dimensional Hamiltonians (\ref{H1D}) and (\ref{H2D}) as functions of $M$, obtained in exactly the same way and using the same values of $L$ for each $M$. Solid lines are best fits of the form $\Delta E = A \, M^{-B}$. }
  \label{fig:gap_vs_M}
\end{figure}

For the supersymmetric Hamiltonian (\ref{HS}), for each value of the truncation parameter $M$ we choose the value of the length parameter $L$ that minimizes the gap $\Delta E = E_3 - E_1$ between its lowest and next-to-lowest energy levels $E_3$ and $E_1$, obtained by exact diagonalization. As explained in Appendix~\ref{apdx:symmetries} below, the energy levels of the supersymmetric system are all doubly degenerate, therefore $E_2 = E_1$ and the first nonzero energy gap is $\Delta E = E_3 - E_1 = E_3 - E_2$. The dependence of $\Delta E$ on $M$ and $L$ is illustrated on Fig.~\ref{fig:gap_supersymmetric_vs_M_and_L}. We found that the dependence of the optimal value of $L$ on the truncation parameter $M$ can be well described by the formula
\begin{eqnarray}
\label{L_vs_M_supersymmetric}
 L = 0.961624 + 0.0409431 (2 M - 1)^{0.377368} .
\end{eqnarray}
With this choice of $L$, the dependence of the energy gap $\Delta E$ on $M$ is illustrated on Fig.~\ref{fig:gap_vs_M}. Fitting suggests that it can be well described by the power law
\begin{eqnarray}
\label{dE_vs_M_supersymmetric}
 \Delta E = 4.36466 \, M^{-1.16619} ,
\end{eqnarray}
which is shown on Fig.~\ref{fig:gap_vs_M} as a solid red line. For comparison, on Fig.~\ref{fig:gap_vs_M} we also show the dependence of the gap $\Delta E$ on $M$ for the free particle Hamiltonians in one and two dimensions:
\begin{eqnarray}
\label{H1D}
 \hat{H}_{1D} = \hat{p}^2 ,
 \\
 \label{H2D}
 \hat{H}_{2D} = \hat{p}_1^2 + \hat{p}_2^2 ,
\end{eqnarray}
which are subject to the same truncations of the Hilbert space as the supersymmetric Hamiltonian. Namely, we consider the one-dimensional Hamiltonian (\ref{H1D}) on the Hilbert space spanned by all one-dimensional basis states (\ref{basis_wavefunc_1D}) with $k < 2 M$ and with the length parameter that depends on $M$ as in (\ref{L_vs_M_supersymmetric}). The power-law dependence of the energy gap of $\hat{H}_{1D}$ on $M$ appears to be quite close to the result (\ref{dE_vs_M_supersymmetric}) for the supersymmetric Hamiltonian: $\Delta E = 2.54431 \, M^{-1.17463}$. This power-law fit is shown on Fig.~\ref{fig:gap_vs_M} with a solid line.

For the free two-dimensional Hamiltonian (\ref{H2D}), the Hilbert space is truncated to states $\ket{k_1} \ket{k_2}$ with $k_1 + k_2 < 2 M$, as in (\ref{truncation_prescription}), and the length parameter is again given by (\ref{L_vs_M_supersymmetric}). In this case, we also obtain a power law dependence of $\Delta E$ on $M$, although with a somewhat different power: $\Delta E = 1.6393 \, M^{-0.987156}$.

In what follows, we will often compare the results obtained for the supersymmetric and the bosonic Hamiltonians (\ref{HS}) and (\ref{HB}) with results for the one- and two-dimensional free Hamiltonians (\ref{H1D}) and (\ref{H2D}) at the same temperature and for the same Hilbert space truncation. We will see that at low temperatures the eigenstates of the supersymmetric Hamiltonian (\ref{HS}) are with a good precision one-dimensional, and a meaningful comparison can be made with the one-dimensional free Hamiltonian (\ref{H1D}). Similarly, the two-dimensional free-particle Hamiltonian (\ref{H2D}) qualitatively reproduces some features of OTOCs at early times and high temperatures, which are most likely the artifacts of a finite infrared cutoff.

The largest eigenvalues of the Hamiltonian matrices (\ref{HB_matrix}) and (\ref{HS_matrix}) also grow with $M$. For example, for the matrix of the bosonic Hamiltonian (\ref{HB_matrix}) for $M=70$ and $M=100$ the largest eigenvalues are $E_{max}\lr{M = 70} = 27073.5$ and $E_{max}\lr{M = 100} = 56734.3$. For the matrix of the supersymmetric Hamiltonian (\ref{HS_matrix}), we have $E_{max}\lr{M = 70} = 38426.4$ and $E_{max}\lr{M = 100} = 91055.4$, respectively. The truncation parameter $M$ thus provides both the infrared and the ultraviolet cutoff for the continuous Hamiltonians (\ref{HB}) and (\ref{HS}). The OTOCs appear to be not very sensitive to the ultraviolet cutoff at high energies, which allows us to obtain reliable results using only a relatively small number $n \ll M^2$ of lowest energy levels.

On the other hand, the infrared cutoff due to the truncation to $k_1 + k_2 \leq 2 \, M$ is crucial to obtain a numerically tractable approximation for the supersymmetric Hamiltonian (\ref{HS}) with a continuous energy spectrum. As discussed in the Introduction, the flat directions $x_1 = 0$ and $x_2 = 0$ of the classical Hamiltonian remain flat also for the supersymmetric Hamiltonian, so that the system can still escape to infinity, and at least some of the wave functions have infinite spatial extent. Our truncation of the Hilbert space to a finite number of basis states (\ref{basis_wavefuncs}) with $k_1 + k_2 \leq 2 M$ limits the spatial extent of wave functions, thus effectively introducing soft boundaries on spatial coordinates $x_1$ and $x_2$. Indeed, using at most $2 M$ lowest harmonic oscillator eigenstates (\ref{basis_wavefunc_1D}), we can construct wave functions with spatial extent that does not exceed $\vev{\hat{x}^2} \simeq 2 M \, L^2$. This is obvious from equalities $\bra{n} \, \hat{x}^2/L^2 \, \ket{n} = \bra{n} \, \hat{p}^2 \, L^2 \, \ket{n} = \bra{n} \, \hat{H}_0 \, \ket{n} = \lr{n + 1/2}$ for the one-dimensional oscillator Hamiltonian $\hat{H}_0 = \frac{L^2 \, \hat{p}^2}{2} + \frac{\hat{x}^2}{2 \, L^2}$ with eigenstates (\ref{basis_wavefunc_1D}). Hence the subspace spanned on eigenstates with $n \leq 2 M$ can only contain states with $\vev{\hat{x}^2} \lesssim 2 M \, L^2$. Our approach is therefore similar to numerical regularization of any other unbounded system with continuous spectrum by putting it in a finite box.

\begin{figure}[h!tpb]
  \centering
  \includegraphics[width=0.48\textwidth]{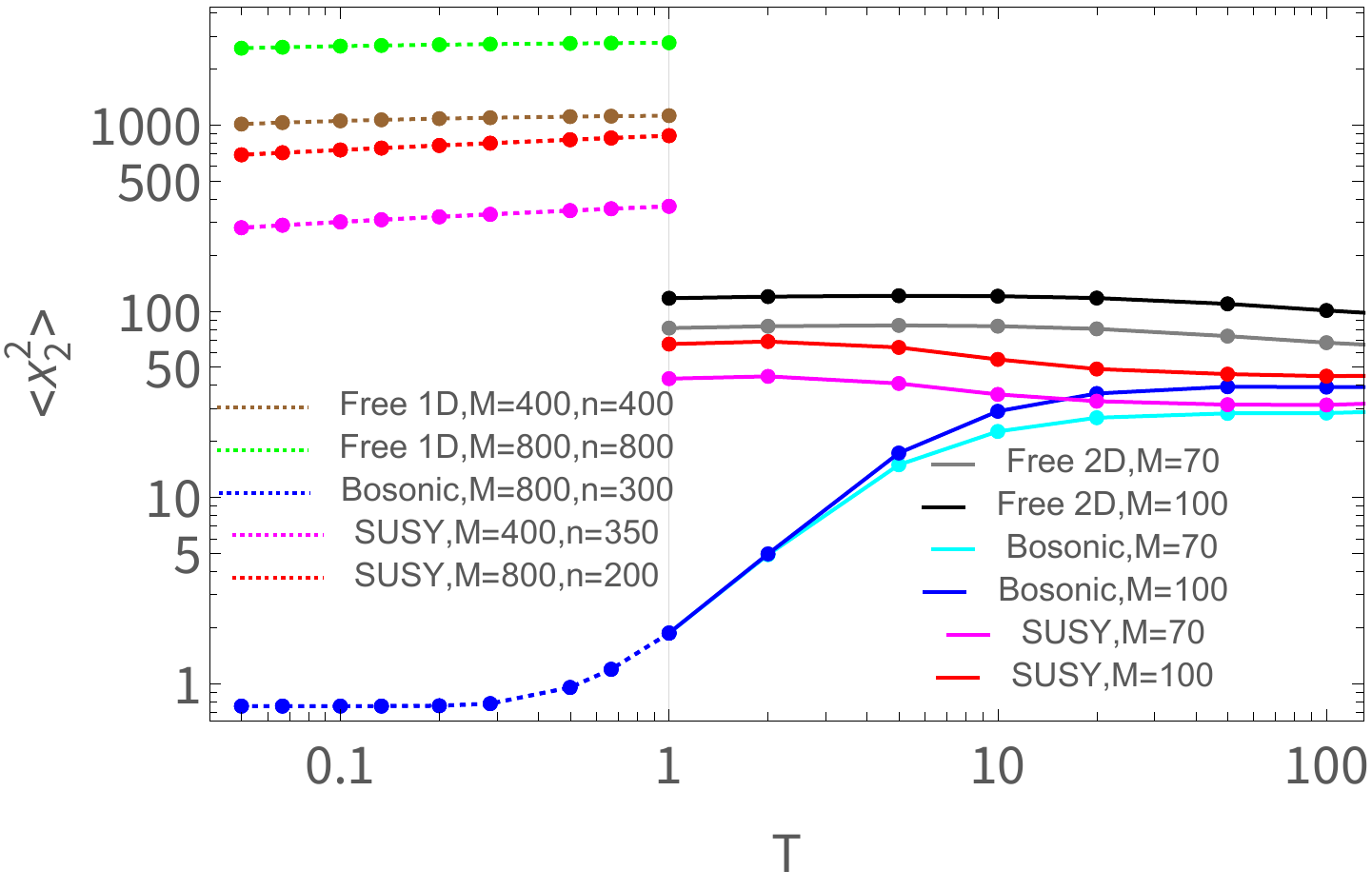}\\
  \caption{Temperature dependence of the thermal expectation values $\vev{\hat{x}_2^2} = \tr\lr{\hat{\rho} \, \hat{x}_2^2}$ for the supersymmetric and bosonic Hamiltonians (\ref{HS}) and (\ref{HB}) at different values of truncation parameters $M$. For $T \geq 1$, we show the results for $M = 70$ and $M = 100$ with all eigenvalues taken into account. For $T \leq 1$, we show the results for $M = 400$ and $M = 800$, with only $n \ll M^2$ lowest energy levels taken into account. For comparison, we also show $\vev{\hat{x}_2^2}$ for the free one- and two-dimensional Hamiltonians (\ref{H1D}) and (\ref{H2D}) with the same truncations of the Hilbert space.}
  \label{fig:x2_vs_T}
\end{figure}

To illustrate how the finite values of $M$ impose a cutoff on spatial coordinates, on Fig.~\ref{fig:x2_vs_T} we show thermal expectation values $\vev{\hat{x}_2^2} = \tr\lr{\hat{\rho} \hat{x}_2^2}$ as functions of the temperature $T$ at different values of $M$ for supersymmetric, bosonic, and free one- and two-dimensional Hamiltonians (\ref{HS}), (\ref{HB}), (\ref{H1D}) and (\ref{H2D}). As discussed above, the data for the free one- and two-dimensional Hamiltonians (\ref{H1D}) are used for comparison at low and high temperatures, respectively.

In the high-temperature regime, we use all eigenstates of the truncated supersymmetric, bosonic and two-dimensional free Hamiltonians (\ref{HS}), (\ref{HB}) and (\ref{H2D}) to calculate the expectation value. In the low-temperature regime, we use only $n \ll M^2$ eigenstates that correspond to $n$ lowest energy levels. We check that our values of $n$ are big enough by comparing the results obtained with $n$ and $n/2$ lowest energy levels, which are found to be very close to each other.

Fig.~\ref{fig:x2_vs_T} shows that for the supersymmetric Hamiltonian (\ref{HS}) as well as for the one- and two-dimensional free Hamiltonians, thermal expectation values $\vev{\hat{x}_2^2} = \tr\lr{\hat{\rho} \, \hat{x}_2^2}$ strongly depend on $M$ but have weak temperature dependence. This is an expected behavior for a free particle confined within a region of space with size $\sim \sqrt{M \, L^2}$ determined by the infrared cutoff scale. For the free Hamiltonians (\ref{H1D}) and (\ref{H2D}) in one and two dimensions, the expectation values $\vev{\hat{x}_2^2}$ are indeed reasonably close to the estimates $\vev{\hat{x}_2^2} \approx 2 M \, L^2$. For the supersymmetric Hamiltonian, $\vev{\hat{x}_2^2}$ is considerably smaller than these estimates. This suggests that potential energy terms have a significant effect on the spatial structure of wave functions at high energies, despite the flatness of the $x_1 = 0$ and $x_2 = 0$ directions.

On the other hand, for the bosonic Hamiltonian (\ref{HB}) the expectation value $\vev{\hat{x}_2^2}$ exhibits strong temperature dependence and significantly decreases at low temperatures. This is an expected behavior for a gapped system with localized wave functions. The dependence on the truncation parameter $M$ is negligible for sufficiently small temperatures, and only becomes important at high temperatures. Only at $T \gtrsim 20$, the expectation values $\vev{\hat{x}_2^2}$ for the bosonic and the supersymmetric Hamiltonians (\ref{HB}) and (\ref{HS}) become reasonably close to each other, and exhibit very weak temperature dependence and strong $M$ dependence. This suggests that at such temperatures the lifting of the flat directions due to quantum effects becomes negligible for the bosonic Hamiltonian, and the infrared cutoff scale becomes important. At the same time, the effect of fermionic terms also becomes smaller, and the supersymmetric system exhibits classical chaotic dynamics that is similar to the dynamics of the bosonic Hamiltonian.

Finally, we should note that the truncation of the full Hilbert space of the supersymmetric Hamiltonian (\ref{HS}) to the subspace spanned by a finite number $2 M \lr{M+1}$ of basis states of the form (\ref{basis_wavefuncs}) breaks the exact supersymmetry of the model. In particular, the equality $\hat{H}_S = \hat{Q}^2$ is violated at the upper edge of the energy spectrum if we truncate the matrices of the Hamiltonian $\hat{H}_S$ and the supersymmetry generator $\hat{Q}$ to have finite dimensions. In practice, we find that all the energy levels of $\hat{H}_S$ remain positive upon the truncation, so the truncation preserves the cancellation of the negative unbounded terms in the fermionic operators in $\hat{H}_S$. Furthermore, the effect of truncation becomes negligible for thermal expectation values in the limit of large $M$, as anti-periodic boundary conditions for fermions on the thermal circle break supersymmetry anyway, and the upper edge of the spectrum is suppressed at finite temperatures. Exact supersymmetry also implies that the ground state of the supersymmetric Hamiltonian should not be doubly degenerate, in contrast to all higher energy levels. On the other hand, with our truncation of the Hilbert space all energy levels are doubly degenerate, as discussed in detail in Appendix~\ref{apdx:symmetries}. Since the energy spectrum of the supersymmetric Hamiltonian is continuous (which we recover in the limit $M \rightarrow 0$), there are infinitely many energy levels that are infinitely close to the ground state, and a single non-degenerate energy level should likewise have a vanishing contribution to thermal expectation values.

\section{Eigenstates of the supersymmetric Hamiltonian}
\label{sec:energy_spectrum}

\begin{figure}[h!tpb]
  \centering
  \includegraphics[width=0.48\textwidth]{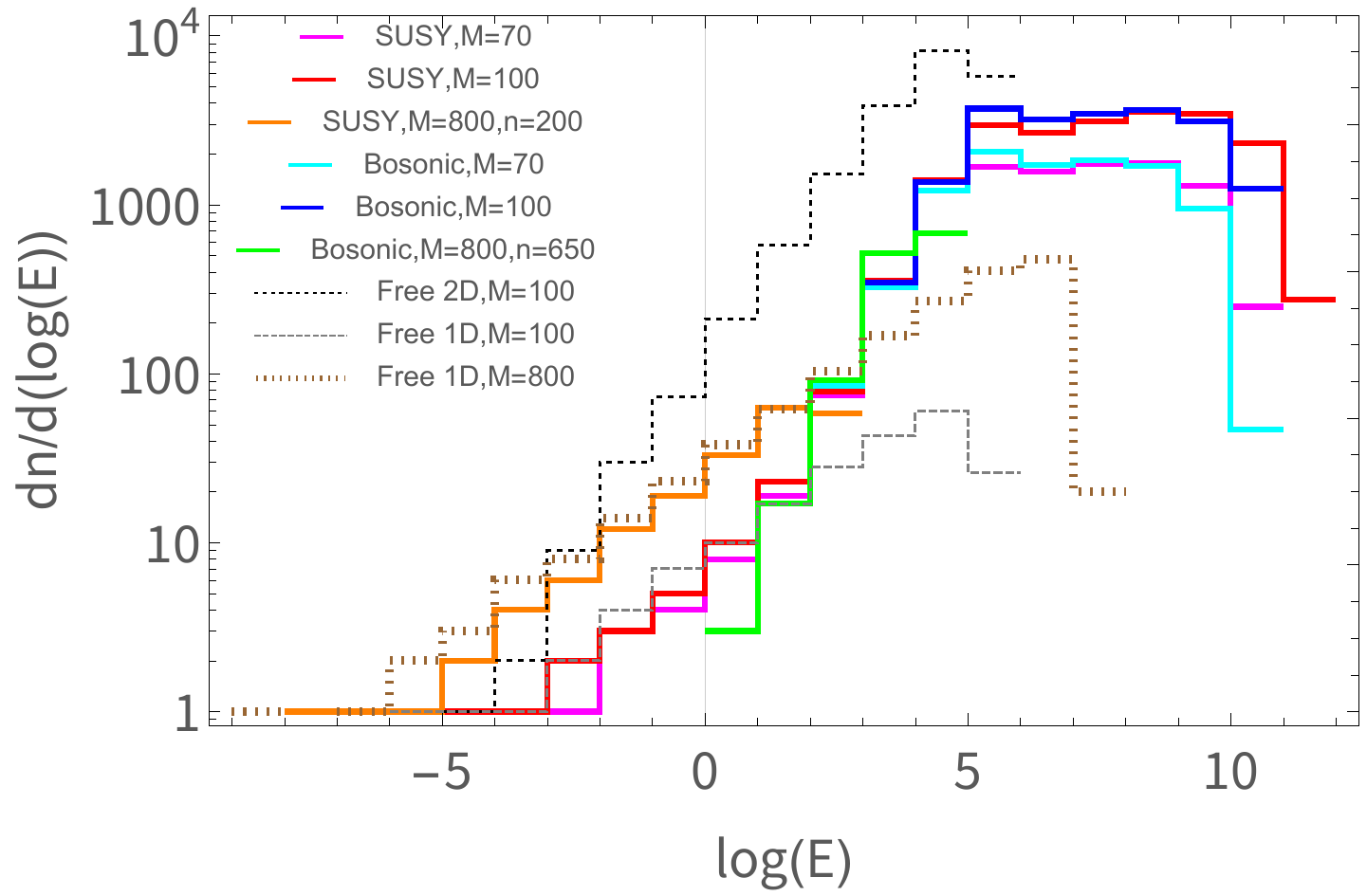}
  \caption{Histograms approximating the log-scale energy level density $\frac{d n\lr{E}}{d \, \log\lr{E}} = E \, \frac{d n\lr{E}}{d E}$ for the supersymmetric and bosonic Hamiltonians (\ref{HS}) and (\ref{HB}) for different values of the truncation parameter $M$. For comparison, we also show similar histograms for the free one- and two-dimensional Hamiltonians (\ref{H1D}) and (\ref{H2D}) with same Hilbert space truncations. For the supersymmetric and bosonic Hamiltonians with truncation parameter $M = 800$, we only show the histograms of $n \ll M^2$ lowest eigenvalues.}
  \label{fig:global_level_hist}
\end{figure} 

Preparing to consider out-of-time-order correlators, in this Section we consider the spectral properties of the supersymmetric Hamiltonian (\ref{HS}) and compare them with those of the bosonic Hamiltonian (\ref{HB}) and the free one- and two-dimensional Hamiltonians (\ref{H1D}) and (\ref{H2D}) (at low and high temperatures, respectively). We start with the global distribution of energy levels. For both the supersymmetric and the bosonic Hamiltonians, histogramming all eigenvalues of the Hamiltonian matrices (\ref{HB_matrix}) and (\ref{HS_matrix}) suggests that the level density $\frac{d n\lr{E}}{d E}$ falls off as $1/E$. However, this analysis is not very informative, as all of the interesting low-lying eigenvalues are counted within a single near-zero bin. To properly resolve the distribution of eigenvalues at all scales, we analyze the histograms of $\log\lr{E}$. Such histograms approximate the distributions $\frac{d n\lr{E}}{d \, \log\lr{E}} = E \, \frac{d n\lr{E}}{d E}$ and are shown on Fig.~\ref{fig:global_level_hist} in logarithmic scale.

The log-scale histograms reveal two different scaling regimes of the level density. At high energies with $\log\lr{E} \gtrsim 5$ (or, equivalently, $E \gtrsim 150$), $\frac{d n\lr{E}}{d \, \log\lr{E}}$ appears to be almost constant up to the sharp UV cutoff. This corresponds to the density of energy levels $\frac{d n\lr{E}}{d E} \sim E^{-1}$ that does not contain any dimensionful parameter.

On the other hand, at low energies all the histogram plots have an almost constant slope in our double log scale, which corresponds to the power-law scaling of the form $\frac{d n\lr{E}}{d \log\lr{E}} \sim E^{\alpha} = e^{\alpha \, \log\lr{E}}$, or, equivalently, $\log\lr{\frac{d n\lr{E}}{d \log\lr{E}}} = \alpha \, \log\lr{E}$. It is instructive to compare our results with the scaling law
\begin{eqnarray}
\label{DoS_scaling_free_1D}
 \frac{d n\lr{E}}{d \log\lr{E}} = E \, \frac{d n\lr{E}}{d E} \sim \sqrt{E}
\end{eqnarray}
for the one-dimensional free Hamiltonian (\ref{H1D}), where $d n \sim d p \sim d \sqrt{E}$. On the other hand, for the two-dimensional free Hamiltonian (\ref{H2D}) the number of states is $d n \sim 2 \pi p \, dp \sim d E$, which leads to the linear scaling
\begin{eqnarray}
\label{DoS_scaling_free_2D}
 \frac{d n\lr{E}}{d \log\lr{E}} = E \, \frac{d n\lr{E}}{d E} \sim E .
\end{eqnarray}

\begin{figure*}[h!tpb]
  \centering
  \includegraphics[width=0.195\textwidth]{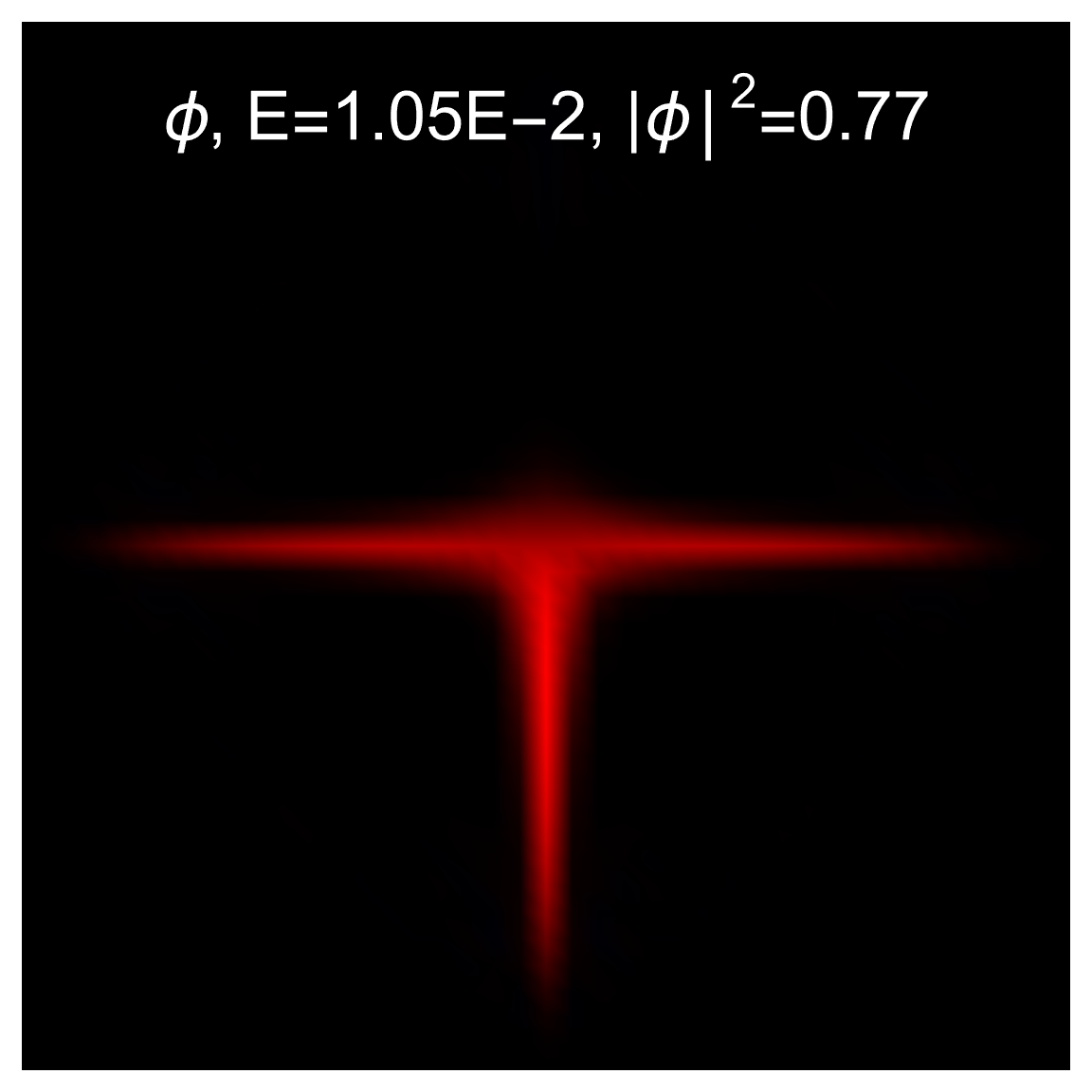}
  \includegraphics[width=0.195\textwidth]{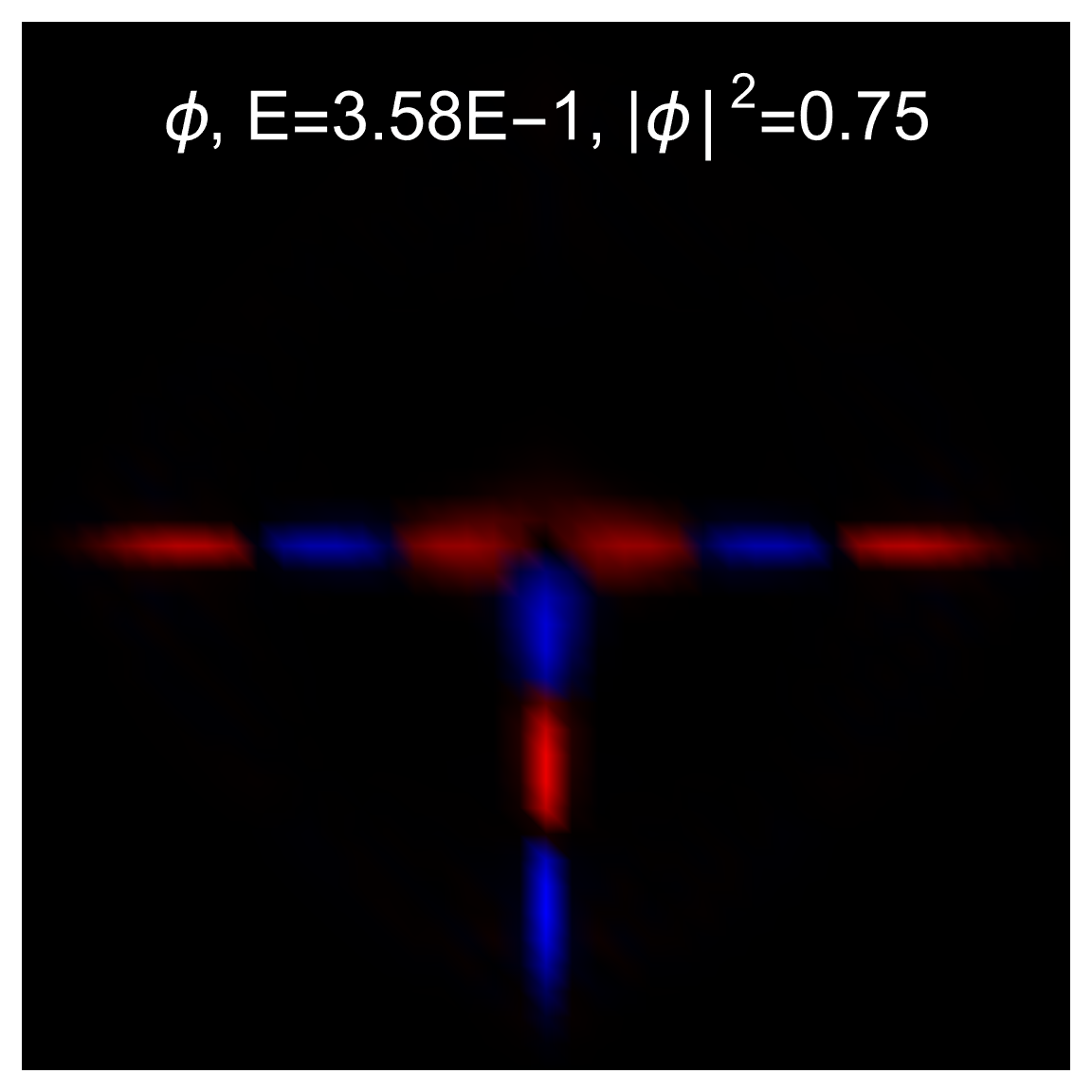}
  \includegraphics[width=0.195\textwidth]{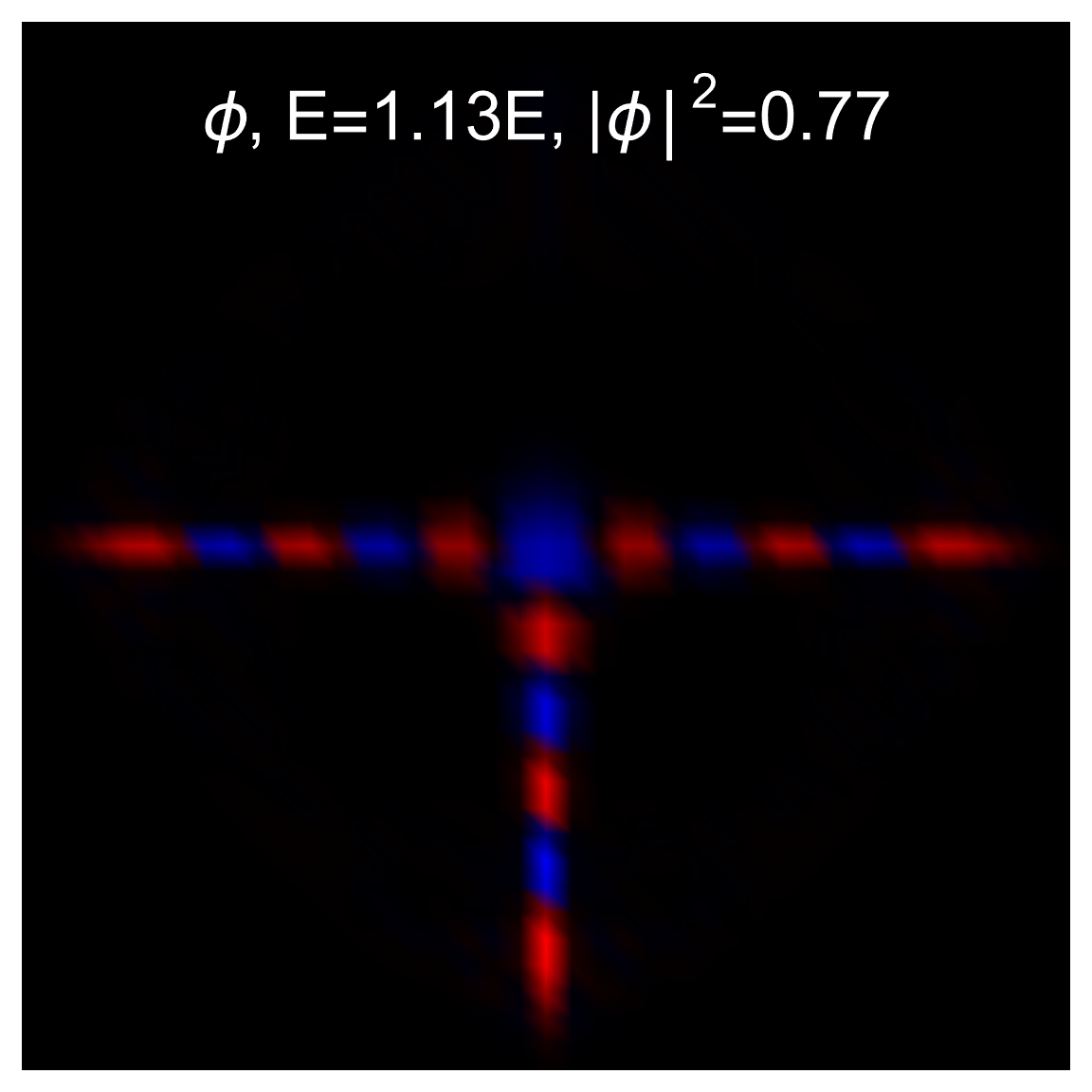}
  \includegraphics[width=0.195\textwidth]{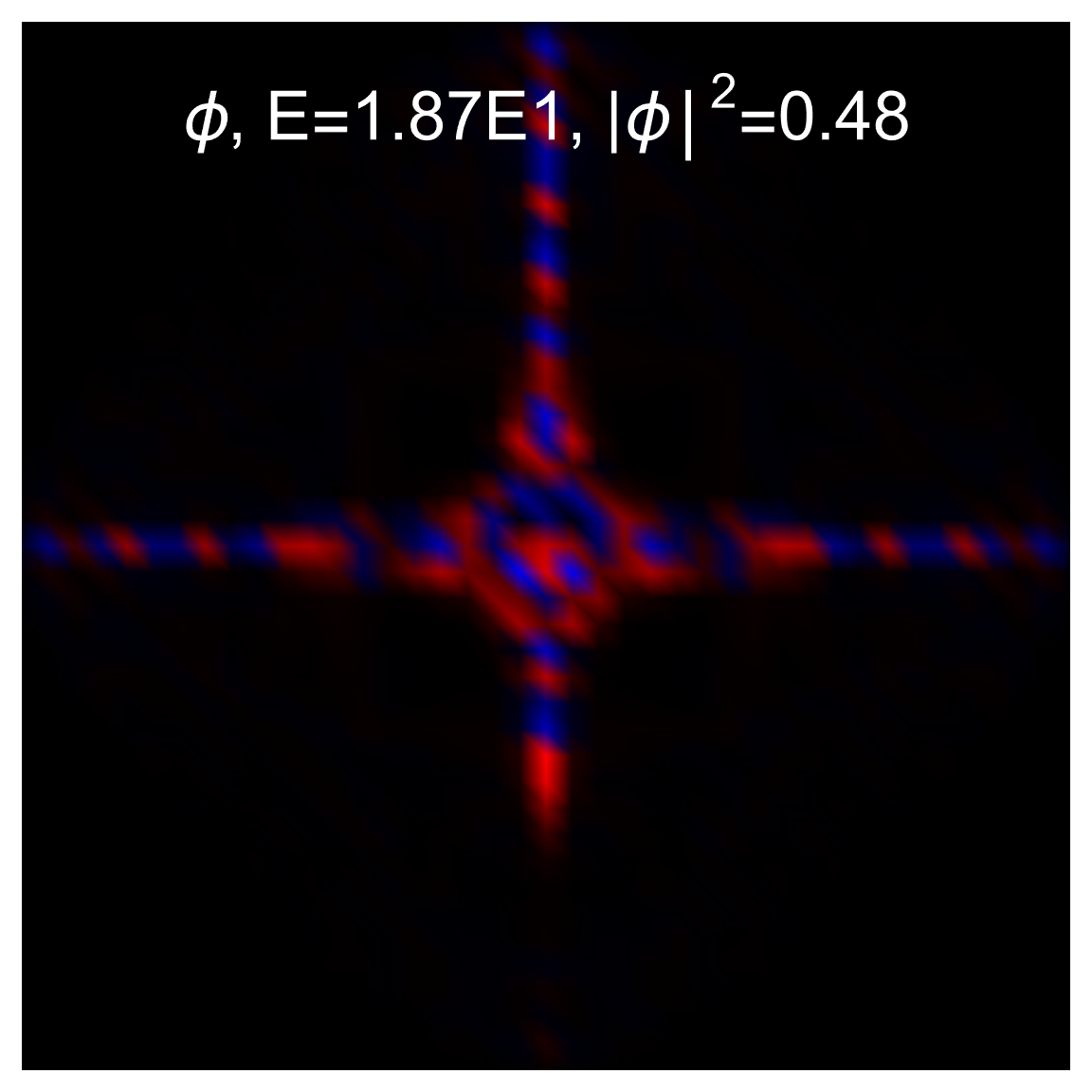}
  \includegraphics[width=0.195\textwidth]{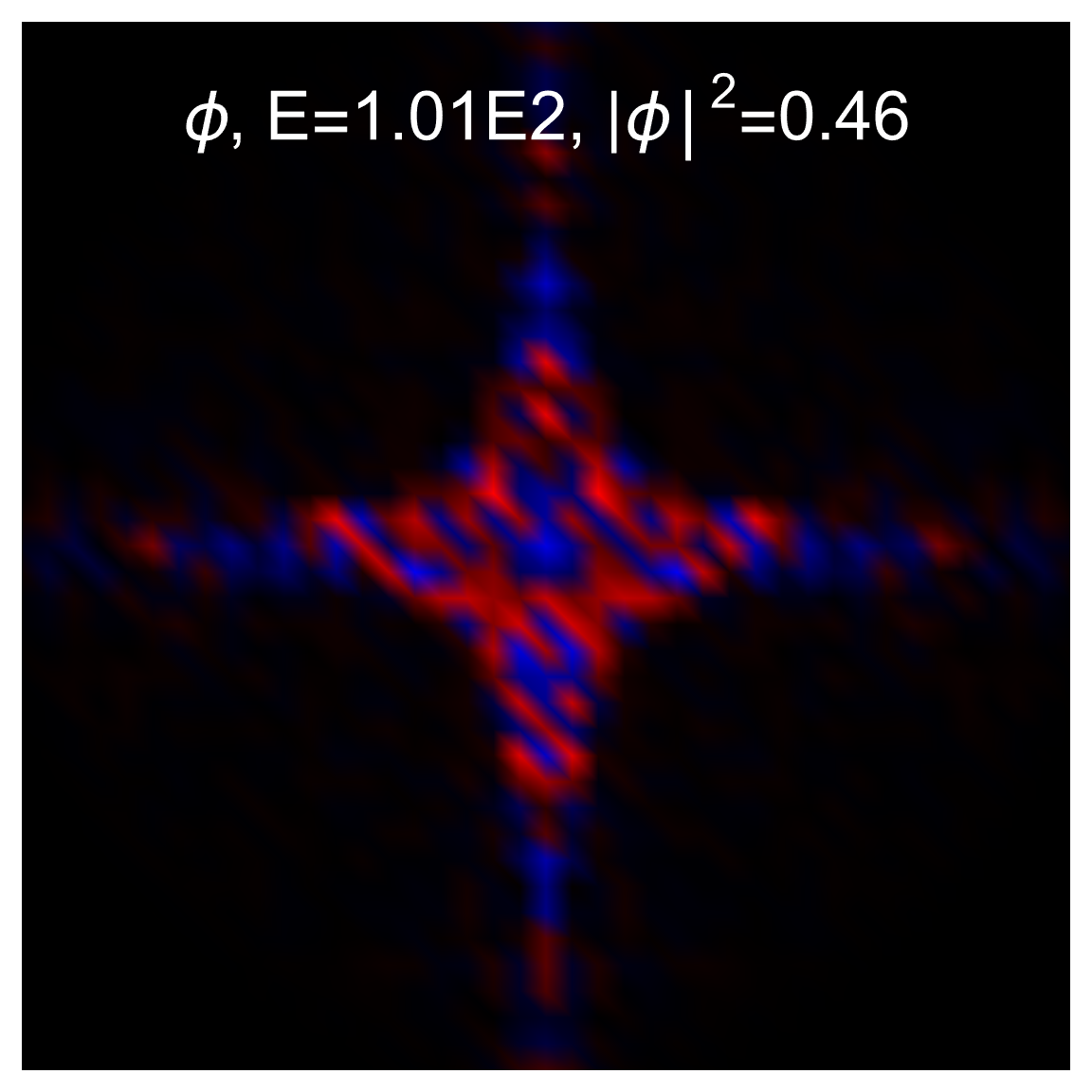}\\
  \includegraphics[width=0.195\textwidth]{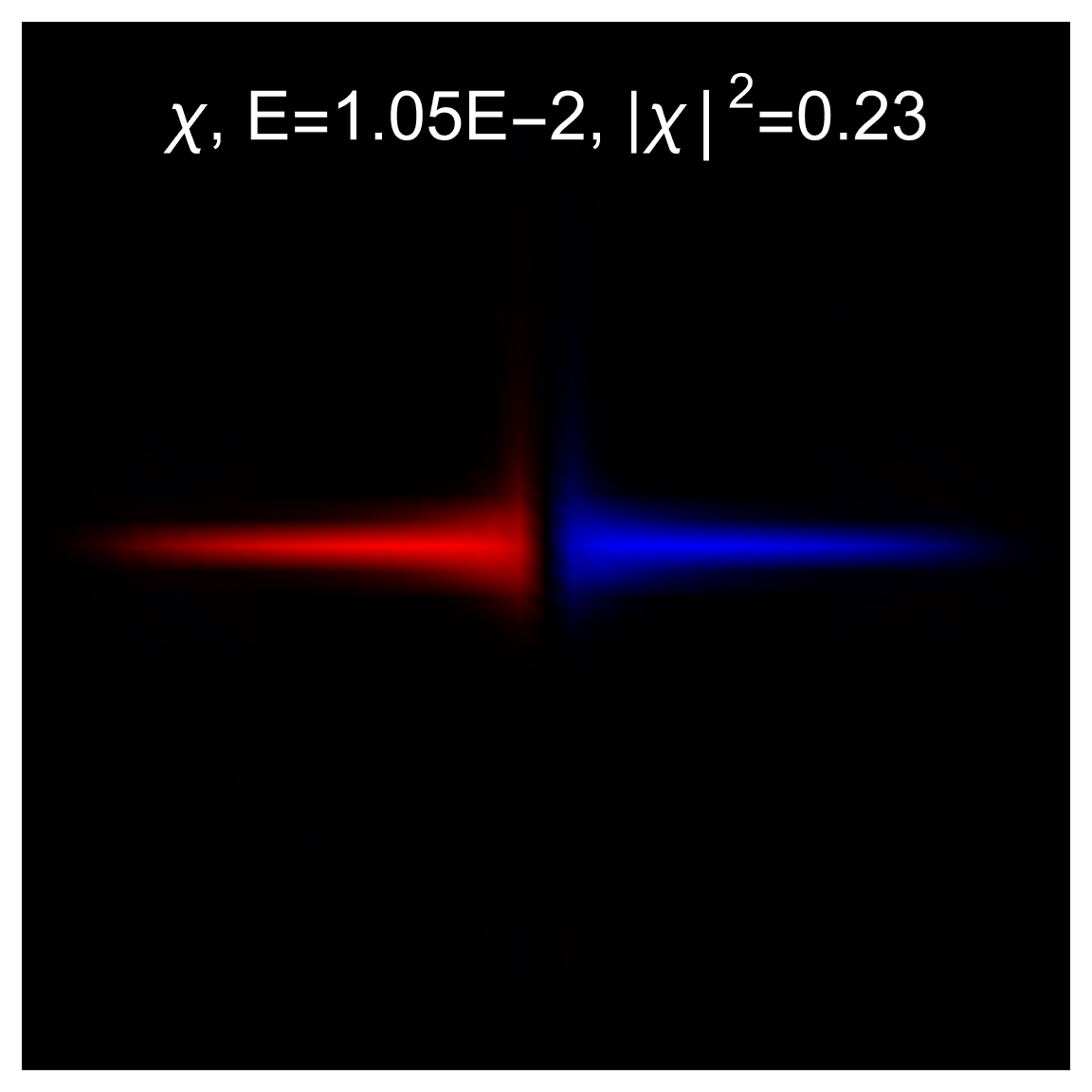}
  \includegraphics[width=0.195\textwidth]{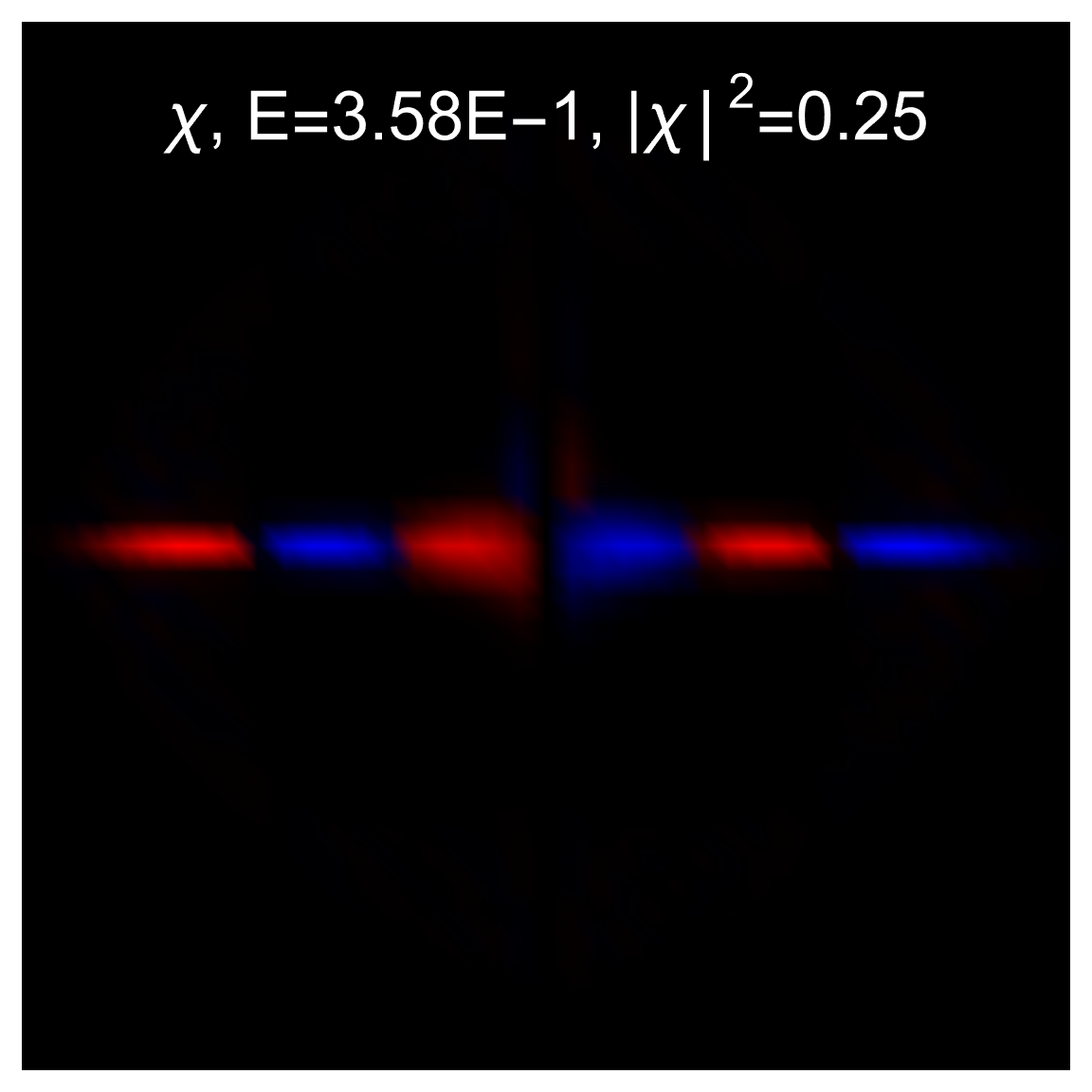}
  \includegraphics[width=0.195\textwidth]{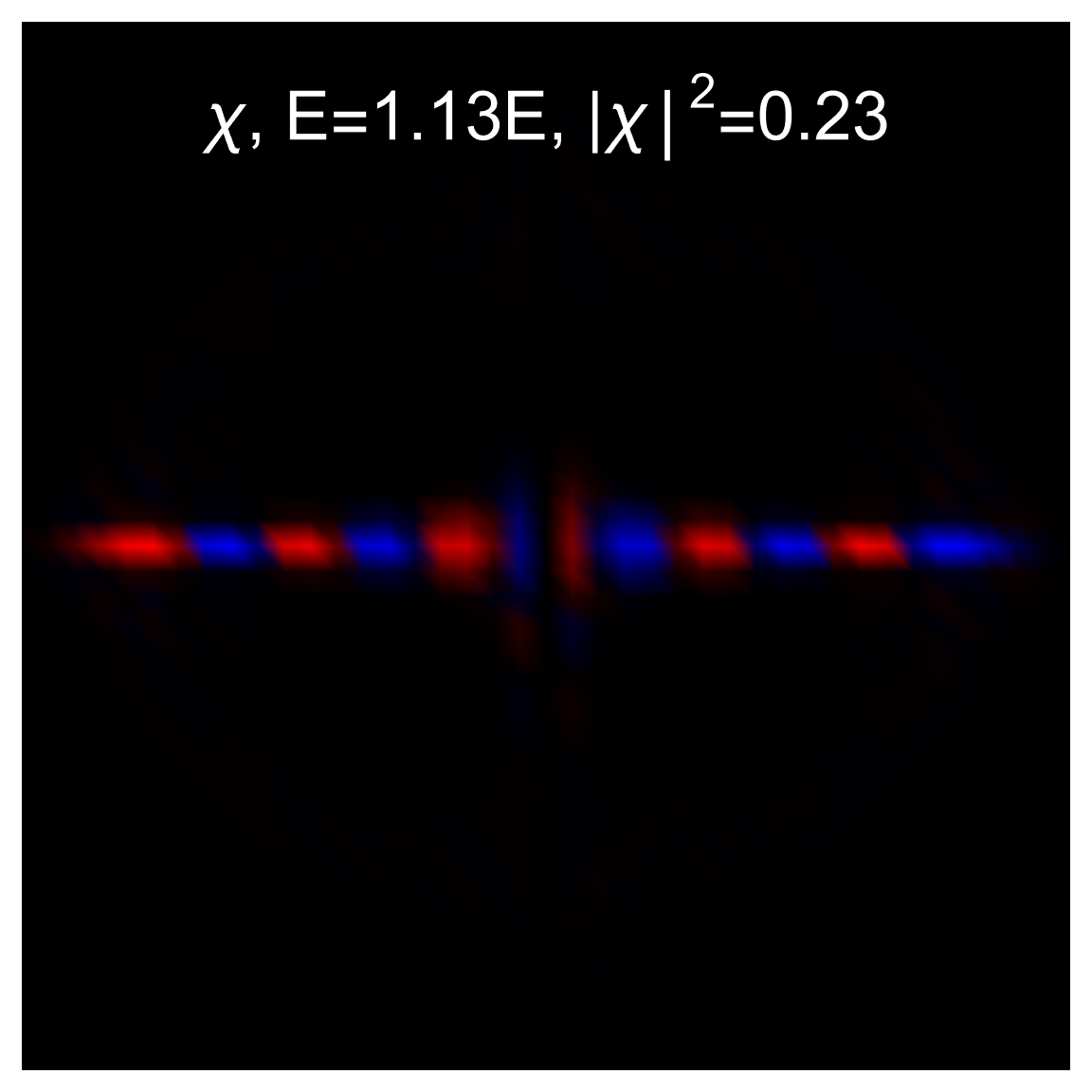}
  \includegraphics[width=0.195\textwidth]{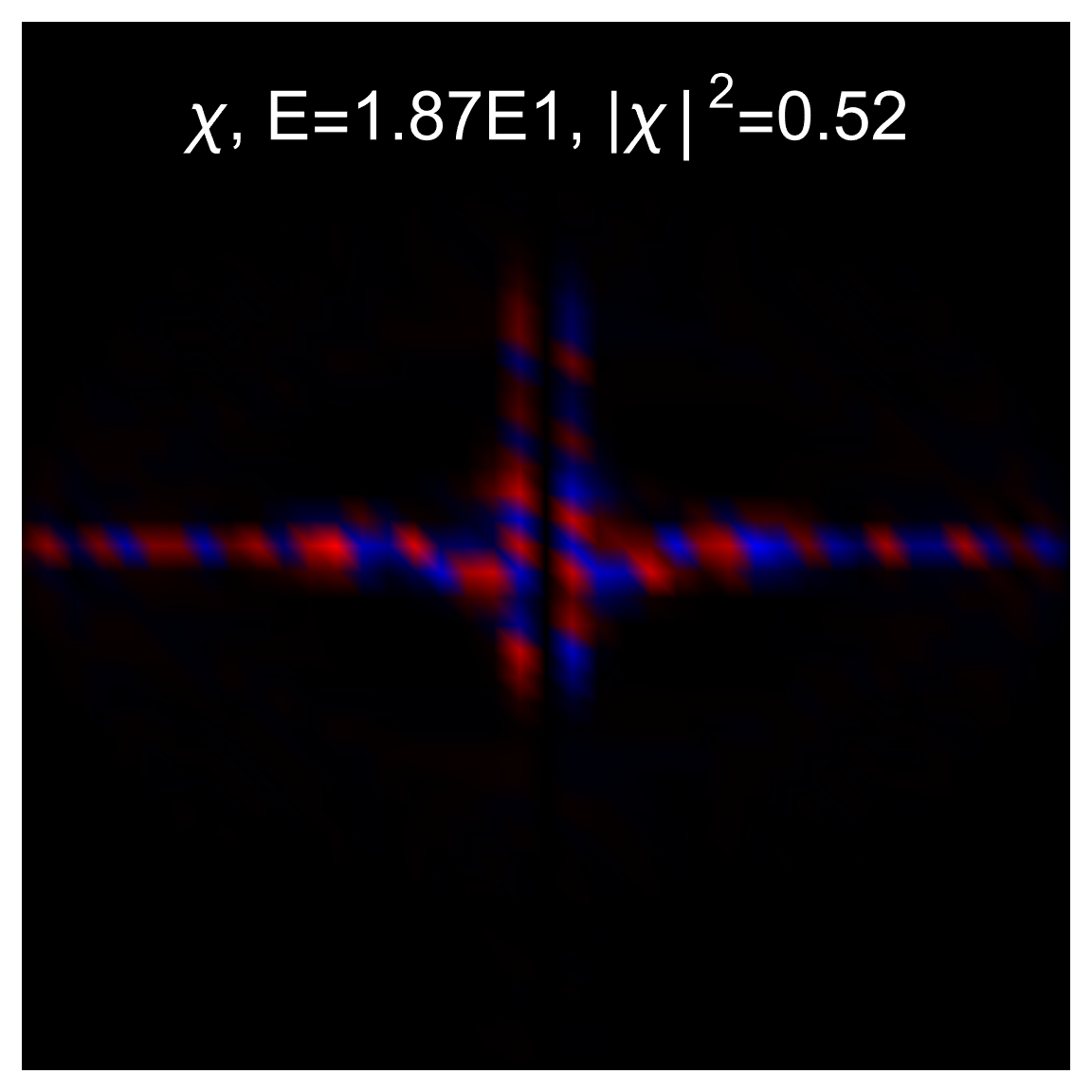}
  \includegraphics[width=0.195\textwidth]{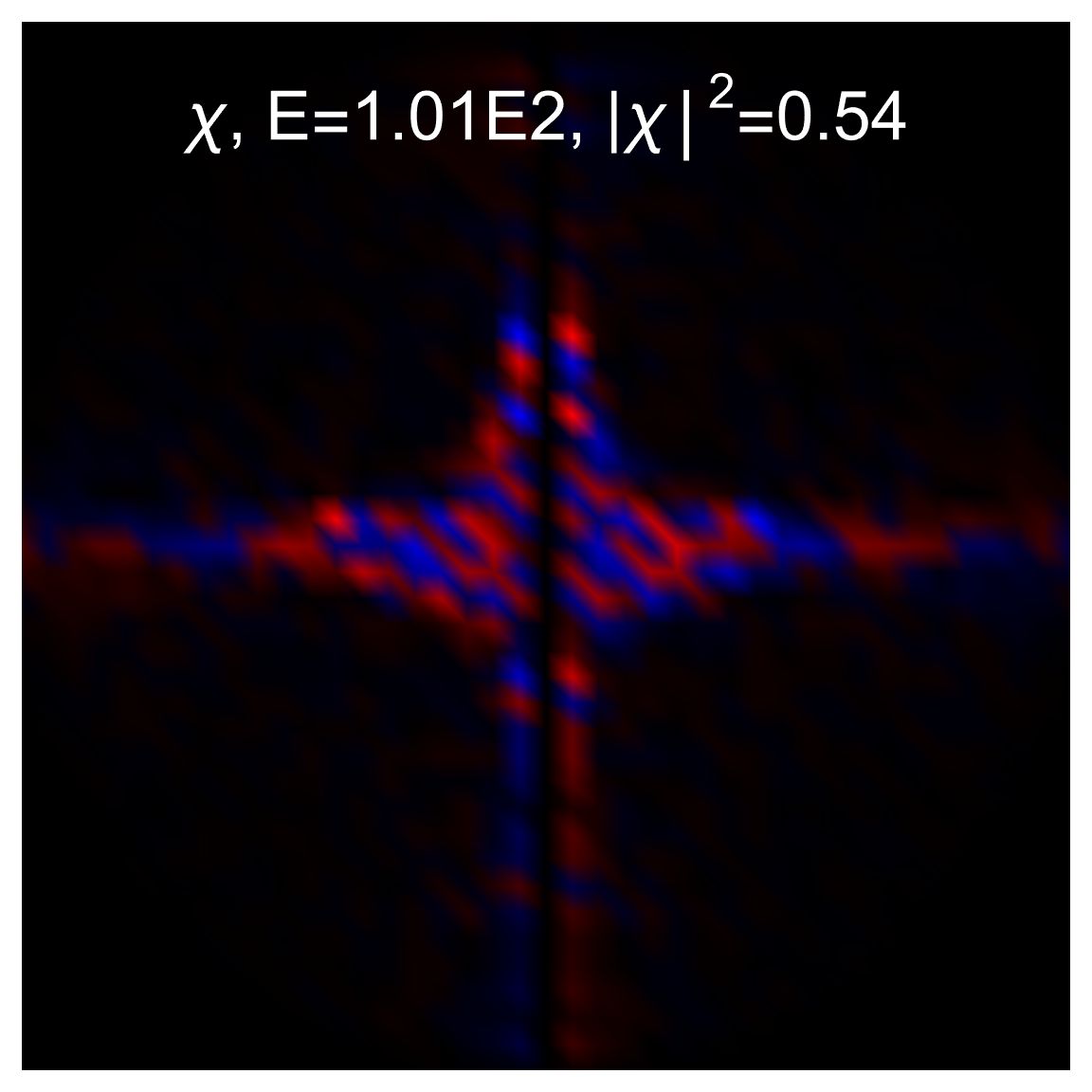}\\
  \caption{Density plots of the wave functions of some of the eigenstates of the supersymmetric Hamiltonian (\ref{HS}), with the corresponding values of energy given in the plot labels. We use truncation parameter $M = 70$ for these plots. For each plot, the color scale is adjusted such that the brightest color corresponds to the maximal absolute value of the wave function. Black background corresponds to zero wave function. Red and blue regions correspond to positive and negative wave function values. The square region for all plots is $20 \leq x_1 \leq 20 $, $20 \leq x_2 \leq 20 $. The plots in the top and in the bottom rows correspond to the components $\phi\lr{x_1, x_2}$ and $\chi\lr{x_1, x_2}$ of the two-component wave function $\Psi\lr{x_1, x_2} = \lrc{\phi\lr{x_1, x_2}, \chi\lr{x_1, x_2}}$. The squared norms $|\phi\lr{x_1, x_2}|^2$ and $|\chi\lr{x_1, x_2}|^2$ are given in the plot labels.}
  \label{fig:evec_density_plots}
\end{figure*} 

From Fig.~\ref{fig:global_level_hist} we can see that for small $E$ the slopes of the histogram plots for the supersymmetric Hamiltonian are quite close to that for the free one-dimensional Hamiltonian (\ref{H1D}) with $M = 100$ and $M = 800$. As $M$ is increased, in both cases the histograms are shifted towards lower energies. This is our first argument in favour of effectively one-dimensional, gapless structure of low-lying eigenstates of the supersymmetric model. For the bosonic Hamiltonian, low-lying energy levels remain discrete even for $M \rightarrow \infty$. Correspondingly, the histogram is shifted towards larger $E$ and has a steeper slope.

To get further insights into the structure of eigenstates of the supersymmetric Hamiltonian at different energy scales, on Fig.~\ref{fig:evec_density_plots} we show density plots of some of the wave functions obtained with the truncation parameter $M = 70$. In the plot labels, we give the corresponding values of energy as well as the squared norms $|\phi\lr{x_1, x_2}|^2$ and $|\chi\lr{x_1, x_2}|^2$ of the two components of the wave function $\Psi\lr{x_1, x_2} = \lrc{\phi\lr{x_1, x_2}, \chi\lr{x_1, x_2}}$. Both squared norms add up to one by virtue of normalization. We see that the low-energy states are indeed strongly localized along the flat directions $x_1 = \pm 0$, $x_2 = \pm 0$ of the potential energy. The number of times the wave functions change sign along the direction of their maximal extent coincides with the serial number of the energy level, which is also an argument in favour of effectively one-dimensional structure.

It might seem that the ``T''-shape structure of the functions $\phi\lr{x_1, x_2}$ violates the parity symmetry of the model. However, here we only show the results for the eigenstates with positive $x_1$ parity (see Appendix~\ref{apdx:harmonic_basis}). Eigenstates with negative $x_1$ parity will have the ``T'' shape turned upside down, so that the full parity symmetry will be restored in the sum over both parity sectors. It turns out that the OTOCs of operators $\hat{x}_2$ and $\hat{p}_2$ take exactly the same values in both $x_1$ parity sectors, and henceforth we only work with eigenstates of the supersymmetric Hamiltonian with positive $x_1$ parity.

\begin{figure}[h!tpb]
  \centering
  \includegraphics[width=0.48\textwidth]{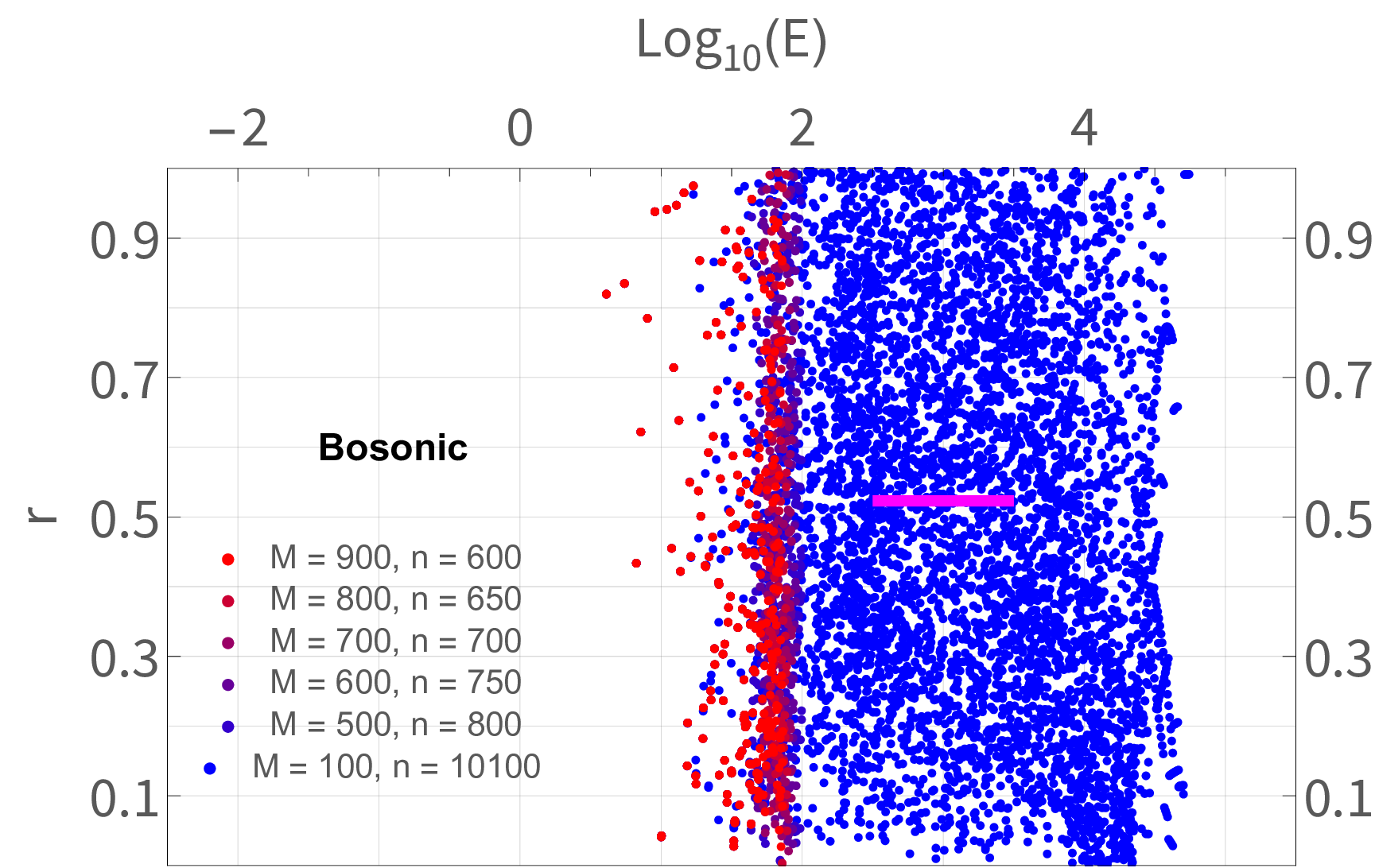}\\
  \includegraphics[width=0.48\textwidth]{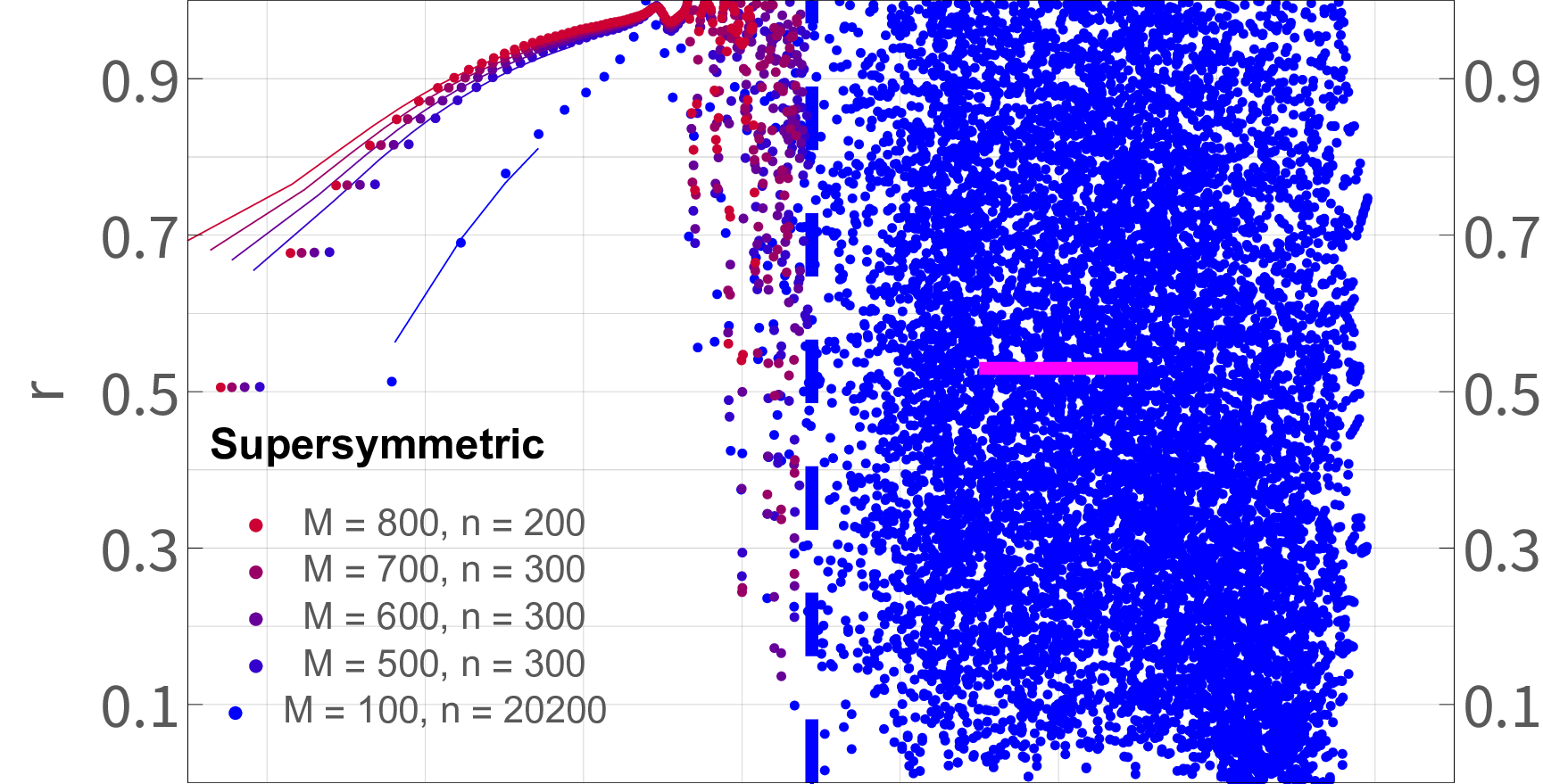}\\
  \includegraphics[width=0.48\textwidth]{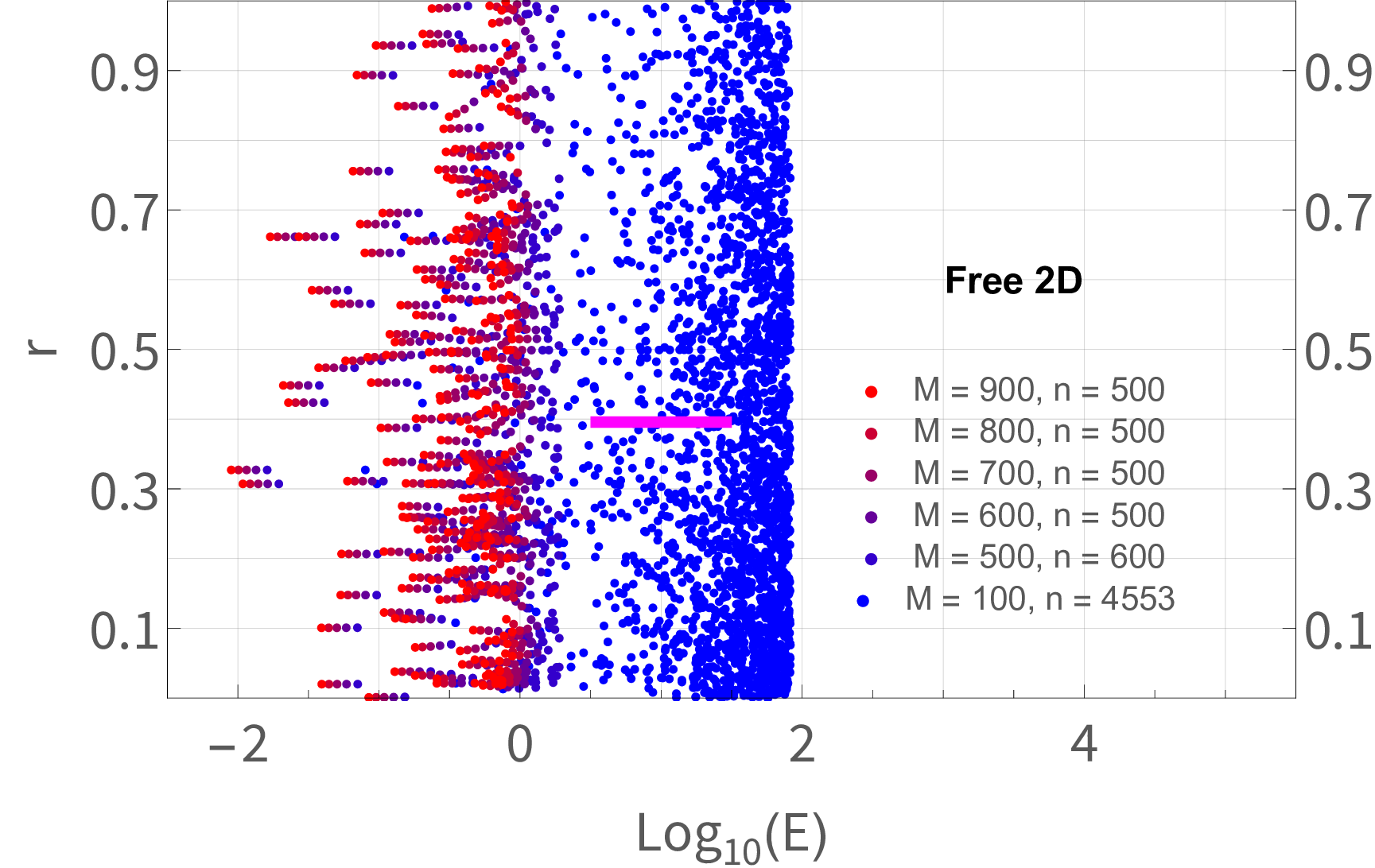}\\
  \caption{Scatter plots of the ratios $r_i$, defined as in (\ref{r_ratio_def}), as functions of the energy $E_i$. From top to bottom, we compare the results for the bosonic, supersymmetric, and free two-dimensional Hamiltonians (\ref{HB}), (\ref{HS}) and (\ref{H2D}), respectively. For the truncation parameter $M = 100$, all energy levels are shown. For $M \geq 500$, we only show between $200$ and $800$ lowest energy levels calculated using the Arnoldi algorithm. For the supersymmetric case, vertical dashed line shows the energy above which the irreps of $D_{4d}$ group associated with each energy level become irregularly ordered, and thin curved lines of the same color as data points correspond to the formula (\ref{r_ratio_1D}) with best-fit parameters $c$. For $M = 100$, horizontal magenta lines show the values of $r_i$ averaged over the energy window given by the line extent. }
  \label{fig:r_ratio}
\end{figure} 

To get further insights into quantum chaos exhibited by the bosonic and supersymmetric Hamiltonians (\ref{HB}) and (\ref{HS}), it is useful to analyze microscopic correlations between energy levels. Quantum chaos is usually characterized by the repulsion between adjacent energy levels. This repulsion forces the energy level spacings $\Delta E_i = E_{i+1} - E_i$ to have one of the few universal statistical distributions, which correspond to ensembles of Gaussian unitary, orthogonal, or symplectic random matrices. For Hamiltonians without any quenched disorder, such analysis assumes that energy levels far away from the edges of the spectrum can be considered as quasi-random quantities belonging to a statistical ensemble.

A convenient measure of level repulsion associated with quantum chaos is the $r$-ratio \cite{Huse:cond-mat/0610854,Luitz:1411.0660}
\begin{eqnarray}
\label{r_ratio_def}
 r_i  = \frac{\min\lr{\Delta E_{i-1}, \Delta E_i}}{\max\lr{\Delta E_{i-1}, \Delta E_i}} .
\end{eqnarray}
For the Gaussian Orthogonal Ensemble (GOE) of random real symmetric matrices, statistical average of $r_i$ over many energy levels in the bulk of the spectrum (or, equivalently, over many random matrices) is $r_{GOE} = 0.53$. On the other hand, for non-chaotic integrable systems the energy levels are typically uncorrelated. Correspondingly, the number of energy levels within a fixed interval is usually well described by Poisson distribution \cite{BerryTaborLevelClustering}, and the average of $r_i$ over many energy levels is close to $r_{Poisson} = 0.39$.

For Hamiltonians that are invariant under a nontrivial symmetry group, the level spacings $\Delta E_i = E_{i+1} - E_i$ that enter the $r$-ratio (\ref{r_ratio_def}) should be the differences between consecutive energy levels that correspond to eigenstates transforming under the same irreducible representation (irrep) of the symmetry group. In other words, we make an ordered list of eigenstates that transform under some fixed irrep of the symmetry group, label the elements of this list by consecutive integer indices $i$, and calculate the $r$-ratio for the elements of this list.

As discussed in detail in Appendix~\ref{apdx:symmetries}, the bosonic Hamiltonian is invariant under a finite non-Abelian group $C_{4v}$ with four Abelian and one non-Abelian irreducible representations (irreps) \cite{GelessusC4v}. The supersymmetric Hamiltonian is invariant under a larger finite non-Abelian group $D_{4d}$ with four Abelian and three non-Abelian irreps \cite{GelessusD4d}. For our analysis of the $r$-ratio for the bosonic Hamiltonian (\ref{HB}), we select the energy levels that transform under the two-dimensional non-Abelian irrep $\mathcal{E}_0$ of $C_{4v}$. For the supersymmetric Hamiltonian (\ref{HS}), the eigenstates belong to one of the two-dimensional non-Abelian irreps $\mathcal{E}_1$ or $\mathcal{E}_2$ of $D_{4d}$, so we pick the states that transform under $\mathcal{E}_1$. For comparison, we also calculate the $r$-ratio for the free two-dimensional Hamiltonian (\ref{H2D}) with the same Hilbert space truncation. While the symmetry group of this free Hamiltonian is the full $O\lr{2}$ group, the energy levels can still be classified according to irreps of $C_{4v}$ which is a subgroup of $O\lr{2}$. To calculate the $r$-ratio in this case, we use energy levels that transform under irrep $\mathcal{E}_0$ of $C_{4v}$.

Scatter plots of $r_i$ versus the energy $E_i$ are shown on Fig.~\ref{fig:r_ratio}. For the truncation parameter $M = 100$, we show all eigenvalues of the Hamiltonian matrices (\ref{HB_matrix}) and (\ref{HS_matrix}). For $M \geq 500$, we use between $200$ and $800$ smallest eigenvalues of the Hamiltonian matrices (\ref{HB_matrix}) and (\ref{HS_matrix}) obtained using the Arnoldi algorithm.

To set the stage, we first discuss the results for the purely bosonic Hamiltonian (\ref{HB}), shown in the top plot on Fig.~\ref{fig:r_ratio}. In this case, lowest energy levels that transform under $\mathcal{E}_0$ irrep are of order of $10^1$, and for energies $E \gtrsim 10^2$ the $r$ ratio is fluctuating almost randomly between $0$ and $1$, filling the entire plot area almost uniformly. This is the expected behavior for a system that is chaotic in this energy range, both in quantum and in classical mechanics. Averaging $r_i$ over many energy levels in the window between $E = 10^{2.5} \approx 316.2$ and $E = 10^{3.5} \approx 3162.3$, we obtain the value $\bar{r} = 0.523 \pm 0.006$ that is very close to the universal value $r_{GOE} = 0.53$ for the Gaussian Orthogonal Ensemble. This expectation value as well as the extent of the energy window for which it was obtained are shown on Fig.~\ref{fig:r_ratio}. In full agreement with previous studies \cite{Haller:PRL1984,Akutagawa:2004.04381}, we therefore conclude that the energy spectrum of the bosonic Hamiltonian (\ref{HB}) exhibits chaotic behavior at least for energies of order $E \gtrsim 10^1 \ldots 10^2$.

\begin{figure}
  \centering
  \includegraphics[width=0.48\textwidth]{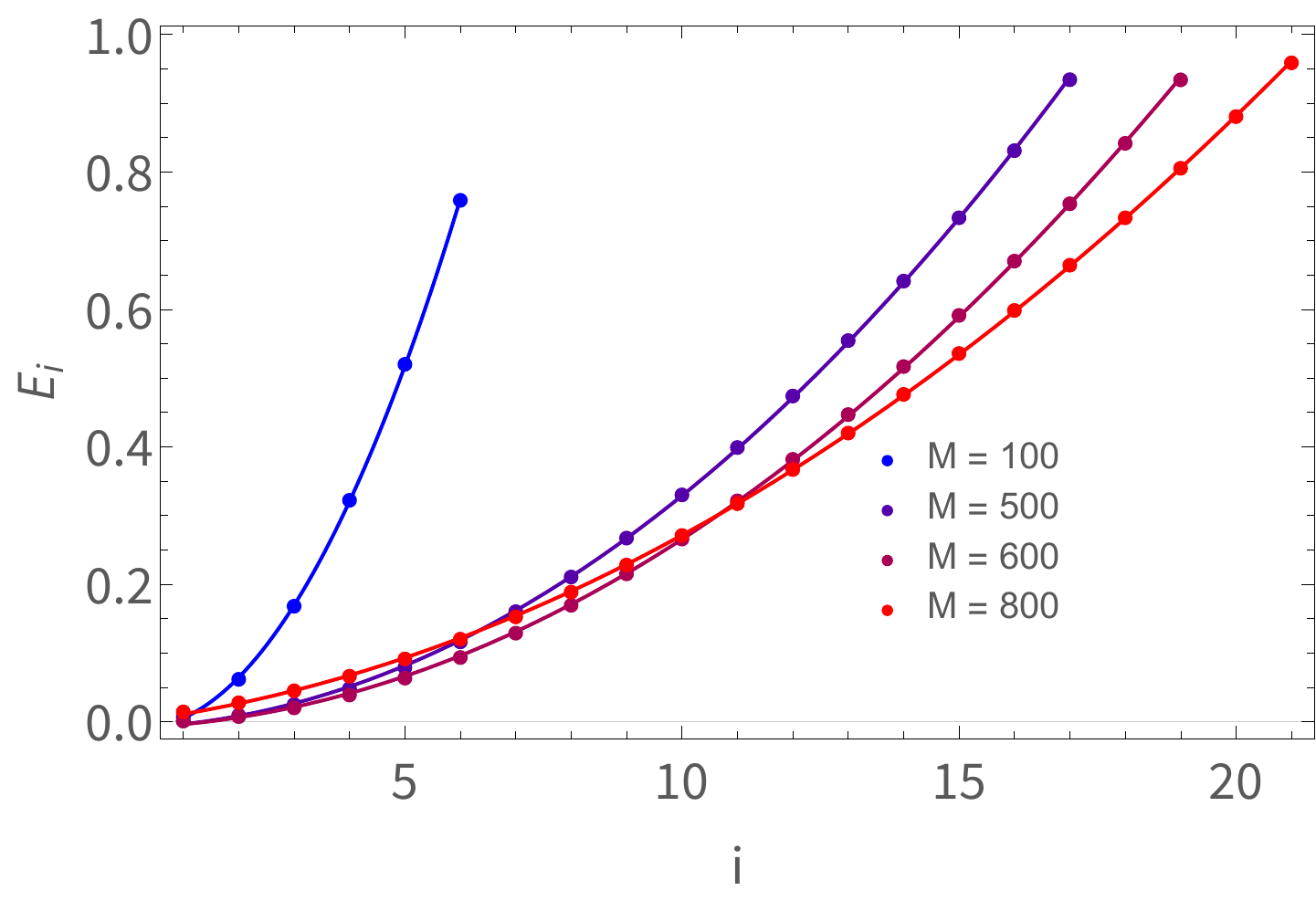}
  \caption{Low-lying energy levels $E_i$ of the supersymmetric Hamiltonian (\ref{HS}) as a function of their serial number $i$ for different values of the truncation parameter $M$. Only eigenstates that transform under the irrep $\mathcal{E}_1$ of the symmetry group $D_{4d}$ of the Hamiltonian and have $E_i \leq 1$ are considered. Solid lines are fits of the form $E_i = a + b \, \lr{i + c}^2$.}
  \label{fig:susyqm_low_energy_states}
\end{figure} 

For energies $E \gtrsim 10^2$, the behavior of the $r$-ratio for the supersymmetric Hamiltonian (\ref{HS}) appears to be very similar to the one for the bosonic Hamiltonian (\ref{HB}). Namely, the $r$ ratio also fluctuates randomly between $0$ and $1$, averaging to $\bar{r} = 0.530 \pm 0.004$ in the same window of energies $E = 10^{2.5} \ldots 10^{3.5}$ that we considered for the bosonic Hamiltonian. The behavior for the low-energy part of the spectrum is, however, completely different. There is a family of low-lying energy levels with $E \lesssim 10^{0.5} \approx 3.2$ for which the $r$-ratio $r_i$ behaves in a smooth and regular way as a function of energy $E_i$, rising from $r_i \approx 0.5$ up to $r_i \approx 1.0$. This behavior is observed in the energy range that agrees well with the extent of the low-energy ``tail'' with $\frac{d n}{d \log\lr{E}} = \sqrt{E}$ in histograms of global energy level density in Fig.~\ref{fig:global_level_hist}. Interestingly, the change between the non-chaotic low-energy states and the chaotic high-energy states, where the $r$-ratio approaches unit value, appears to be quite sharp.

Such a smooth rising behavior of the $r$-ratio can be expected for finite-size one-dimensional systems. As we show on Fig.~\ref{fig:susyqm_low_energy_states}, the dependence of the energy levels $E_i$ on their serial number $i$ with a good precision can be described by a quadratic expression
\begin{eqnarray}
\label{low_energy_fit}
 E_i = a + b \, \lr{i + c}^2, \quad i = 0, 1, 2, \ldots ,
\end{eqnarray}
which leads to
\begin{eqnarray}
\label{r_ratio_1D}
 r_i = \frac{i + c + 1/2}{i + c + 3/2} .
\end{eqnarray}
Solid lines on the middle plot on Fig.~\ref{fig:r_ratio} correspond to the expression (\ref{r_ratio_1D}) plotted as a function of $E_i$ given by (\ref{low_energy_fit}). The parameters $a$, $b$, $c$ are extracted from the fits shown on Fig.~\ref{fig:susyqm_low_energy_states}. We find that the best fit parameters $b$ are with a good precision inversely proportional to the truncation parameter $M$. The parameter $a$ is very small, and the shift parameter $c$ quite quickly grows with $M$: $c = 0.79$ for $M = 100$, $c = 1.40$ for $M = 500$, and $c = 4.05$ for $M = 800$. For few lowest eigenstates, small deviations of data points that are hardly noticeable on Fig.~\ref{fig:susyqm_low_energy_states} get amplified (compare thin lines with low-energy data points on the middle plot on Fig.~\ref{fig:r_ratio}), and our fits appear to be not as good as for somewhat larger energies with $E_i < 1$. It is instructive to compare these findings with similar analysis for the free one-dimensional Hamiltonian (\ref{H1D}). In this case, $b$ is also inversely proportional to $M$, and the fit parameters $a \approx 0.25$ and $c \approx 1$ are almost independent of $M$. Another instructive case is a one-dimensional, infinitely deep potential well of width $L$, for which $E_i = \frac{\pi^2 \, \lr{i+1}^2}{L^2}$ and $r_i = \frac{2 i + 1}{2 i + 3}$ are completely independent of $L$. Strong dependence of the parameter $c$ on the infrared cutoff set by the parameter $M$ for the supersymmetric Hamiltonian (\ref{HS}) is quite different from these one-dimensional models. For one-dimensional models, this parameter is controlled by boundary conditions at the end points of the region to which one-dimensional motion is effectively confined.

It is interesting to note that for the supersymmetric system the change between the regular, quasi-one-dimensional spectrum and the chaotic, random-matrix-like spectrum can be also observed from the ordering of irreps under which the eigenstates transform. Namely, for the low-energy part of the spectrum, eigenstates with even and odd serial numbers transform under the non-Abelian irreps $\mathcal{E}_1$ and $\mathcal{E}_2$ of $D_{4d}$, respectively. For energies above some threshold level, this ordering is violated, and now and then there appear several eigenstates in a row that transform under one and the same irrep. For $M=100$, we show the position of the first ordering irregularity of this kind in the middle plot on Fig.~\ref{fig:r_ratio} with a vertical dashed thick line.

For comparison, on the plot in the bottom of Fig.~\ref{fig:r_ratio} we show the energy dependence of the $r$-ratio (\ref{r_ratio_def}) for the free two-dimensional Hamiltonian (\ref{H2D}), subject to the same truncation as the supersymmetric and bosonic Hamiltonians (\ref{HS}) and (\ref{HB}). Here all the energies are significantly smaller than for the interacting systems. While $r_i$ ratio does exhibit some quasi-random behavior towards the higher end of the spectrum, averaging $r_i$ over the energy window between $E = 10^{0.5} \approx 3.16$ and $E = 10^{1.5} \approx 31.6$, we obtain $\bar{r} = 0.40 \pm 0.01$, in good agreement with the expected value $r = 0.39$ for spectra of integrable systems with Poisson distribution of energy levels.

\section{Out-of-time-order correlators}
\label{sec:otocs}

\begin{figure*}[h!tpb]
  \centering
  \includegraphics[width=0.49\textwidth]{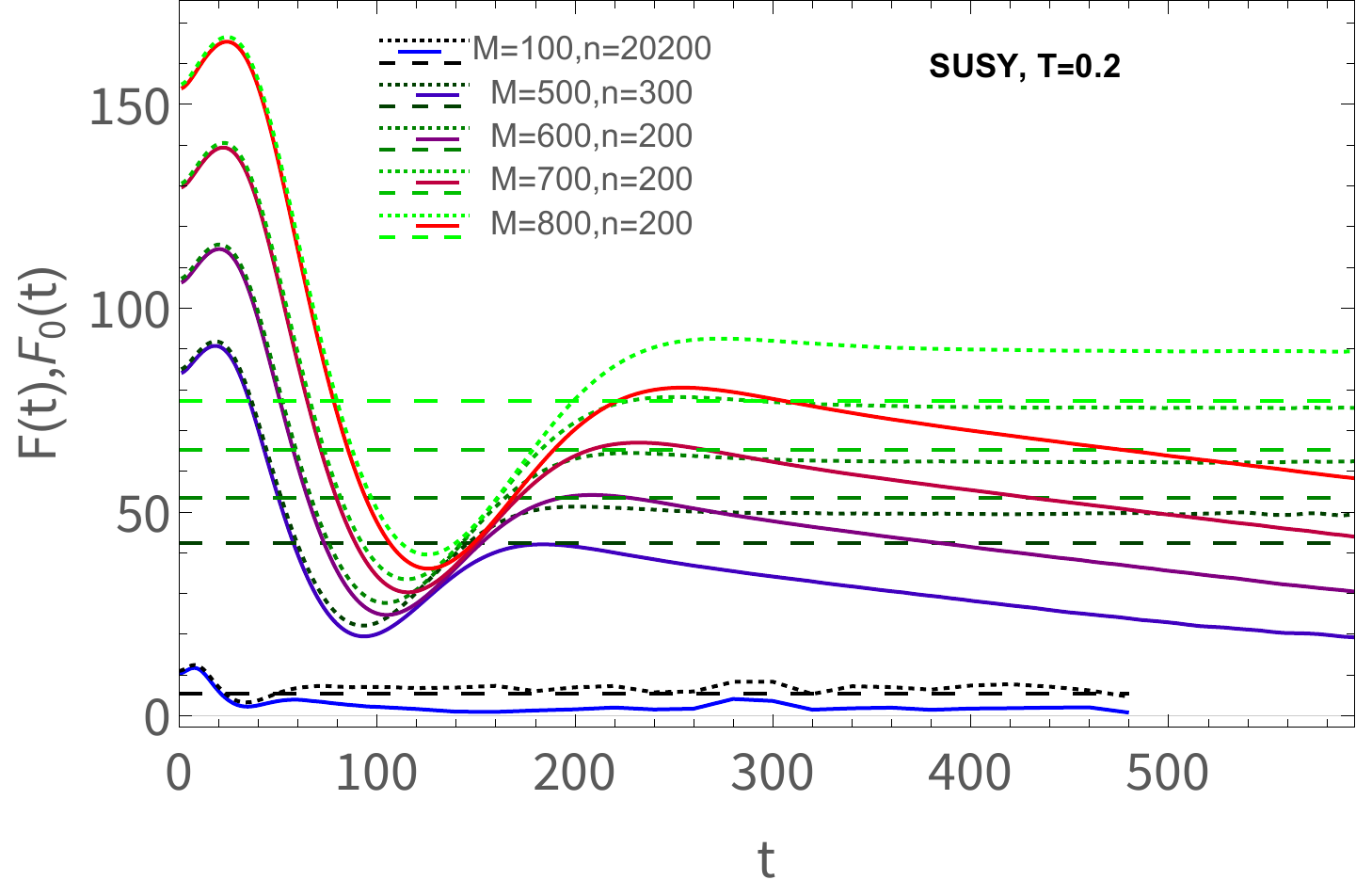}
  \includegraphics[width=0.49\textwidth]{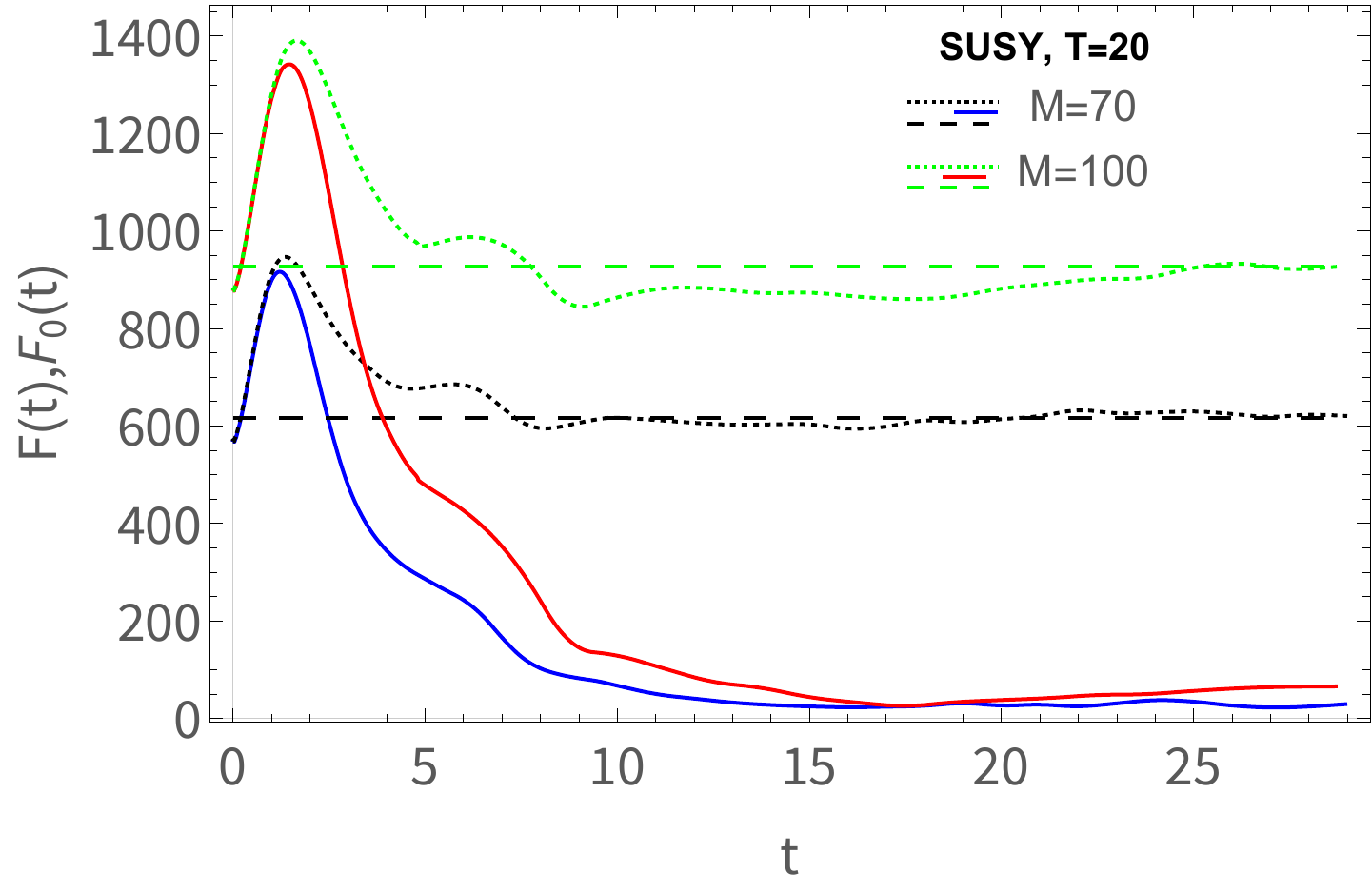}\\
  \includegraphics[width=0.49\textwidth]{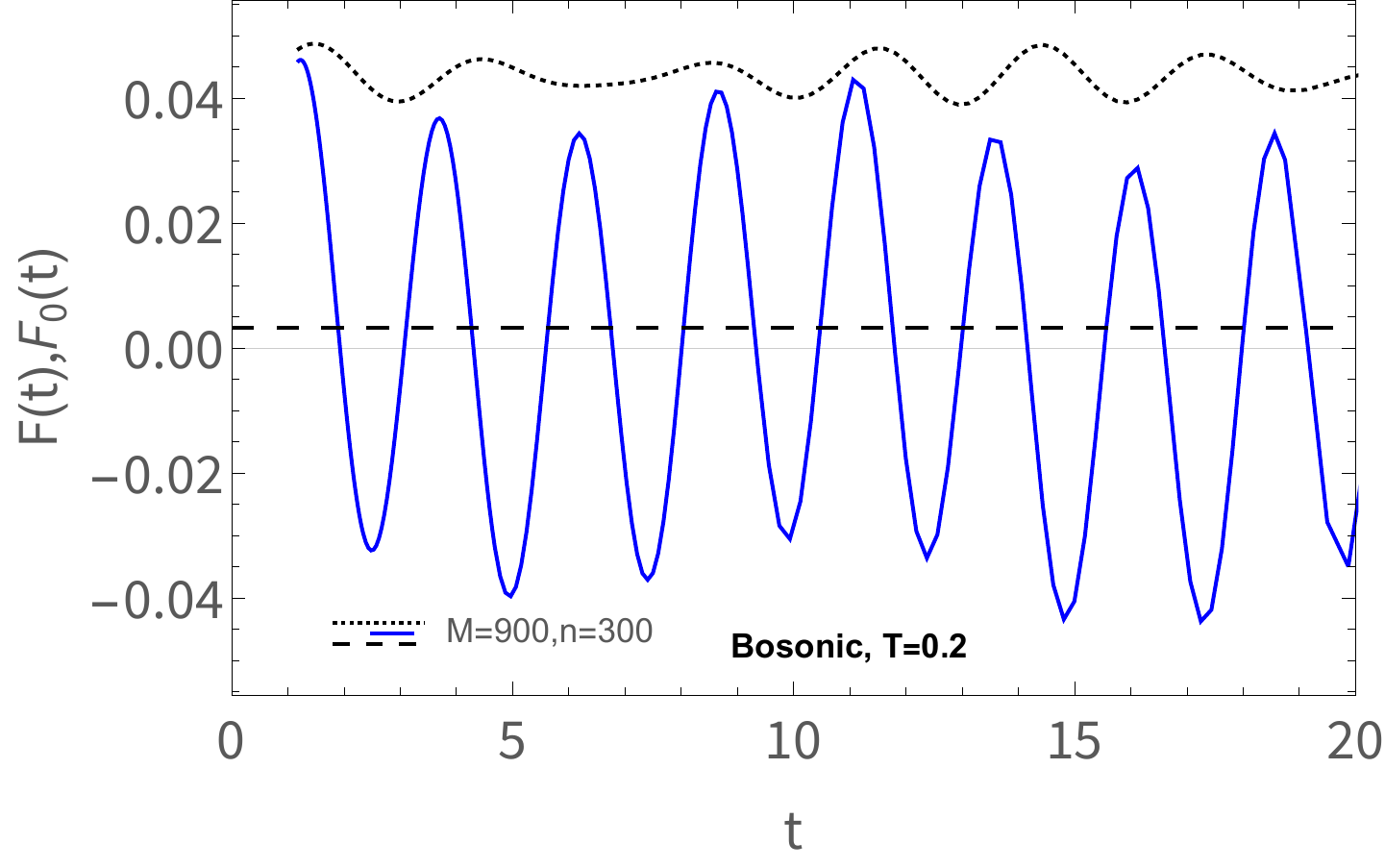}
  \includegraphics[width=0.49\textwidth]{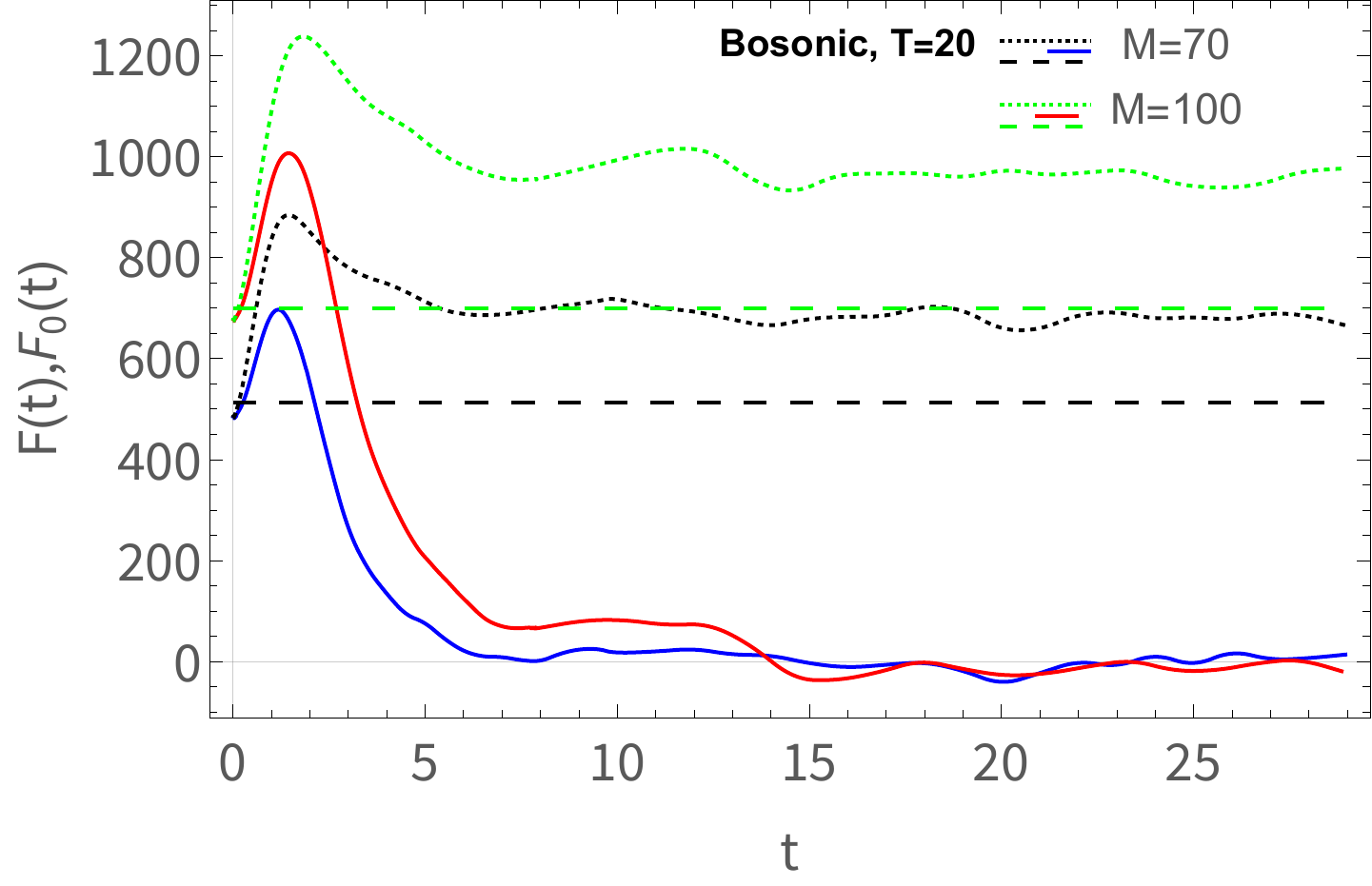}\\
  \includegraphics[width=0.49\textwidth]{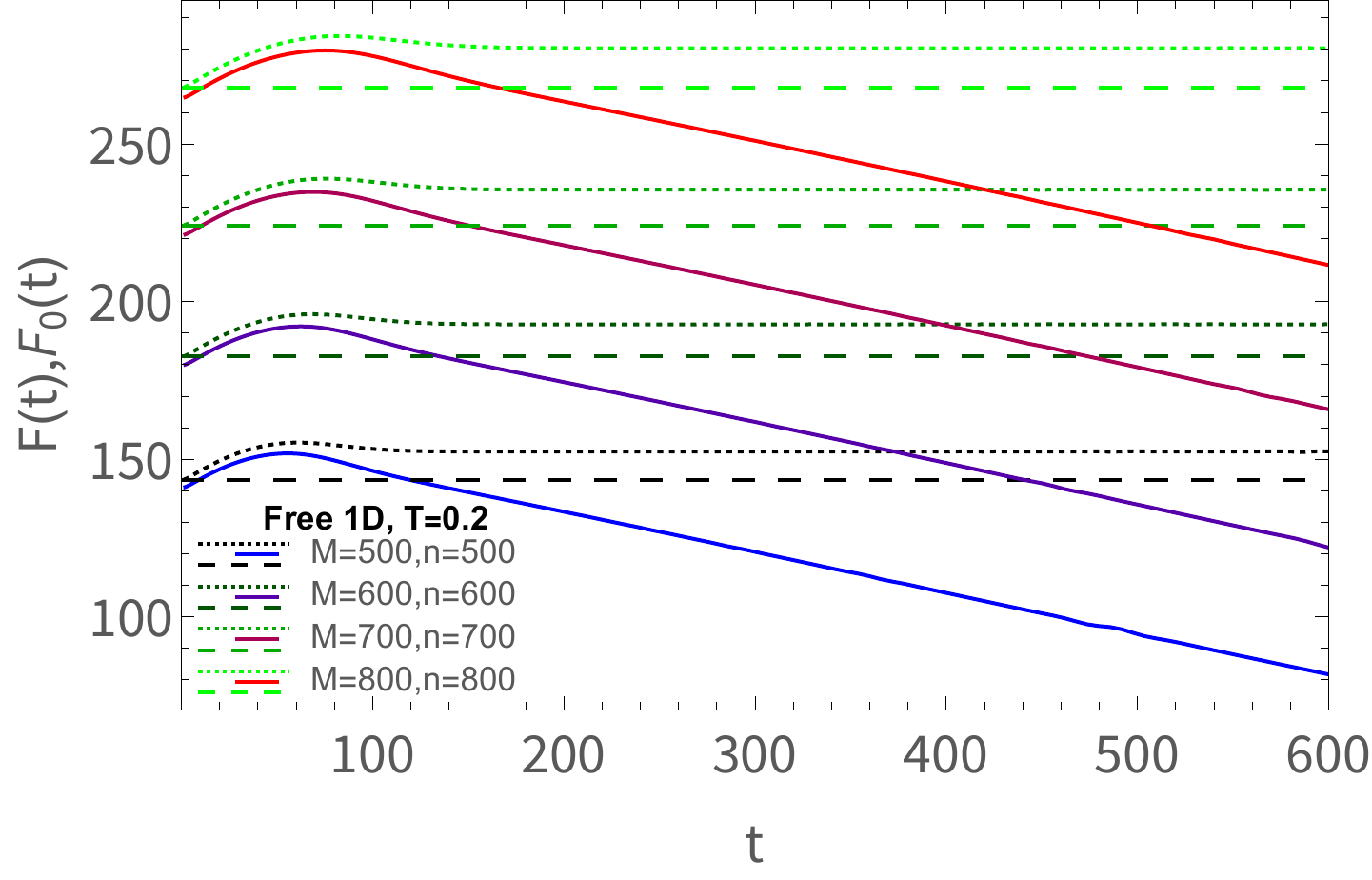}
  \includegraphics[width=0.49\textwidth]{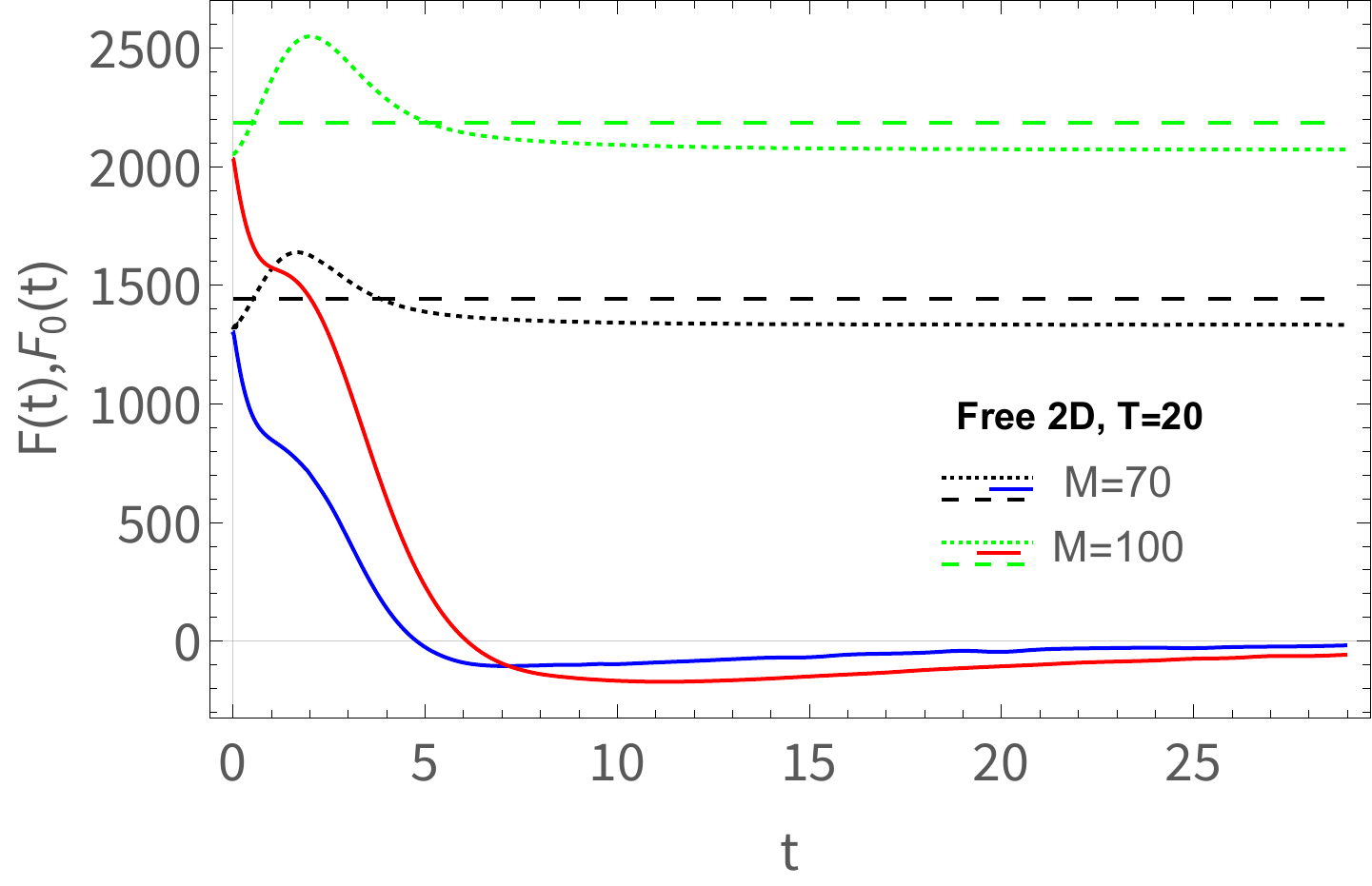}\\
  \caption{Time dependence of the functions $F\lr{t}$ (solid lines with colors changing from blue to red) and $F_0\lr{t}$ (dotted lines with colors changing from black to green), defined in (\ref{otoc_decomposition}), for different Hamiltonians and for different values of the truncation parameter $M$. Horizontal dashed lines with colors changing from black to green show the corresponding products of regularized thermal expectation values $2 \vev{\hat{x}_2^2}_r \vev{\hat{p}_2^2}_r = 2 \tr\lr{\hat{\rho}^{\frac{1}{2}} \, \hat{x}_2 \, \hat{\rho}^{\frac{1}{2}} \, \hat{x}_2 } \, \tr\lr{\hat{\rho}^{\frac{1}{2}} \, \hat{p}_2 \, \hat{\rho}^{\frac{1}{2}} \, \hat{p}_2 }$.}
  \label{fig:FF0}
\end{figure*} 

In this Section we consider out-of-time-order correlators of operators $\hat{x}_2\lr{t}$ and $\hat{p}_2\lr{0}$, which correspond to the classical Poisson brackets (\ref{class_Poisson}) defining the classical Lyapunov exponents\footnote{We use the operators $\hat{x}_2$ and $\hat{p}_2$ because they are invariant with respect to $x_1$ parity transformations (reflections of $x_1$). We use $x_1$ parity to block-diagonalize the matrix (\ref{HS_matrix}) of the supersymmetric Hamiltonian (\ref{HS}) and reduce the computational cost of exact diagonalization (see Appendix~\ref{apdx:harmonic_basis}). The OTOCs of the operators $\hat{x}_1$ and $\hat{p}_1$ are of course equivalent to that of $\hat{x}_2$ and $\hat{p}_2$, but are more complicated to calculate with our choice of the basis states.}. While the expression (\ref{OTOCs_def}) in the introductory Section is the most straightforward quantum generalization of Poisson brackets of the form (\ref{class_Poisson}), in practice it is more convenient to work with regularized OTOCs that have the same classical limit but exhibit less singular quantum behavior \cite{Maldacena:1503.01409}:
\begin{eqnarray}
\label{otoc_reg}
 C\lr{t}
 =
 -\tr\lr{ \lrs{\hat{\rho}^{\frac{1}{8}} \hat{x}_2\lr{t} \hat{\rho}^{\frac{1}{8}}, \hat{\rho}^{\frac{1}{8}} \hat{p}_2\lr{0} \hat{\rho}^{\frac{1}{8}}}^2 }  ,
\end{eqnarray}
where $\hat{\rho} = e^{-\beta \, \hat{H}}/Z$ is the thermal density matrix, $\hat{\rho}^{\frac{1}{8}} = e^{-\beta \, \hat{H}/8}/Z^{\frac{1}{8}}$ is its fractional power, and $Z = \tr \, e^{-\beta \, \hat{H}}$ is the thermal partition function. We will substitute either the supersymmetric, the bosonic, or the free one- or two-dimensional Hamiltonians for the abstract Hamiltonian $\hat{H}$ in (\ref{otoc_reg}).

To discuss the time dependence of OTOCs, it is also often convenient to represent $C\lr{t}$ as a difference of two contributions
\begin{eqnarray}
\label{otoc_decomposition}
 C\lr{t}   = F_0\lr{t} - F\lr{t} ,
 \nonumber \\
 F\lr{t}   = \nonumber \\ =
  2 \re \tr\lr{\hat{\rho}^{\frac{1}{4}} \hat{x}_2\lr{t} \hat{\rho}^{\frac{1}{4}} \hat{p}_2\lr{0} \hat{\rho}^{\frac{1}{4}} \hat{x}_2\lr{t} \hat{\rho}^{\frac{1}{4}} \hat{p}_2\lr{0}} ,
 \nonumber \\
 F_0\lr{t} = \nonumber \\ =
  2 \re \tr\lr{\hat{\rho}^{\frac{1}{4}} \hat{x}_2\lr{t} \hat{\rho}^{\frac{1}{4}} \hat{p}_2\lr{0} \hat{\rho}^{\frac{1}{4}} \hat{p}_2\lr{0} \hat{\rho}^{\frac{1}{4}} \hat{x}_2\lr{t}} .
\end{eqnarray}
All operators are time-ordered in $F_0\lr{t}$, and for sufficiently ergodic systems this function is expected to have a conventional behavior of a finite-temperature, time-ordered correlator. In particular, if the finite-temperature two-point correlators $\tr\lr{\hat{\rho} \, \hat{x}_2\lr{t} \, \hat{p}_2\lr{0}}$ decay sufficiently quickly with time $t$, $F_0\lr{t}$ is expected to approach a constant value $f_0 = 2 \vev{\hat{p}_2^2}_r \, \vev{\hat{x}_2^2}_r$, where $\vev{\hat{O}^2}_r = \tr\lr{\hat{\rho}^{\frac{1}{2}} \, \hat{O} \, \hat{\rho}^{\frac{1}{2}} \, \hat{O}}$ is a regularized thermal expectation value \cite{Kundu:2109.08693,Maldacena:1503.01409}.

The functions $F\lr{t}$ and $F_0\lr{t}$, calculated according to (\ref{otoc_decomposition}), are plotted on Fig.~\ref{fig:FF0} for different Hamiltonians and for different values of the truncation parameter $M$. Free one- and two-dimensional Hamiltonians are used for comparisons at low and at high temperatures, respectively.  The temperatures are $T = 0.2$ and $T = 20.0$ for the plots on the left and on the right, which are well in the low- and high-temperature regimes.

As discussed above, our truncation imposes an infrared cutoff, limiting the system size to $\vev{\hat{x}^2} \lesssim 2 \, M \, L^2$ and hence regularizing the flat directions $x_1 = 0$ and $x_2 = 0$ of the classical and the supersymmetric Hamiltonians. For the free one- and two-dimensional Hamiltonians (\ref{H1D}), the truncation parameter $M$ effectively puts our free particle in a box of size $\sim \sqrt{M} \, L$. It is therefore not surprising that on Fig.~\ref{fig:FF0} we observe a strong dependence of the correlators $F\lr{t}$ and $F_0\lr{t}$ on the truncation parameter $M$ for all Hamiltonians and at all temperatures. The only exception is the low-temperature regime of the bosonic Hamiltonian, where only strongly localized wave functions contribute, and the effect of infrared cutoff is negligible.

To some extent, the $M$ dependence of the OTOCs $C\lr{t}$ is due to the strong $M$ dependence of regularized thermal expectation values $\vev{\hat{x}_2^2}_r$ and $\vev{\hat{p}_2^2}_r$. As discussed above, the asymptotic values of the time-ordered correlator $F_0\lr{t}$ and the OTOCs $C\lr{t}$ are proportional to the product of these two thermal expectation values. As illustrated on Fig.~\ref{fig:x2_vs_T}, the expectation value $\vev{\hat{x}_2^2}_r$ grows linearly with $M$ as $M \rightarrow +\infty$, and $F\lr{t}$ and $F_0\lr{t}$ exhibit similar growth.

Except for the case of the bosonic Hamiltonian at low temperatures, on all plots on Fig.~\ref{fig:FF0} we observe the main expected features of $F\lr{t}$ and $F_0\lr{t}$: the two correlators are very close to each other at $t = 0$ and diverge at later times. Exhibiting some initial oscillations, the function $F_0\lr{t}$ approaches a constant value, while $F\lr{t}$ keeps decreasing.

Furthermore, we see that as a result of our Hilbert space truncation, the correlators $F\lr{t}$ and $F_0\lr{t}$ do exhibit the same characteristic features of OTOC growth even for the free-particle Hamiltonians (\ref{H1D}) and (\ref{H2D}). Clearly, for free-particle Hamiltonians without any truncation of the Hilbert space, the OTOC is just constant:
\begin{eqnarray}
\label{OTOC_free}
 C\lr{t}
 =
 - \vev{\lrs{\hat{x}\lr{t}, \hat{p}\lr{0}}^2 }
 = \nonumber \\ =
 - \vev{\lrs{\hat{x}\lr{0} + 2 \,\hat{p}\lr{0} \, t \, , \hat{p}\lr{0}}^2 }
 = \nonumber \\ =
 \vev{\lrs{\hat{x}\lr{0}, \hat{p}\lr{0}}^2 }
 =
 1 .
\end{eqnarray}
Therefore, any nontrivial time dependence of OTOCs for the free-particle Hamiltonians (\ref{H1D}) and (\ref{H2D}) is a truncation artifact at finite $M$.

\begin{figure*}
  \centering
  \includegraphics[width=0.49\textwidth]{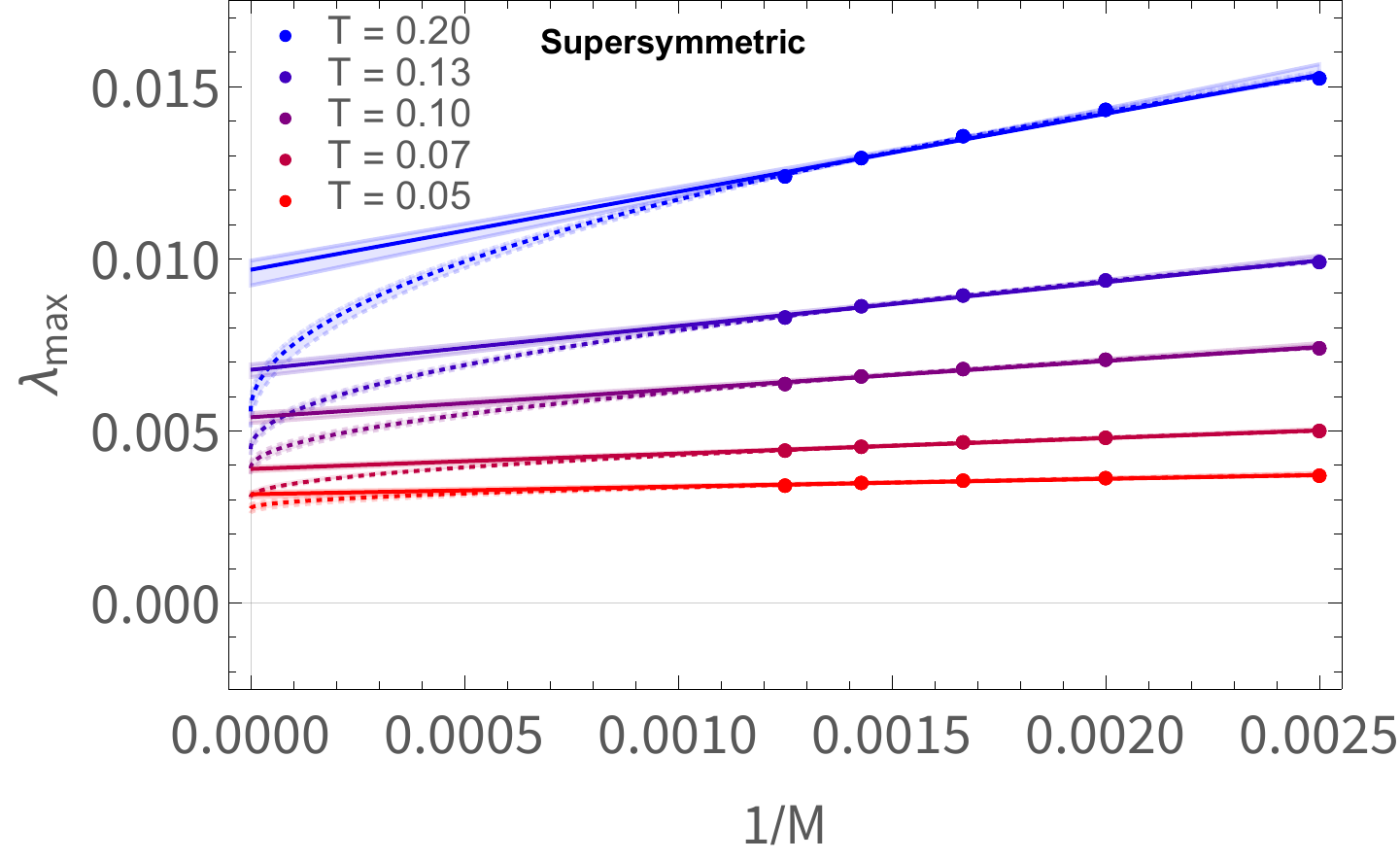}
  \includegraphics[width=0.49\textwidth]{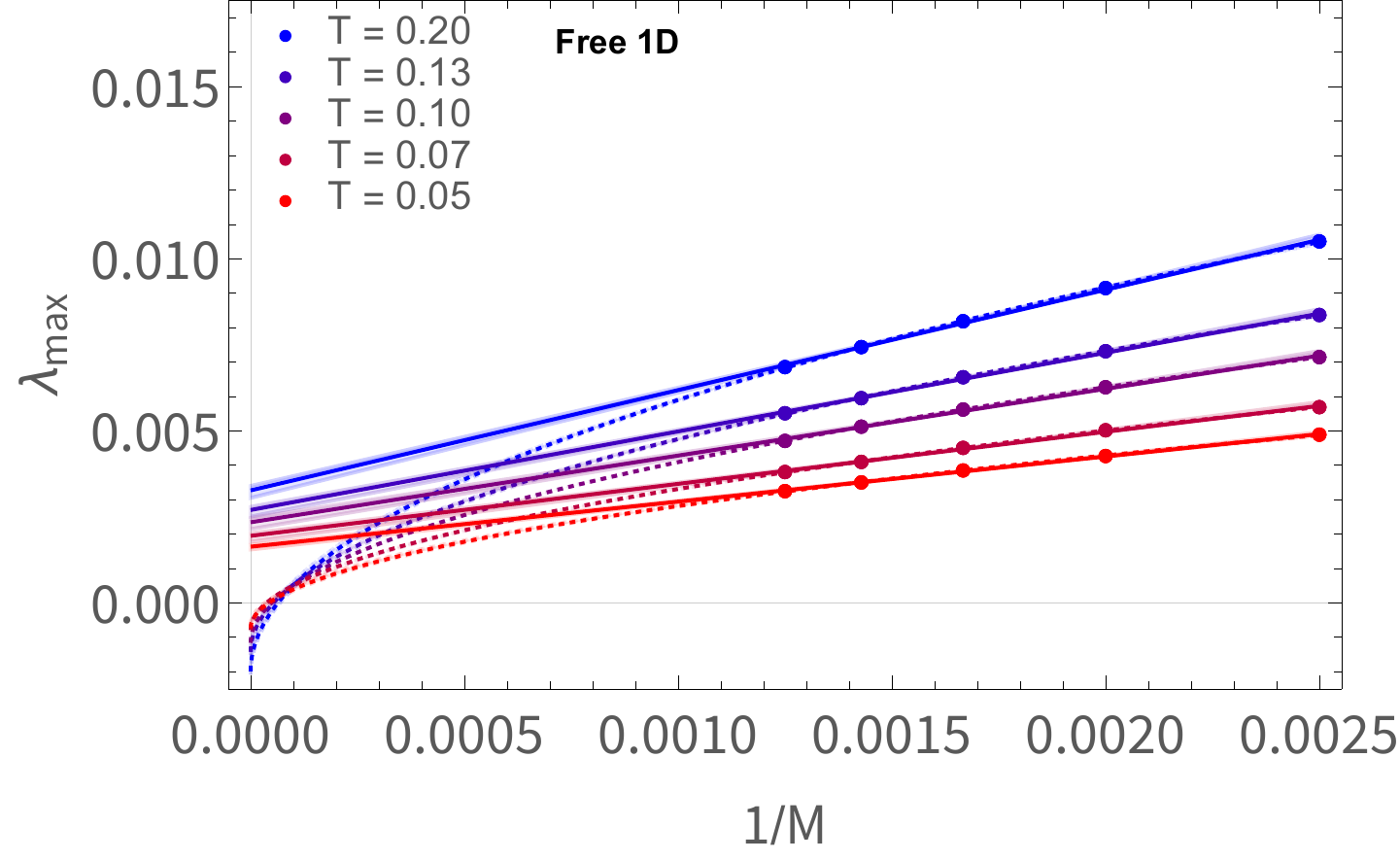} \\
  \includegraphics[width=0.49\textwidth]{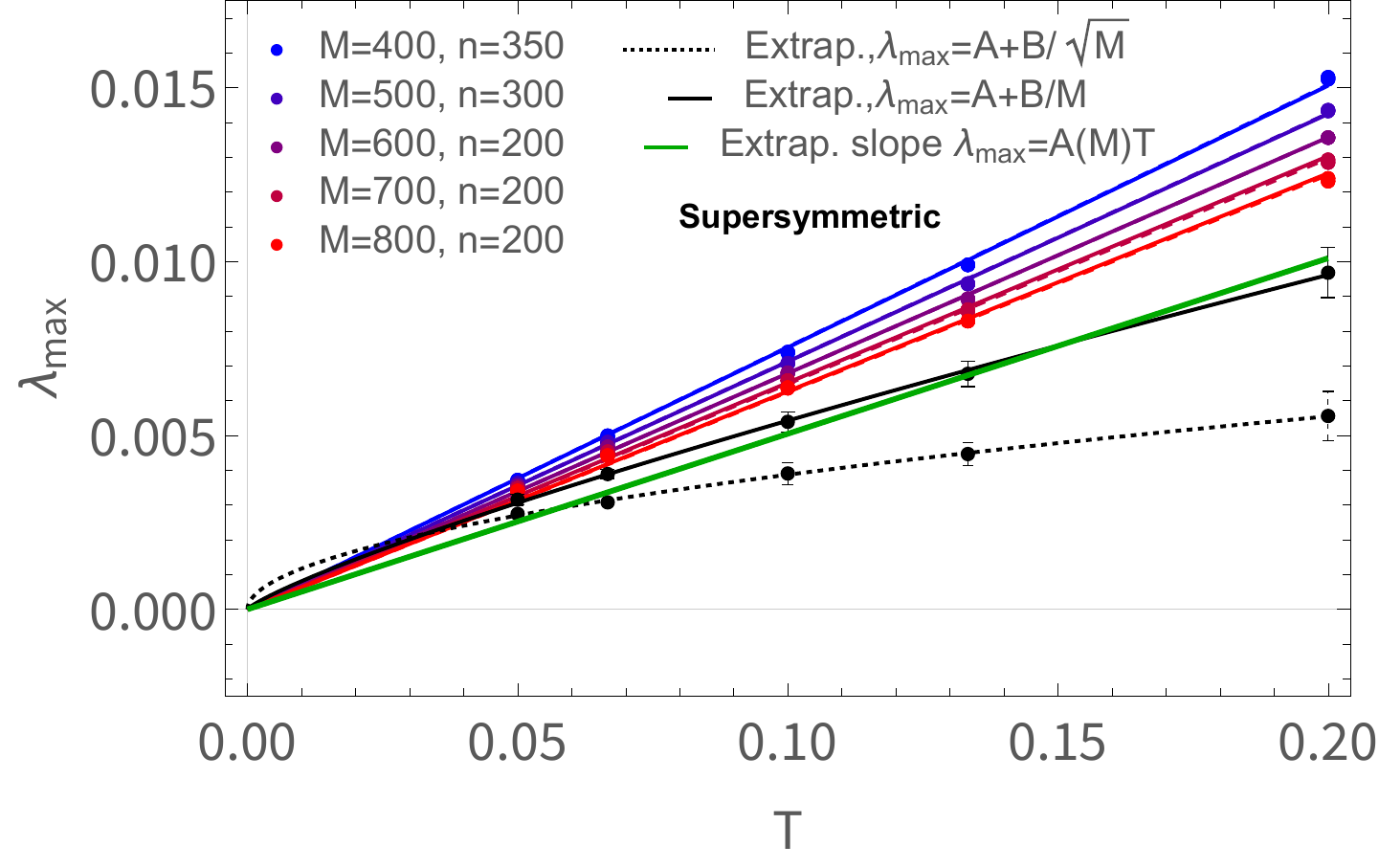}
  \includegraphics[width=0.49\textwidth]{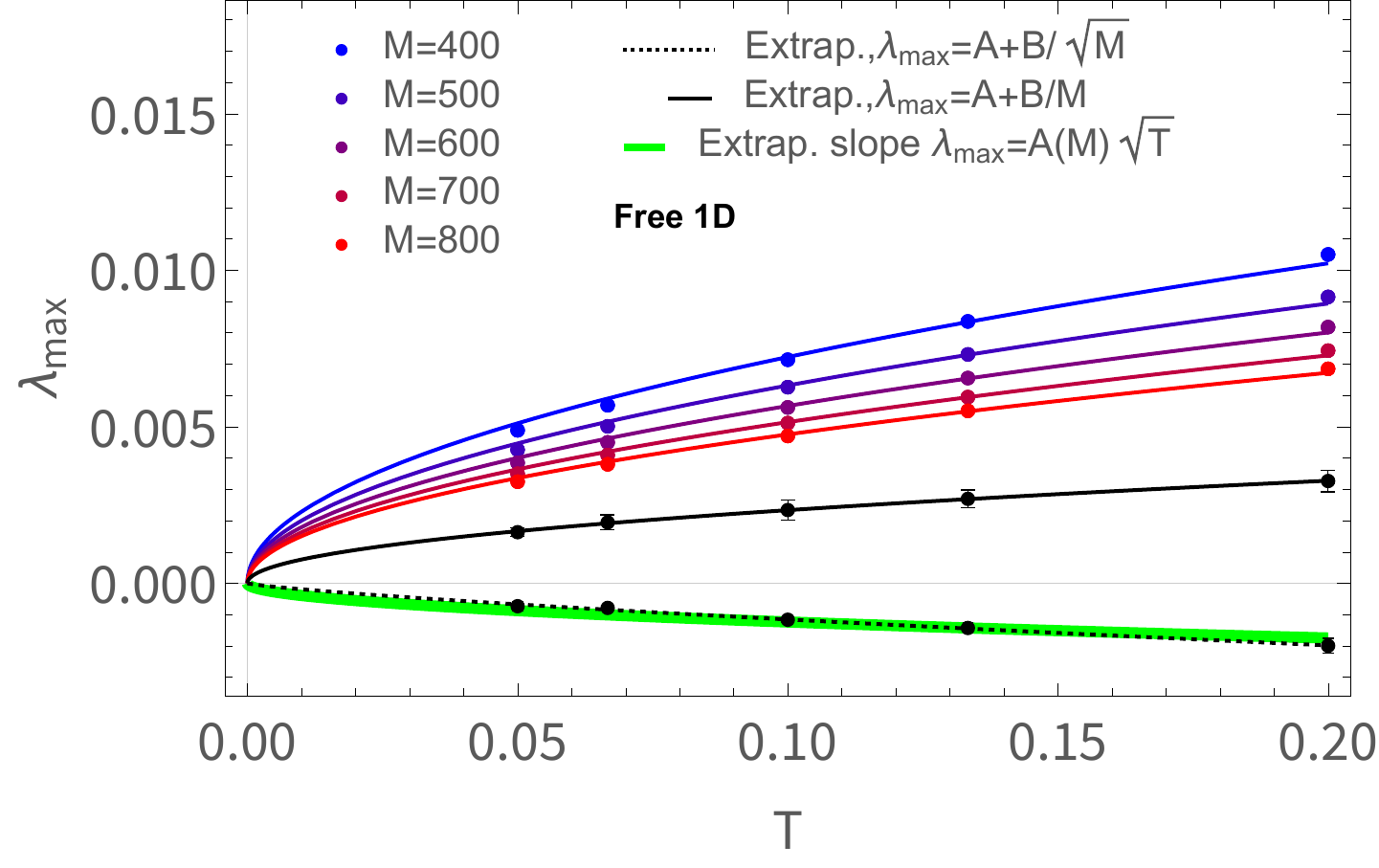} \\
  \caption{Extrapolations of the upper bounds $\lambda_{max}$ on Lyapunov exponents to $M \rightarrow +\infty$ for the supersymmetric system (on the left) and for the one-dimensional free Hamiltonian with the same Hilbert space truncation (on the right). Solid and dashed lines show extrapolations to $M \rightarrow +\infty$ that use fits of the form $\lambda_{max}\lr{M} = A + B/M$ and $\lambda_{max}\lr{M} = A + B/\sqrt{M}$, respectively. Shaded areas around these lines show error estimates obtained by excluding the data points with either smallest or largest value of $M$ from the fits.}
  \label{fig:lambda_vs_M}
\end{figure*}

The timescales characterizing the time dependence of $F\lr{t}$ and $F_0\lr{t}$ appear to be not too different for the free-particle and the supersymmetric cases. This could be expected based on our previous observation that low-energy eigenstates with a finite infrared cutoff are quite similar to the ones for a free particle in a finite one-dimensional box. The fact that even a free particle can exhibit OTOC growth in the presence of spatial boundaries was also noted in \cite{Hashimoto:1703.09435}. For this reason, our strategy will be to carefully extrapolate the OTOCs growth rates to the limit $M \rightarrow +\infty$ in order to distinguish the behavior of OTOCs for the supersymmetric Hamiltonian and the free-particle one-dimensional Hamiltonian (\ref{H1D}) (see Fig.~\ref{fig:lambda_vs_M} for the final results).

In most cases, however, we also significant deviations from the expected behavior of $F\lr{t}$ and $F_0\lr{t}$: except for the high-temperature regime of the supersymmetric Hamiltonian, the late-time value of $F_0\lr{t}$ is noticeably different from the product $f_0 = 2 \vev{\hat{p}_2^2}_r \, \vev{\hat{x}_2^2}_r$ of thermal expectation values. This suggests that our low-dimensional Hamiltonians are not entirely ergodic, and some kind of regular motion prevents the finite-temperature correlators of the form $\tr\lr{\hat{\rho} \, \hat{x}_2\lr{t} \, \hat{p}_2\lr{0}}$ from decaying sufficiently quickly. This is expectable for a system with just a few degrees of freedom. For example, such a finite-temperature correlator would exhibit oscillations instead of decay for a harmonic oscillator or a two-level system. In fact, the functions $F\lr{t}$ and $F_0\lr{t}$ calculated at low temperatures with the bosonic Hamiltonian (\ref{HB}) provide a nice example of such an oscillatory behavior (middle left plot on Fig.~\ref{fig:FF0}). Since the energy spectrum of the bosonic Hamiltonian is discrete (and hence gapped), only the two lowest energy levels will contribute to OTOCs for sufficiently low temperatures. This results in non-decaying oscillations of OTOCs and other correlators, as illustrated in the middle left plot on Fig.~\ref{fig:FF0}.

Interestingly, even though the bosonic and the supersymmetric Hamiltonian are both expected to be chaotic at high temperatures, only for the supersymmetric Hamiltonian the late-time asymptotic value of $F_0\lr{t}$ agrees with $f_0 = 2 \vev{\hat{p}_2^2}_r \, \vev{\hat{x}_2^2}_r$ (see upper right plot on Fig.~\ref{fig:FF0}). As one can see on the middle right plot on Fig.~\ref{fig:FF0}, for the bosonic Hamiltonian the function $F_0\lr{t}$ at large $t$ has a noticeable deviation from $f_0$ even at $T = 20$. This difference might be related to the observation that only for the supersymmetric Hamiltonian the OTOCs exhibit an expected agreement with classical dynamics at high temperatures. In contrast, for the bosonic Hamiltonian, we never observe a good agreement between quantum OTOCs and classical Lyapunov distances. This is demonstrated on Fig.~\ref{fig:DOTOCs} and explained in more details below.

For large-$N$ gauge theories and matrix models, $F\lr{t}$ is expected to behave as $F\lr{t} \approx f_0 - c \, e^{2 \lambda t}$ for sufficiently late times $t$ at which $F_0\lr{t}$ is already sufficiently close to its expected asymptotic value $f_0$ \cite{Maldacena:1503.01409,Kundu:2109.08693}. Clearly, with $F_0\lr{t} = f_0$ and $F\lr{t} = f_0 - c \, e^{2 \lambda t}$ we immediately get an exponential OTOC growth $C\lr{t} = F_0\lr{t} - F\lr{t} = c \, e^{2 \lambda t}$, with $\lambda$ being the Lyapunov exponent. Such growth, however, cannot continue forever, and at even later times $F\lr{t}$ approaches zero and stops decreasing. Correspondingly, the OTOC $C\lr{t}$ saturates at a value close to $f_0$ \cite{Kundu:2109.08693}.

For our numerical data we indeed see how $F\lr{t}$ decreases towards zero in all cases except for the low-temperature limit of the bosonic Hamiltonian. However, we cannot identify any clear exponential decay of $F\lr{t}$ of the form $f_0 - c \, e^{2 \lambda t}$. In the low-temperature regime of the supersymmetric Hamiltonian, the late-time decay of $F\lr{t}$ seems to be linear in $t$ (see upper left plot on Fig.~\ref{fig:FF0}). Such behavior is even more obvious for the one-dimensional free-particle Hamiltonian\footnote{In fact, for the one-dimensional free-particle Hamiltonian, the linearly decreasing function $F\lr{t}$ re-bounces after approaching zero, and exhibits a linear growth until approaching $f_0$. This behavior resembles a classical motion of a particle that bounces between the two walls of a one-dimensional potential well. In our case, a one-dimensional particle is confined to an interval of finite width because of the truncation of Hilbert space which limits the coordinate values to $\vev{\hat{x}^2} < 2 M L^2$} (\ref{H1D}), see the lower left plot on Fig.~\ref{fig:FF0}. At high temperatures, the decay of $F\lr{t}$ towards zero is faster than at low temperatures, but it is still difficult to identify a region with a clear exponential decay.

\begin{figure*}[h!tpb]
  \centering
  \includegraphics[width=0.49\textwidth]{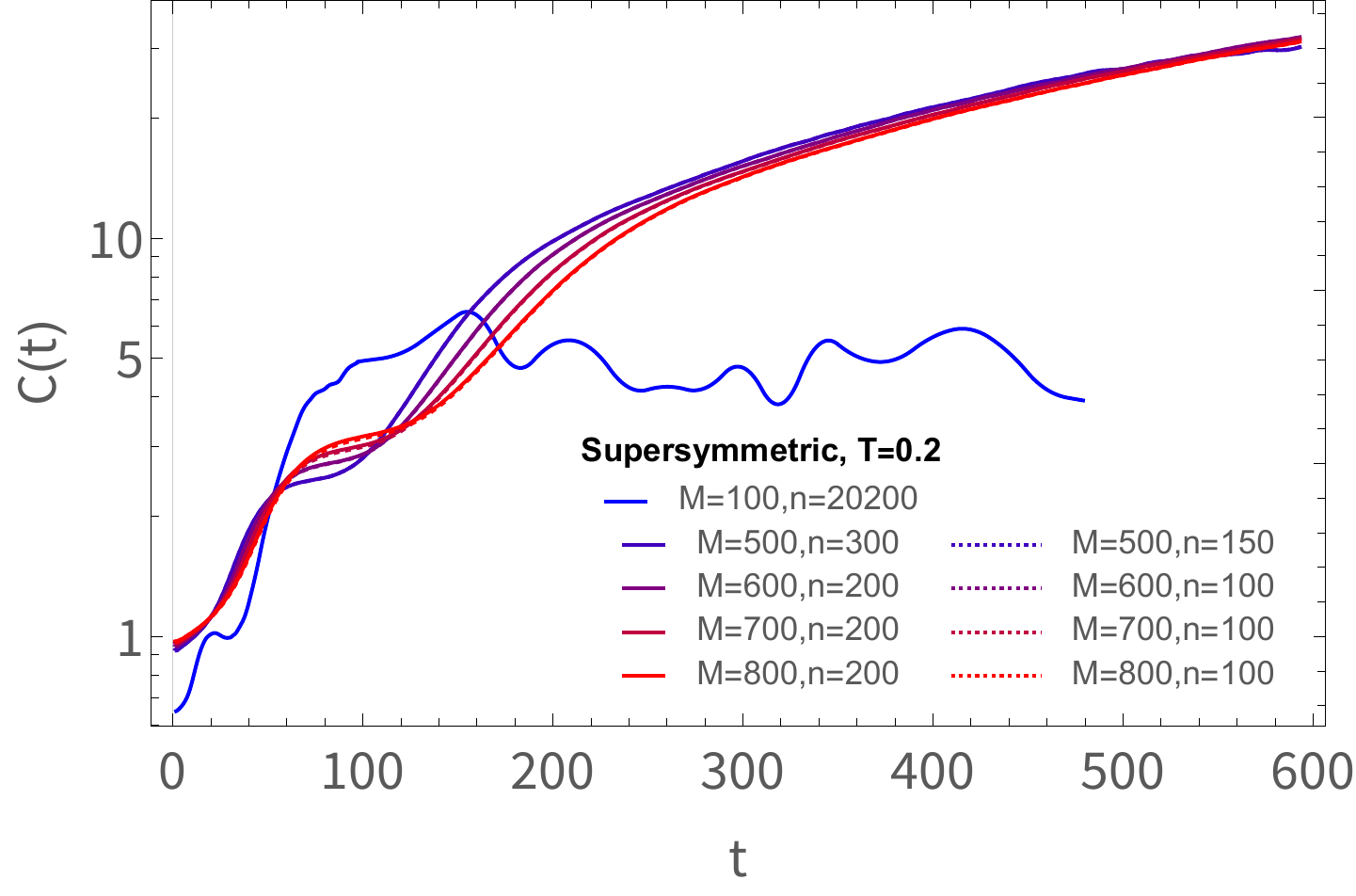}
  \includegraphics[width=0.49\textwidth]{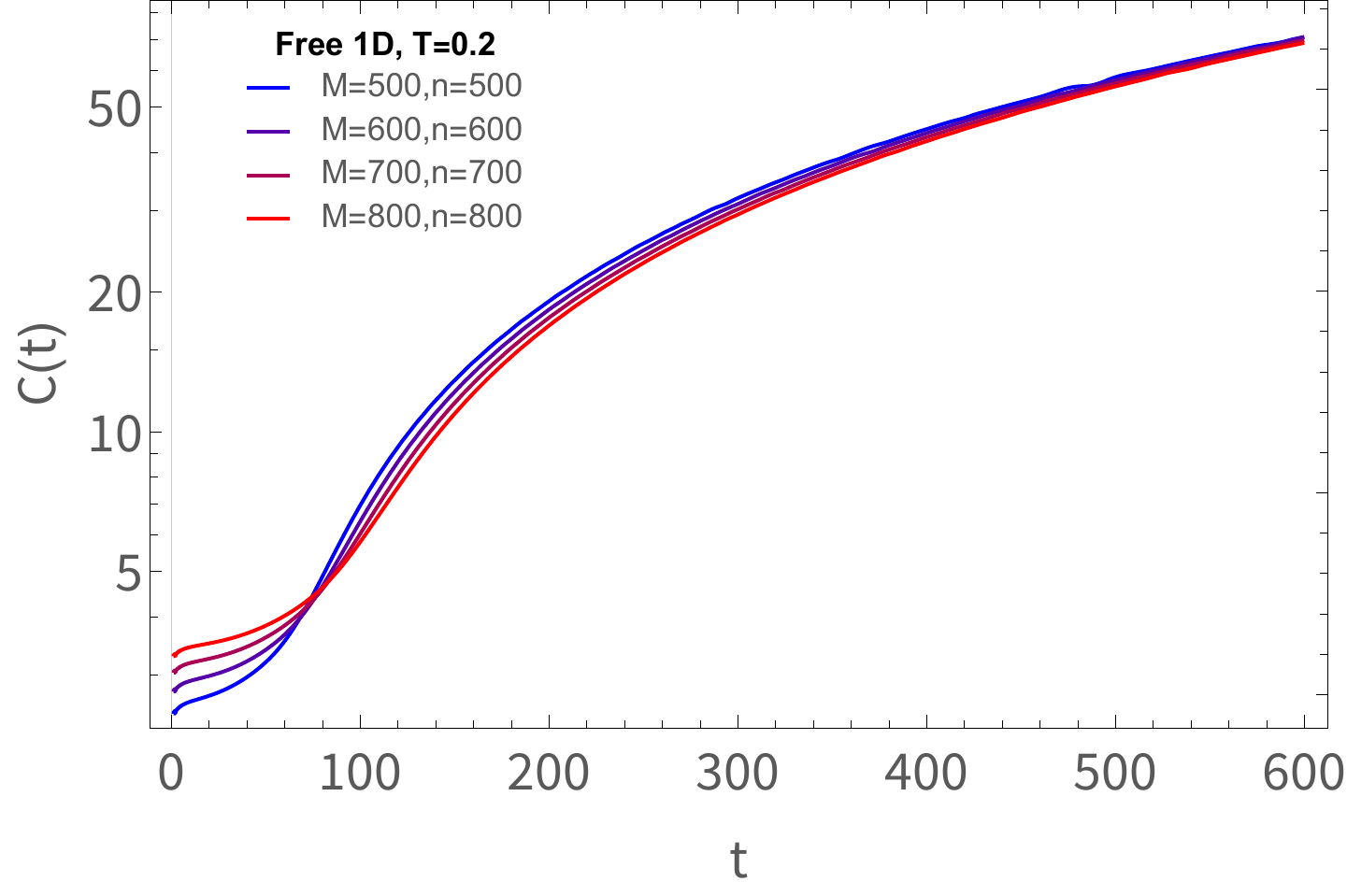}\\
  \includegraphics[width=0.49\textwidth]{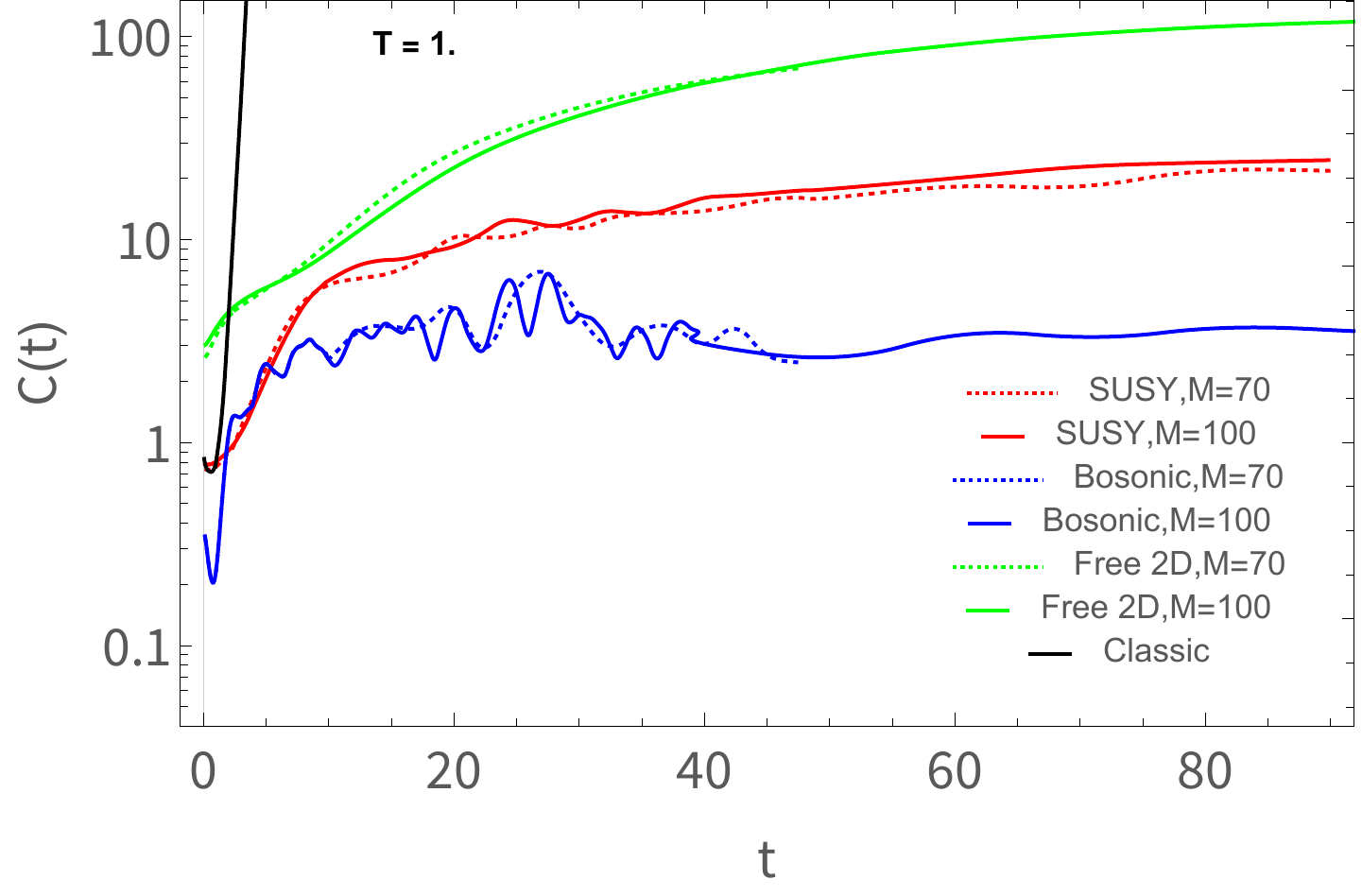}
  \includegraphics[width=0.49\textwidth]{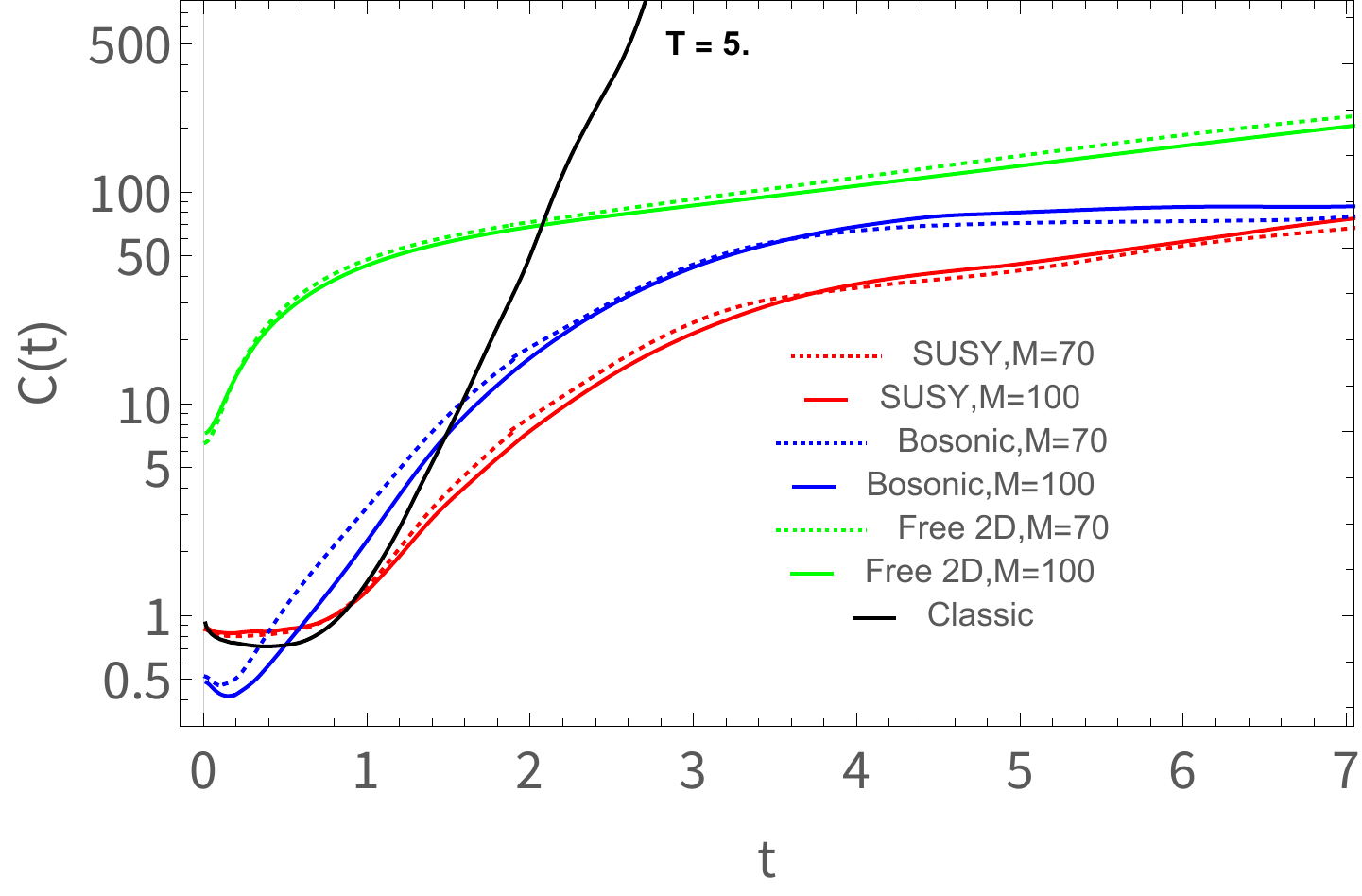}\\
  \caption{Time dependence of the OTOC $C\lr{t}$ for the supersymmetric, bosonic, and free one- and two-dimensional Hamiltonians, for different values of the truncation parameter $M$ and at different temperatures. Solid black line on the plots in the high-temperature regime shows the expectation value $\vev{\lr{\frac{\partial x_2\lr{t}}{\partial x_2\lr{0}}}^2}$ of the squared classical Lyapunov distance calculated at the same temperature. For the plots at $T = 0.2$, the legend also shows the number $n$ of lowest eigenstates that were used to calculate $C\lr{t}$.}
  \label{fig:OTOCs}
\end{figure*}

We illustrate the time dependence of the OTOCs $C\lr{t} = F_0\lr{t} - F\lr{t}$ on Fig.~\ref{fig:OTOCs}, comparing the results obtained with the supersymmetric and the bosonic Hamiltonians (\ref{HS}) and (\ref{HB}) and with the free-particle one- and two-dimensional Hamiltonians. Fig.~\ref{fig:OTOCs} shows that in all cases the OTOCs grow and eventually saturate. The growth rate appears to be non-uniform. In particular, for sufficiently high temperatures $T \gtrsim 1$ and at early times all OTOCs exhibit some initial decrease. Also for the bosonic Hamiltonian the OTOCs feature some oscillations at $T \lesssim 1$, which become stronger and completely dominate the OTOCs at lower temperatures (see the middle left plot on Fig.~\ref{fig:FF0}).

A common expectation is that for sufficiently high temperatures the real-time dynamics of the bosonic Hamiltonian (\ref{HB}) can be described in terms of the corresponding classical equations of motion $\ddot{x}_1 = - 4 \, x_1 \, x_2^2$, $\ddot{x}_2 = - 4 \, x_2 \, x_1^2$. This statement is also expected to apply to the supersymmetric Hamiltonian (\ref{HS}), for which the role of ``fermionic'' terms should become negligible at high temperatures.

To check whether this is indeed the case, we also compare our quantum data for $T \geq 1$ with the classical analogue of OTOCs. It is defined as a square of the relevant Poisson bracket $\lrc{x_2\lr{t}, p_2\lr{0}} = \frac{\partial x_2\lr{t}}{\partial x_2\lr{0}}$, averaged over the thermal distribution of the initial conditions $\lrc{x_1\lr{0}, x_2\lr{0}, p_1\lr{0}, p_2\lr{0}}$ for the classical evolution:
\begin{eqnarray}
\label{class_Lyapunov}
 C_{cl}\lr{t} = \vev{ \lr{\frac{\partial x_2\lr{t}}{\partial x_2\lr{0}} }^2 } .
\end{eqnarray}
The calculation of $C_{cl}\lr{t}$ is discussed in more details in Appendix~\ref{apdx:lyapunov_classical}. For $T = 5$ and $T = 1$, $C_{cl}\lr{t}$ is plotted on Fig.~\ref{fig:OTOCs} with a solid black line.

Quantum OTOCs only appear to be reasonably close to $C_{cl}\lr{t}$ for the case of the supersymmetric Hamiltonian (\ref{HS}) at sufficiently high temperature. At $T = 1$, the OTOCs for the supersymmetric Hamiltonian and the classical OTOCs are only close to each other for a short initial period of time $t \lesssim 1$. The agreement is somewhat better at $T = 5$, where the classical and the quantum OTOCs are reasonably close to each other for $t \lesssim 1.5$. In both cases, the OTOCs exhibit some initial decrease and start growing at $t \gtrsim 0.5$. Interestingly, the agreement between the classical and the quantum OTOCs is lost exactly when the classical OTOCs enters the regime of clear exponential growth at $t \gtrsim 1.5$. For these times, the quantum OTOCs grows considerably slower. On the other hand, our data suggests that OTOCs calculated with the bosonic Hamiltonian (\ref{HB}) are never quite close to the classical OTOCs. This agrees with the observation made in \cite{Hashimoto:1703.09435} that the classical limit is never reached for four-point OTOCs in simple quantum mechanical systems because quantum interference effects become important earlier than the exponential OTOC growth sets in. The fact that for the supersymmetric Hamiltonian we get a considerably better agreement with the classical dynamics suggests that supersymmetry might effectively suppress or cancel out the wave packet spread that is responsible for deviations from classical behavior \cite{Hashimoto:1703.09435}.

Overall, our plots for the quantum OTOCs do not show regions of clear exponential growth that would be similar to the exponential divergence of the classical Lyapunov distance $C_{cl}\lr{t}$. It is not surprising, as the exponential OTOCs growth is in general very difficult to detect for finite-size systems amenable to exact diagonalization studies \cite{Bohrdt:1612.02434}, even for the renowned SYK model which is a paradigmatic example of maximal quantum chaos \cite{Sonner:1707.08013}. We do, however, see a clear difference in the low-temperature behavior of the OTOCs for the bosonic Hamiltonian (\ref{HB}) with gapped energy spectrum and for the supersymmetric and free Hamiltonians which all have continuous energy spectra in the limit $M \rightarrow 0$. For the bosonic Hamiltonian, the spectrum is discrete and OTOCs are oscillatory at low temperatures. For the supersymmetric and free Hamiltonians, OTOCs grow with time down to the lowest temperatures, although this growth is much slower than for the classical dynamics. Furthermore, there is also an observable difference between the behavior of low-temperature OTOCs for the free one-dimensional Hamiltonian and the supersymmetric Hamiltonian. The latter has more nontrivial features and a weaker dependence on the truncation parameter $M$.

To arrive at some estimates for the growth rate of OTOCs in the absence of clear exponential growth, we consider the time derivative of the logarithm of the OTOC $C\lr{t}$
\begin{eqnarray}
\label{dlogdt}
 \lambda\lr{t} = \frac{1}{2} \frac{d}{d t} \log\lr{C\lr{t}} ,
\end{eqnarray}
which should be equal the Lyapunov exponent $\lambda$ if $C\lr{t} = c \, e^{2 \, \lambda \, t}$. The time dependence of $\lambda\lr{t}$ is shown on Fig.~\ref{fig:DOTOCs}. At low temperatures, we again compare the data for the supersymmetric Hamiltonian (\ref{HS}) and the free one-dimensional Hamiltonian (\ref{H1D}). At high temperatures, we compare the results for the bosonic, the supersymmetric and the free two-dimensional Hamiltonians, as well as with the classical result $\lambda_{cl}\lr{t} = \frac{1}{2} \frac{d}{d t} \log\lr{C_{cl}\lr{t}}$.

At all temperatures, the functions $\lambda\lr{t}$ feature a number of peaks and sometimes plateaus. The peaks are labelled by ``x'' or ''o'' symbols, where ``x'' is used for peaks that we believe to be dominant/most important features of OTOCs for a particular Hamiltonian, and ``o'' is used for the peaks that we believe to be sub-dominant or artifact. Plateaus are labelled by vertical lines and a letter ``P''. We will use these peak values of $\lambda\lr{t}$, denoted as $\lambda_{max}$, as upper bound estimates on the values of Lyapunov exponent in our system. Since we observe qualitatively different behavior at low and at high temperatures, let us separately discuss the main features of $\lambda\lr{t}$ in these regimes.

\subsection{High temperatures}
\label{subsec:highT_DOTOCs}

\begin{figure*}[h!tpb]
  \centering
  \includegraphics[width=0.49\textwidth]{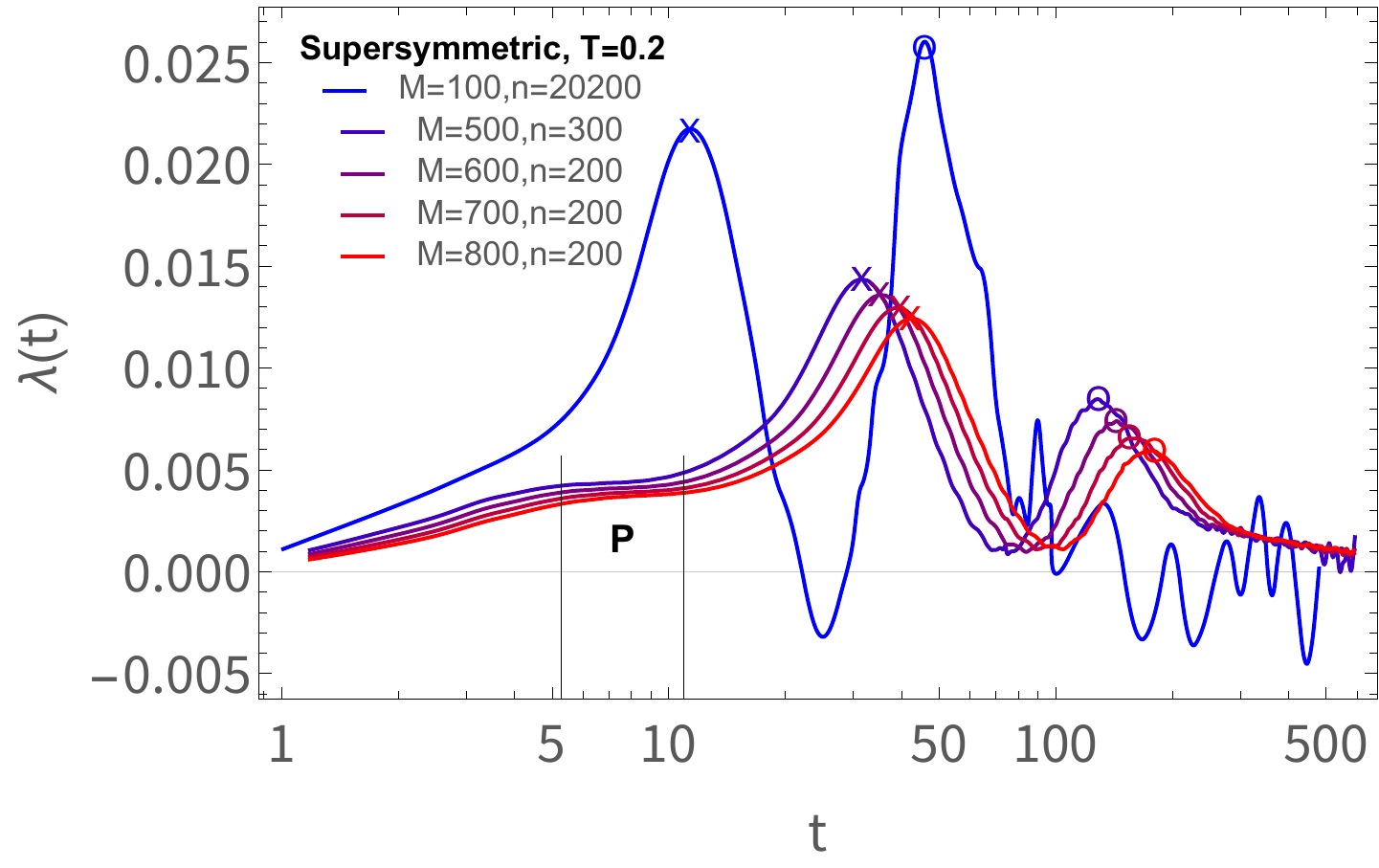}
  \includegraphics[width=0.49\textwidth]{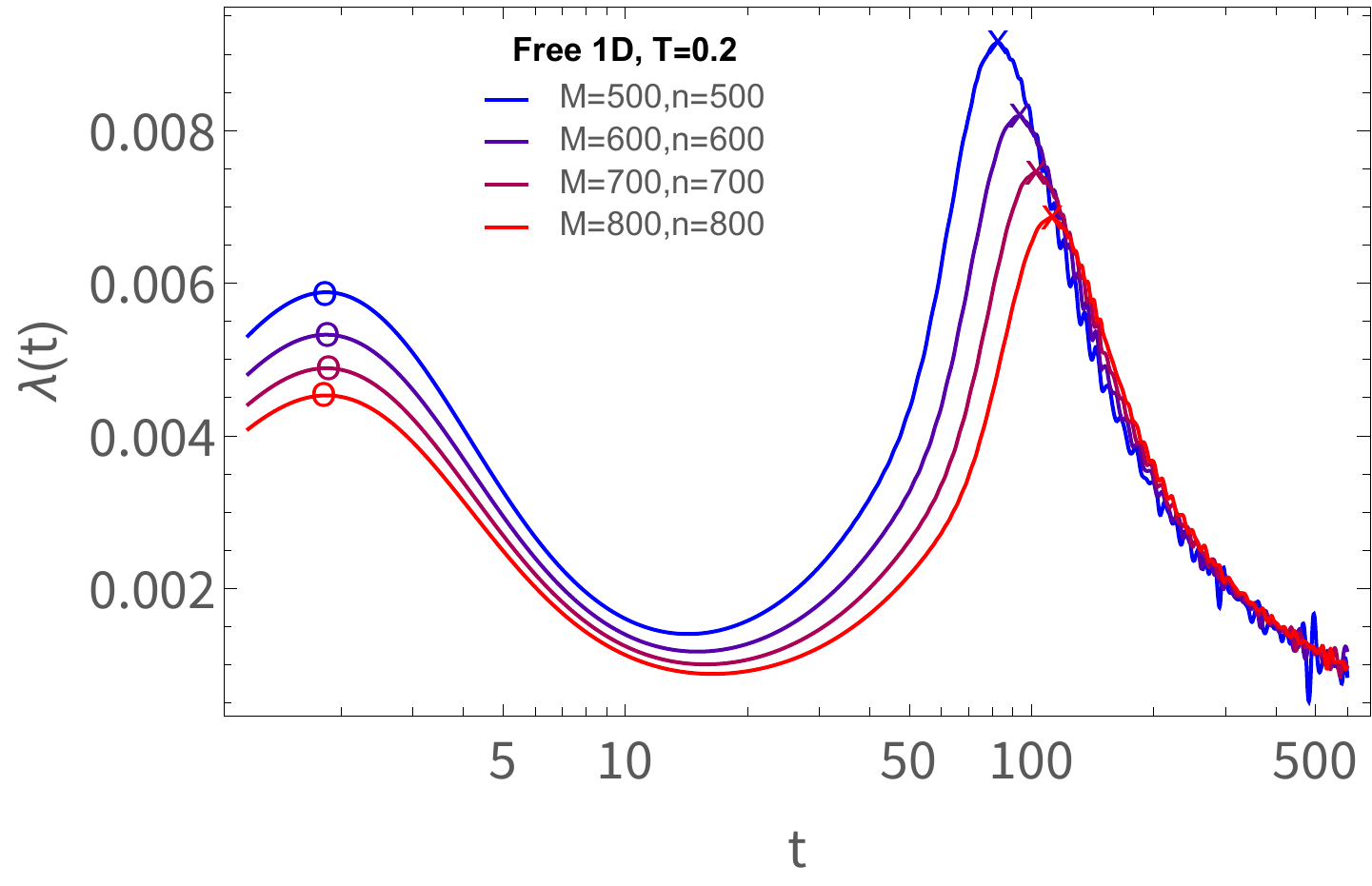}\\
  \includegraphics[width=0.49\textwidth]{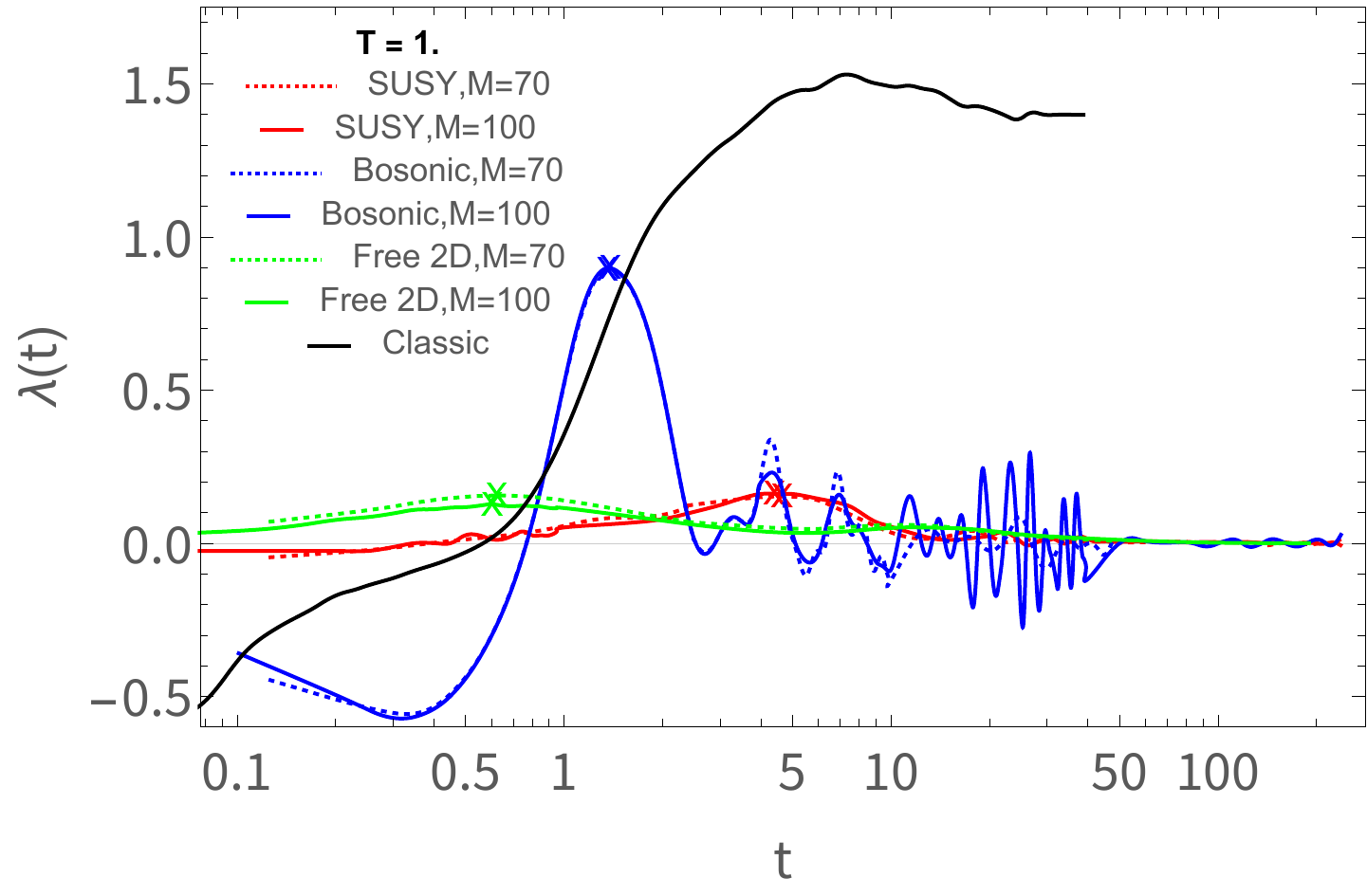}
  \includegraphics[width=0.49\textwidth]{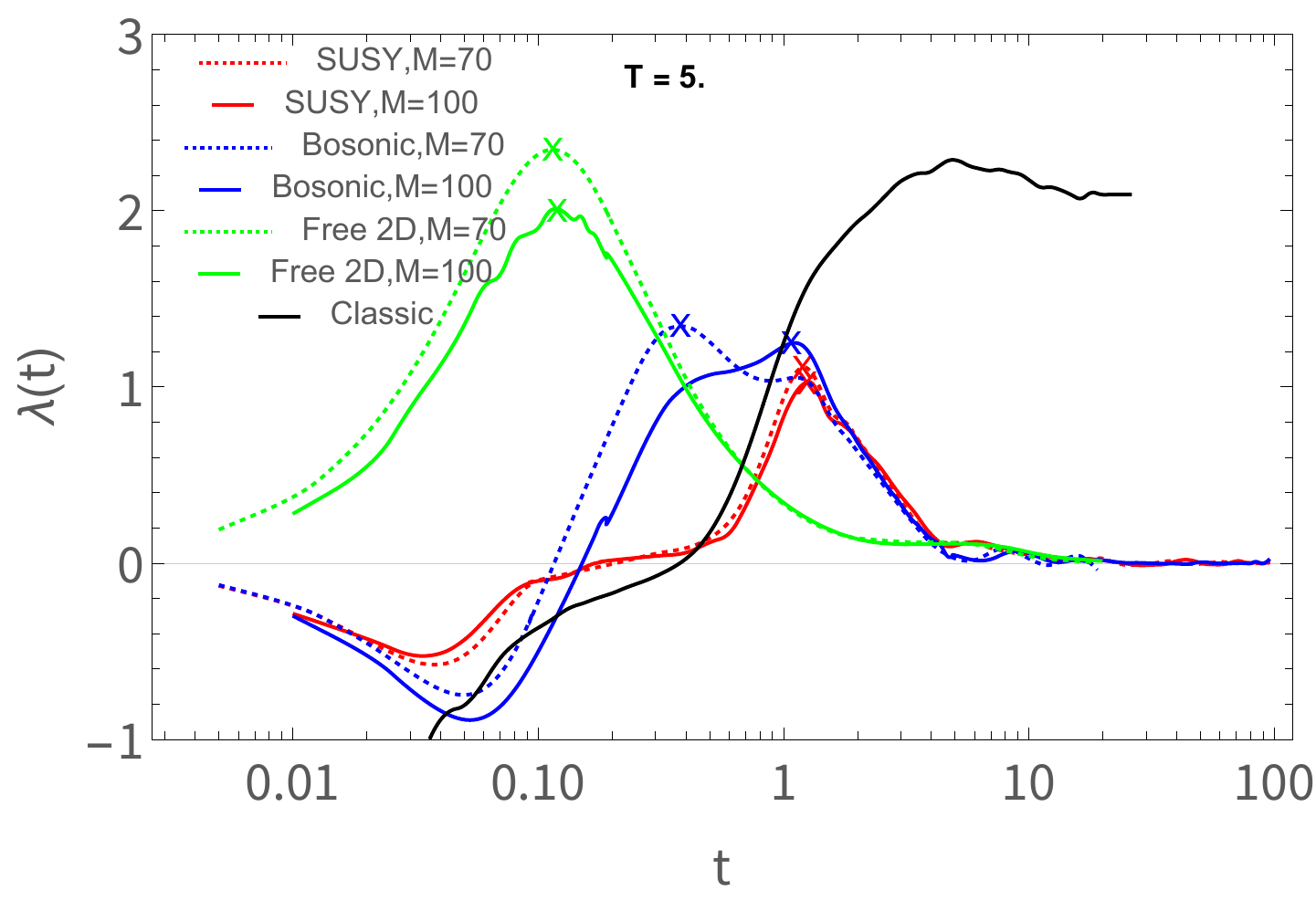}\\
  \includegraphics[width=0.49\textwidth]{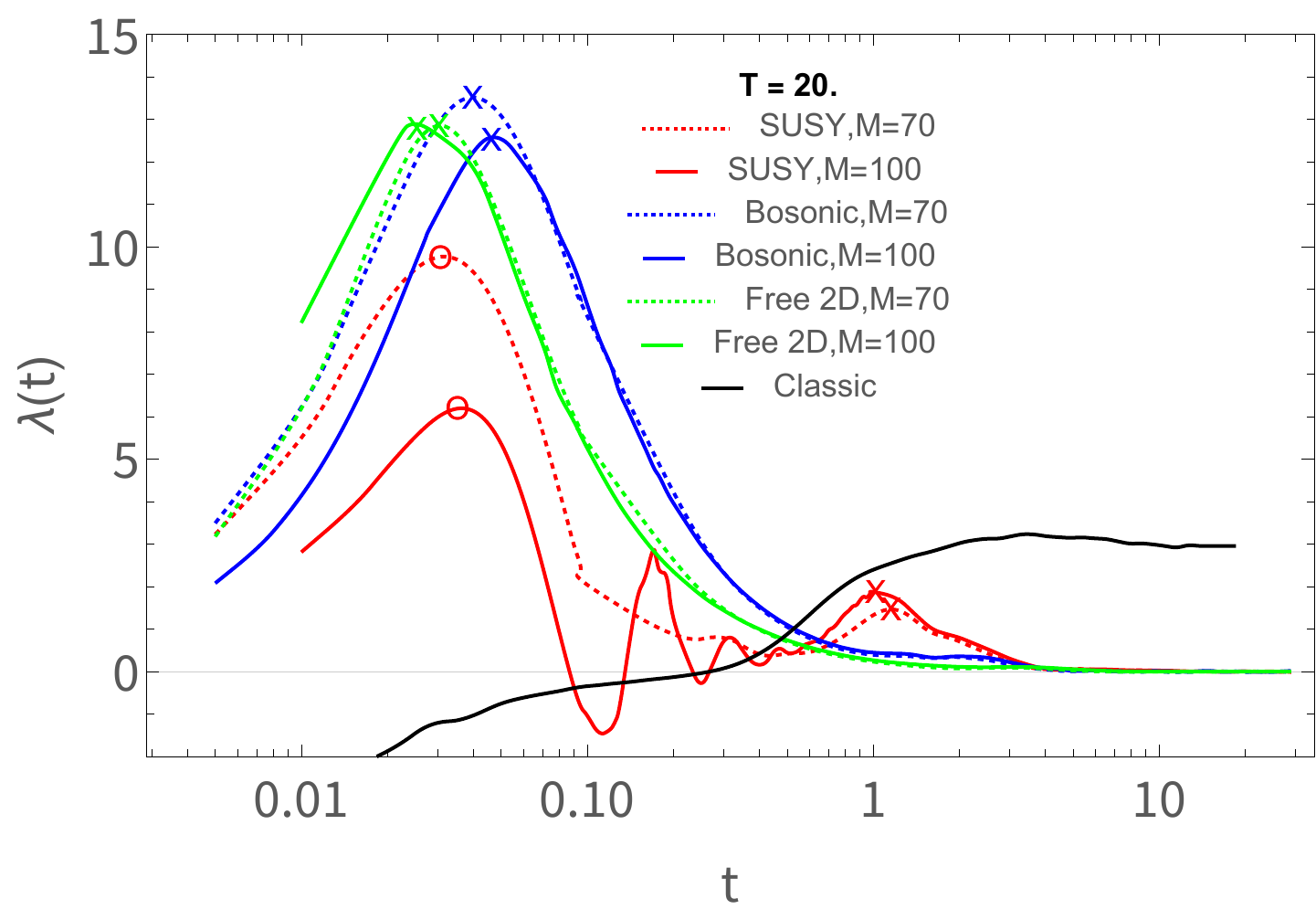}
  \includegraphics[width=0.49\textwidth]{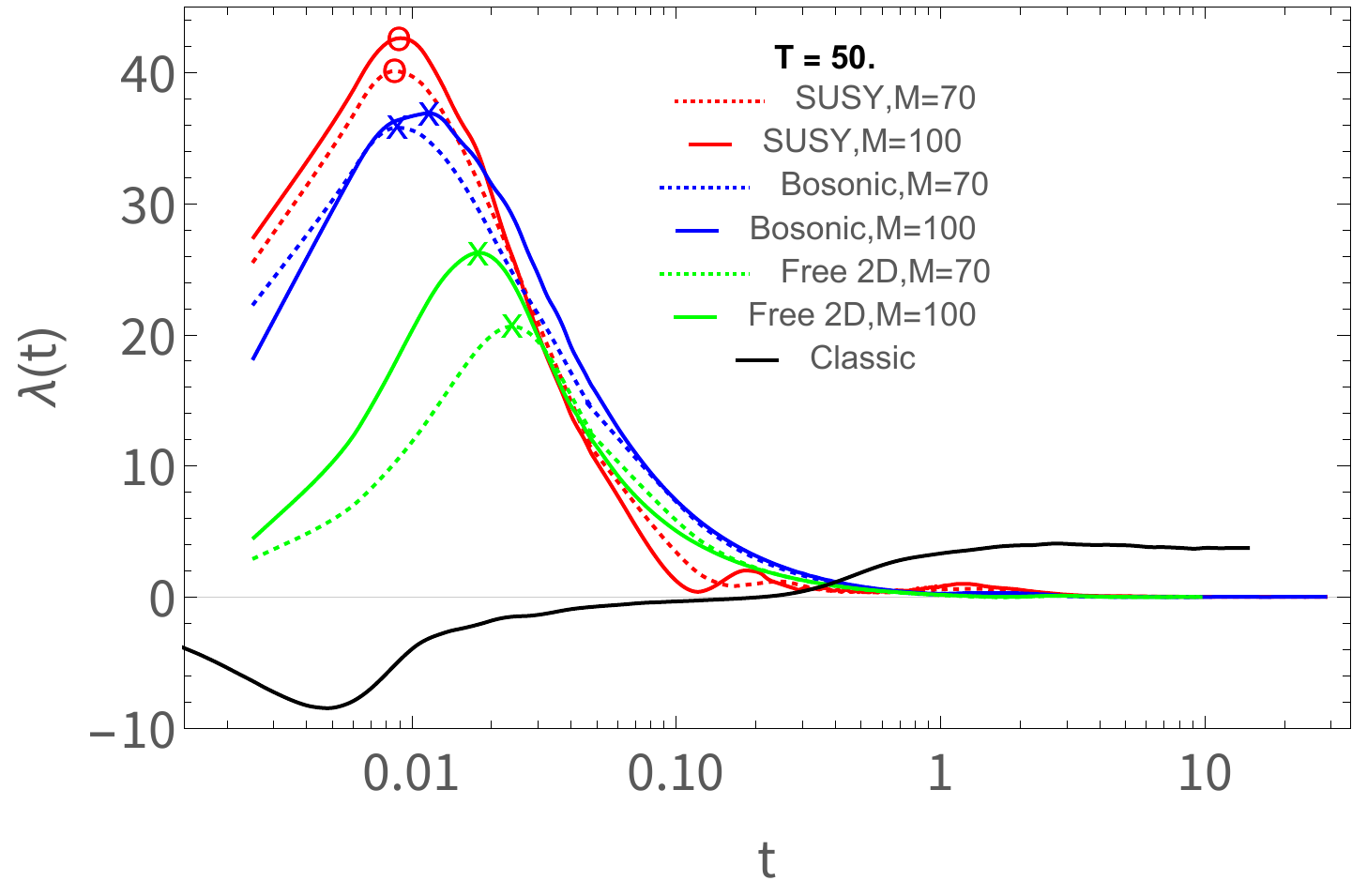}\\
  \caption{Time dependence of the derivatives of the logarithm of OTOC $\lambda\lr{t} = \frac{1}{2} \frac{d}{d t} \log\lr{C\lr{t}}$ for different Hamiltonians and at different temperatures and the values of the truncation parameters $M$ and $n$. Solid black lines show $\lambda_{cl}\lr{t} = \frac{1}{2} \frac{d}{d t} \log\lr{C_{cl}\lr{t}}$  for the classical Lyapunov distance $C_{cl}\lr{t}$, defined in (\ref{class_Lyapunov}). Symbols ``x'' and ''o'' denote the peak values of $\lambda\lr{t}$ that we use as upper bound estimates of the quantum Lyapunov exponents.}
  \label{fig:DOTOCs}
\end{figure*} 

At sufficiently high temperatures, real-time dynamics of both the supersymmetric and the bosonic Hamiltonians is expected to reduce to the classical dynamics. Qualitatively, this regime corresponds to the range of energies $E \gtrsim 10$, for which the energy levels behave similarly to the ones of Gaussian Orthogonal Ensemble (GOE) of random matrices. In this case, we use truncation parameters $M = 70$ and $M = 100$, finding all the eigenvalues and eigenvectors of the corresponding truncated matrices (\ref{HS_matrix}) or (\ref{HB_matrix}). At very high temperatures, the correspondence with the classical dynamics might be violated due to our truncations of the full Hilbert space. As we have already discussed above, we were not able to identify a regime where the OTOCs for the bosonic Hamiltonian would approach the classical OTOCs (see Appendix~\ref{apdx:lyapunov_classical}) sufficiently closely. As conjectured in \cite{Hashimoto:1703.09435}, this might be a generic feature of simple quantum mechanical systems. The data shown on Fig.~\ref{fig:DOTOCs} for $T \gtrsim 1$ supports these observations. We find a limited agreement between the classical and the quantum data only for the case of the supersymmetric Hamiltonian at moderately large temperatures, $5 \lesssim T \lesssim 20$, see the middle right and the bottom left plots on Fig.~\ref{fig:DOTOCs}. The agreement is observed for $0.1 \lesssim t \lesssim 1.5$ at $T = 5$ and for $0.5 \lesssim t \lesssim 1$ at $T = 20$, thus the time range of quantum-classical correspondence shrinks towards high temperatures. Both for $T = 5$ and $T = 20$, $\lambda\lr{t}$ has a distinct peak roughly at the time at which its classical counterpart $\lambda_{cl}\lr{t}$ approaches its plateau value that corresponds to a steady exponential growth of $C_{cl}\lr{t}$. We believe that this peak is a physical feature of a supersymmetric model, therefore we label it with the ``x'' symbol on the plots with $T = 5$ and $T = 20$ on Fig.~\ref{fig:DOTOCs}. Its height is comparable with the plateau value of $\lambda_{cl}\lr{t}$, and its position scales approximately as $t^{\star} \sim T^{-1/2}$. This characteristic ``semi-classical'' peak of $\lambda\lr{t}$ exists also at lower temperatures, down to $T \sim 1$, but the agreement with classical dynamics is gradually lost. A similar peak structure (labelled with ``x'' on Fig.~\ref{fig:DOTOCs}) exists also for the bosonic Hamiltonian at moderately large temperatures $1 \lesssim T \lesssim 10$, but the agreement with the classical dynamics is nowhere sufficiently good. Interestingly, as one can see on the plot with $T = 5$, the data for the bosonic and the supersymmetric Hamiltonians agree very well \emph{after} both peaks, where $\lambda\lr{t}$ disagree with $\lambda_{cl}\lr{t}$ for both Hamiltonians. These moderate-temperature peaks of $\lambda\lr{t}$ are completely absent for the free two-dimensional Hamiltonian, and appear to be a genuine feature of high-temperature, semi-classical chaotic dynamics.

At very high temperatures $T \gtrsim 20$, both the bosonic and the supersymmetric Hamiltonians develop very large peaks of height $\lambda_{max} \sim T$ at early times $t \sim T^{-1}$, labelled with ``o'' for the supersymmetric Hamiltonian. For the bosonic Hamiltonian, this early-time peak seem to continuously transform into the peak near the plateau onset of $\lambda_{cl}\lr{t}$, and is therefore labelled with ``x''. The free two-dimensional Hamiltonian also exhibits a very similar peak, labelled with ``x'' on Fig.~\ref{fig:DOTOCs}. In contrast to the ``semiclassical'' peak discussed above, these early-time peaks appear to have strong dependence on the truncation parameter $M$ for all Hamiltonians. The existence of similar peaks in functions $\lambda\lr{t}$ for free Hamiltonians as well as the strong $M$ dependence of their heights suggest that such early-time peaks are the artifacts of the infrared cutoff due to the Hilbert space truncation. At sufficiently high temperatures $T \gtrsim 50$, these early-time peaks completely dominate the OTOCs, and any agreement with the classical dynamics is lost both for the bosonic and the supersymmetric Hamiltonians.

\subsection{Low temperatures}
\label{subsec:lowT_DOTOCs}

At low temperatures $T \lesssim 1$, the dynamics of the supersymmetric Hamiltonian (\ref{HS}) is dominated by nearly one-dimensional, regular wave functions, as illustrated on Fig.~\ref{fig:evec_density_plots}. We therefore compare the OTOCs for the supersymmetric model with the ones for the free one-dimensional Hamiltonian (\ref{H1D}). The OTOCs for the bosonic Hamiltonian at low temperatures only exhibit oscillations similar to the ones in the middle left plot on Fig.~\ref{fig:FF0}, and we do not consider them. Likewise, comparison with the classical dynamics makes little sense at low temperatures due to the dominance of quantum effects.

The functions $\lambda\lr{t}$ for the supersymmetric and the free one-dimensional systems at a low temperature $T = 0.2$ are shown in the two top plots on Fig.~\ref{fig:DOTOCs} for different values of the truncation parameter $M$. For the supersymmetric case, $\lambda\lr{t}$ develops a plateau (labelled with ``P'' in the top left plot on Fig.~\ref{fig:DOTOCs}) followed by the two peaks (labelled with ``x'' and ``o''). Analyzing the data for the supersymmetric system at different temperatures, we found that the height of the early-time plateau is roughly proportional to the temperature $T$ and its width scales as $\sim T^{-1}$. We also checked that the large-$M$ extrapolation of the plateau height $\lambda_P$ using the ansatz $\lambda_P = A + B/M$ yields a finite result at $M \rightarrow \infty$. The fact that $\lambda\lr{t}$ reaches a plateau means that for this time range the growth of OTOCs is with a good precision exponential. However, a comparison with the behavior of the functions $F\lr{t}$ and $F_0\lr{t}$, shown on the top left plot on Fig.~\ref{fig:FF0}, shows that this growth happens well before the function $F_0\lr{t}$ reaches its constant asymptotic value. In fact, even the positions of both subsequent peaks of $\lambda\lr{t}$ correspond to much earlier times than the saturation of $F_0\lr{t}$. We conclude that whatever time range we choose to define the quantum Lyapunov exponent for the supersymmetric Hamiltonian, we cannot find agreement with the expected behavior of the functions $F\lr{t}$, $F_0\lr{t}$ and $C\lr{t}$ in large-$N$ systems, where $C\lr{t}$ only grows exponentially when $F_0\lr{t}$ saturates. This is not surprising, as there is no natural large parameter like $N$ for our supersymmetric Hamiltonian (\ref{HS}).

The heights of the two peaks of $\lambda\lr{t}$, labelled with ``x'' and ''o'' in the top left plot on Fig.~\ref{fig:DOTOCs}, can be used as upper bounds on the rate of exponential OTOCs growth for the supersymmetric Hamiltonian. While both peaks are changing considerably as the truncation parameter $M$ is increasing, the dependence on $M$ appears to be significantly stronger for the second, subdominant, peak (labelled with ``o''). The dependence of the peak height on $M$ is reasonably well described by the formula $\lambda_{max} = A + B/M$, and extrapolations to $M \rightarrow +\infty$ yield the results for $\lambda_{max}\lr{M \rightarrow +\infty} = A$ that are close to zero. It is therefore likely that this second peak is a truncation artifact.

On the other hand, the first, dominant peak of $\lambda\lr{t}$ (labelled with ``x'' in the top left plot on Fig.~\ref{fig:DOTOCs}) has a weaker dependence on $M$, and might well present a physical feature of OTOCs in the supersymmetric case, together with the plateau that precedes it. The position of this peak scales approximately as $t^{\star} = T^{-1/2}$. We base our final estimates of the upper bound on low-temperature quantum Lyapunov exponents in the supersymmetric case on the height $\lambda_{max}$ of this peak. It is therefore important to ensure that this peak is a physical feature and not an artifact.

To this end, we consider two different methods to extrapolate $\lambda_{max}$ to $M \rightarrow +\infty$. The first method, illustrated on the top left plot on Fig.~\ref{fig:lambda_vs_M}, is to perform extrapolations to $M \rightarrow +\infty$ separately for each value of the temperature. To this end we use least squares fits to two different models:
\begin{eqnarray}
 \label{extrap_linear}
 \lambda_{max}\lr{M} = \lambda_{max}\lr{M \rightarrow + \infty} + \frac{B}{M}
 \\
 \label{extrap_sqrt}
 \lambda_{max}\lr{M} = \lambda_{max}\lr{M \rightarrow + \infty} + \frac{B}{\sqrt{M}}
 \\
\end{eqnarray}
We estimate the extrapolation errors in the values $\lambda_{max}\lr{M \rightarrow + \infty}$ by considering their variation upon the removal of data points with either the smallest or the largest values of $M$ from the fit. These bounds are shown on the top plots on Fig.~\ref{fig:lambda_vs_M} as shaded areas around lines that correspond to the fits with all data points. Let us stress that since our data comes from exact diagonalization, the numbers are not plagued with any statistical uncertainties.

To highlight the difference between the behavior of OTOCs for the supersymmetric and the free one-dimensional Hamiltonians, on the top right plot on Fig.~\ref{fig:lambda_vs_M} we also show similar extrapolations for the maximal value $\lambda_{max}$ of the function $\lambda\lr{t}$ for the free one-dimensional Hamiltonian (\ref{H1D}). For this Hamiltonian, the function $\lambda\lr{t}$ has two distinct peaks with strong dependence on $M$. The first peak, denoted by ``o'' on the top right plot on Fig.~\ref{fig:DOTOCs}, corresponds to very early times, much earlier than the plateau onset for the supersymmetric Hamiltonian. It has no direct counterpart in the supersymmetric case, and we do not consider it in detail. On the other hand, the position of the second peak, labelled with ``x'' on the top right plot on Fig.~\ref{fig:DOTOCs}, is between the two peaks of the function $\lambda\lr{t}$ obtained with the supersymmetric Hamiltonian. The height $\lambda_{max}$ of this peak is comparable to the maximal values of $\lambda\lr{t}$ for the supersymmetric Hamiltonian. We therefore use its height $\lambda_{max}$ as an upper bound for Lyapunov exponent for the free one-dimensional Hamiltonian (\ref{H1D}) with truncated Hilbert space, and illustrate the $M$ dependence of $\lambda_{max}$ on Fig.~\ref{fig:lambda_vs_M}.

From the top left plot on Fig.~\ref{fig:lambda_vs_M} we see that for the supersymmetric Hamiltonian the extrapolations using both fits (\ref{extrap_linear}) and (\ref{extrap_sqrt}) yield nonzero results. The results of linear extrapolations in $1/M$ (\ref{extrap_linear}) are expectably higher than for the square root extrapolations. Fit uncertainties appear to be not very large. The resulting temperature dependence of the extrapolated values $\lambda_{max}\lr{M \rightarrow + \infty}$ is illustrated on the bottom left plot on Fig.~\ref{fig:lambda_vs_M} with black points and solid and dashed black lines. With the linear extrapolation model (\ref{extrap_linear}), this dependence appears to be approximately linear. Namely, a fit of $\lambda_{max}\lr{M \rightarrow + \infty, T}$ with the function $\alpha \, T^{\gamma}$ yields $\gamma = 0.826$ (solid black line), close to the linear scaling of the Lyapunov exponent with temperature, $\lambda\lr{T} \sim T$. With the square root extrapolation model (\ref{extrap_sqrt}), the same power law fit yields $\gamma = 0.520$ (dashed line), thus favoring the dependence of the form $\lambda\lr{T} \sim \sqrt{T}$. The square root extrapolation model in fact yields somewhat smaller values of squared deviations $\chi^2/d.o.f.$, but not significantly smaller.

For the free one-dimensional Hamiltonian, the extrapolations using the square root model (\ref{extrap_sqrt}) yield the results that are either compatible with zero or negative, see the top right plot on Fig.~\ref{fig:lambda_vs_M}. Extrapolations with the linear model (\ref{extrap_sqrt}) yield finite results which are however significantly smaller than the corresponding extrapolated values for the supersymmetric Hamiltonian. Squared deviations $\chi^2/d.o.f.$ appear to be considerably smaller for the square root model (\ref{extrap_sqrt}). The resulting temperature dependence of the extrapolated values $\lambda_{max}$ is shown on the bottom right plot on Fig.~\ref{fig:lambda_vs_M} with black points and solid black (for the linear extrapolations (\ref{extrap_linear})) and dashed (for the square root extrapolations (\ref{extrap_sqrt})) lines.

Another strategy to estimate the temperature dependence of the maximal values of $\lambda\lr{t}$ in the limit $M \rightarrow + \infty$ is to fit the temperature dependence of $\lambda_{max}$ at fixed $M$ with a suitable fitting function, and to extrapolate the parameters of this fit to $M \rightarrow +\infty$ afterwards. For each fixed $M$, the temperature dependence of $\lambda_{max}$ for the supersymmetric Hamiltonian is with a good precision linear. We therefore fit $\lambda_{max}$ at fixed $M$ to the linear function $\lambda_{max}\lr{M, T} = A\lr{M} \, T$ of the temperature $T$. In turn, the dependence of the fit parameter $A\lr{M}$ on $M$ can be well fitted by the formula $A\lr{M} = A\lr{M \rightarrow +\infty} + B/M$, which yields the extrapolated value $A\lr{M \rightarrow +\infty} = 0.05$. The resulting extrapolation of $\lambda_{max}\lr{M \rightarrow +\infty, T}$ is shown on the lower left plot on Fig.~\ref{fig:lambda_vs_M} with solid green line, together with the values of $\lambda_{max}$ at each finite $M$ and the linear fits $\lambda_{max}\lr{M, T} = A\lr{M} \, T$ thereof.

Applying the same procedure to the free one-dimensional Hamiltonian, we find that in this case the data for $\lambda_{max}\lr{M, T}$ at fixed $M$ can be sufficiently well fitted to the function $A \, \sqrt{T}$. In this case, the $M$ dependence of the fit parameter $A\lr{M}$ is better described by the formula $A\lr{M} = A\lr{M \rightarrow +\infty} + B/\sqrt{M}$, rather than $A\lr{M} = A\lr{M \rightarrow +\infty} + B/M$. Using the former formula to extrapolate the fit parameter to $M \rightarrow +\infty$, we find a small negative result. The resulting function $A\lr{M \rightarrow +\infty} \, \sqrt{T}$ is shown on the bottom right plot on Fig.~\ref{fig:lambda_vs_M} as a solid green line. Remarkably, it agrees well with the result of $M \rightarrow +\infty$ extrapolations at fixed temperatures, described above. This agreement suggests that the non-chaotic behavior of OTOCs for the free one-dimensional Hamiltonian (\ref{H1D})  is indeed recovered as the limit $M \rightarrow +\infty$ is taken and the Hilbert space truncation is removed. This is in stark contrast with the case of the supersymmetric Hamiltonian (\ref{HS}), for which all our extrapolation methods yield a positive and finite value of the OTOC growth rate in the limit $M \rightarrow +\infty$.

\begin{figure}[h!tpb]
  \centering
  \includegraphics[width=0.48\textwidth]{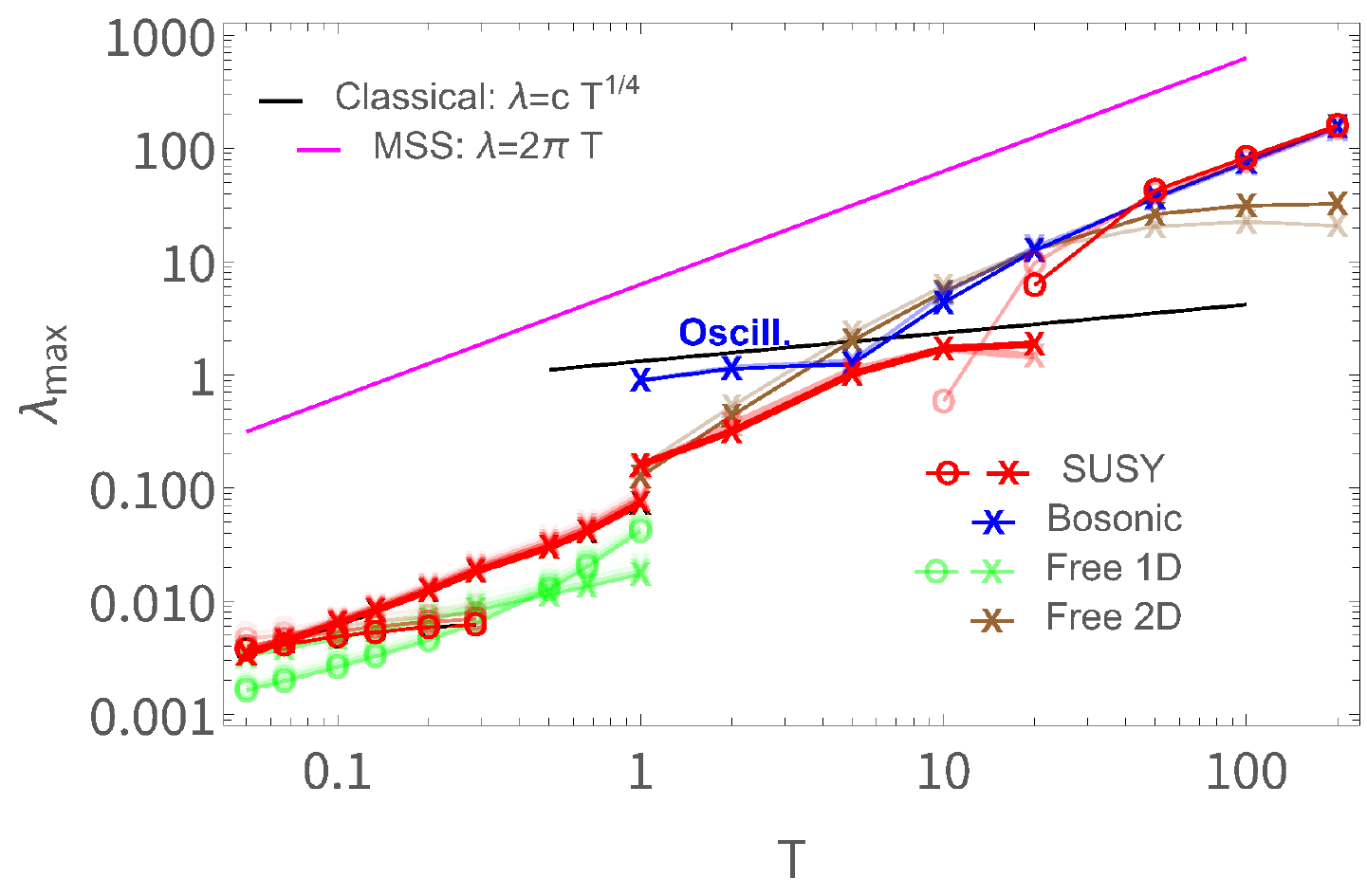} \\
  \caption{A summary plot with our estimates (upper bounds) of the values of quantum Lyapunov exponents for the supersymmetric Hamiltonian (\ref{HS}) (red lines/symbols), compared with similar estimates for the bosonic Hamiltonian (\ref{HB}) (blue lines) and for the free one-dimensional (green lines) and two-dimensional (brown lines) Hamiltonians (\ref{H1D}) and (\ref{H2D}). The estimates are based on the values of $\lambda\lr{t} = \frac{1}{2} \frac{d}{d t}\log\lr{C\lr{t}}$ in most characteristic local maxima. The symbol ``x'' denotes the heights of the maxima that we consider most important for each model. The symbol ``o'' denotes either artifact or sub-dominant but prominent maxima. The use of ``x'' and ``o'' symbols for each model is in one-to-one correspondence with the plots on Fig.~\ref{fig:DOTOCs}. Data points for the bosonic Hamiltonian labelled as ``oscill.'' correspond to oscillatory behavior of OTOCs, with $\lambda_{max}$ corresponding to the maximal value of the oscillating function. We collate the data for all $M$ without any extrapolations. More opaque lines/symbols correspond to larger values of $M$.}
  \label{fig:final_summary_plots}
\end{figure}

Finally, we illustrate the temperature dependence of all our estimates of $\lambda_{max}$ as a function of temperature on Fig.~\ref{fig:final_summary_plots}. We show the data obtained for all values of the truncation parameter $M$, and use more opaque lines to distinguish larger values of $M$. To avoid any extrapolation ambiguities, on this plot we do not show any extrapolations, only the actual numerical data for finite $M$ values. Different line colors correspond to different Hamiltonians. The data points for the peak values of $\lambda\lr{t}$ that we believe to be most important for each model are labelled with ``x'' symbols. ``o'' symbols denote the peak values for either artifact or sub-dominant but nevertheless prominent maxima. The same ``x'' or ``o'' symbols are used to label the corresponding peaks of $\lambda\lr{t}$ on Fig.~\ref{fig:DOTOCs}.

\section{Discussion and conclusions}
\label{sec:conclusions}

In this work we considered the simplest supersymmetric extension $\hat{H}_S = \hat{H}_B \otimes I + \hat{x}_1 \otimes \sigma_1 + \hat{x}_2 \otimes \sigma_3$ of the bosonic Hamiltonian $\hat{H}_B = \hat{p}_1^2 + \hat{p}_2^2 + \hat{x}_1^2 \, \hat{x}_2^2$. The latter is known to feature both quantum and classical chaos, and is closely related to the Hamiltonian of spatially compactified pure $SU\lr{2}$ Yang-Mills theory. We focused on the energy level statistics and out-of-time order correlators (OTOCs) $C\lr{t} = - \vev{\lrs{\hat{x}_2\lr{t}, \hat{p}_2\lr{0}}^2}$. The OTOCs are the quantum counterparts of the classical Lyapunov distance $C_{cl}\lr{t} = \vev{\lrc{x_2\lr{t}, p_2\lr{0}}^2} = \vev{\lr{\frac{\partial x_2\lr{t}}{\partial x_2\lr{0}}}^2}$, which characterizes the sensitivity of one of the coordinates $x_2\lr{t}$ to its initial value $x_2\lr{0}$.

Since the energy spectrum of the bosonic Hamiltonian is gapped, its real-time dynamics is completely regular at low temperatures. In particular, the OTOCs exhibit regular oscillations and show no signature of exponential growth (see the middle left plot on Fig.~\ref{fig:FF0}). For bosonic matrix models and Yang-Mills theory, this regime would correspond to the low-energy confinement regime.

On the other hand, the energy spectrum of the supersymmetric Hamiltonian (\ref{HS}) is known to be continuous \cite{Nicolai:NPB1989}. One of our main results is that the low-energy spectrum of the supersymmetric model is completely regular, and shows no signatures of random-matrix-type energy level statistics. The corresponding wave functions are effectively one-dimensional and are localized along the flat directions of the supersymmetric Hamiltonian (see Fig.~\ref{fig:evec_density_plots}).

Nevertheless, these low-energy states produce a monotonous OTOC growth down to the lowest temperatures. We presented numerical evidence that this OTOC growth is not an artifact of our Hilbert space truncation. All our extrapolations towards the physical limit suggest that the time-dependent Lyapunov divergence rate $\lambda\lr{t} = \frac{1}{2} \frac{d}{d t} \log\lr{C\lr{t}}$ remains finite. We observed that $\lambda\lr{t}$ reaches a plateau of a finite time extent at intermediate times $t \sim T^{-1/2}$ (labelled with ``P'' on the upper left plot on Fig.~\ref{fig:DOTOCs}), which suggests an exponential growth of the OTOCs $C\lr{t}$ over a finite time range. However, in contrast to large-$N$ gauge theories and matrix model, in our system there is no intrinsic large parameter and no parametric scale separation that would allow to clearly identify the regime of exponential OTOC growth.

The plateau is followed by a distinct peak of $\lambda\lr{t}$ (labelled with ``x'' on the upper left plot on Fig.~\ref{fig:DOTOCs}). We base our estimates on the quantum Lyapunov exponent $\lambda$, shown on Fig.~\ref{fig:final_summary_plots}, on the height of this peak. While for a finite Hilbert space truncation even the free particle Hamiltonian exhibits a monotonous OTOC growth, we demonstrated that in the free case the Lyapunov divergence rate $\lambda\lr{t}$ is likely to extrapolate to zero as the truncation is removed.

We found limited evidence that the estimates of quantum Lyapunov exponent in the supersymmetric model, based either on the plateau or peak heights, scale linearly with temperature, $\lambda = c \, T$ with $c \approx 0.05$. Such a linear scaling would ensure that the model does not violate the MSS bound $\lambda < 2 \pi T$. We cannot however completely exclude a scaling $\lambda \approx c' \, T^{\gamma}$ with a fractional power $1/2 \lesssim \gamma \lesssim 1$. While this scaling would formally violate the MSS bound $\lambda < 2 \pi T$ at sufficiently low temperatures, there is no deep reason why the MSS bound should not be violated for our simple Hamiltonians. Indeed, the derivation of the MSS bound \cite{Maldacena:1503.01409} is based on the assumption that the OTOCs exhibit exponential growth after the time-ordered correlator $F_0\lr{t}$ (defined in (\ref{otoc_decomposition})) saturates at a constant value $f_0 = 2 \vev{\hat{x}_2^2} \vev{\hat{p}_2^2}$. This assumption is justifiable for large-$N$ matrix models and non-Abelian gauge theories. However, as one can observe by comparing the data on Figs.~\ref{fig:FF0},~\ref{fig:OTOCs} and \ref{fig:DOTOCs}, in our case $F_0\lr{t}$ reaches its asymptotic value at much later times than the times at which the OTOCs exhibit maximal growth. Again, this could be expected in the absence of any parametric scale separation.

Our estimates of the quantum Lyapunov exponents are summarized on Fig.~\ref{fig:final_summary_plots} for the whole range of temperatures, and for all Hamiltonians and all values of the truncation parameters that we consider. The data on Fig.~\ref{fig:final_summary_plots} suggests that the quantum Lyapunov exponent for the supersymmetric Hamiltonian (\ref{HS}) continuously interpolates between an approximately linear scaling $\lambda \sim T$ at low temperatures and the fractional power scaling $\lambda \sim T^{1/4}$ that is characteristic for the classical dynamics (see Eq.~(\ref{class_Lyapunov})). This is also an expected behavior for the BFSS model, which, being holographically dual to black branes \cite{Hanada:2110.01312,Costa:1411.5541}, is suspected to saturate the MSS bound at low temperatures, and reduce to the classical Yang-Mills dynamics at high temperatures. The temperature dependence of $\lambda$ at low and high temperatures is also quite different from the results obtained with the free one- and two-dimensional Hamiltonians (\ref{H1D}) and (\ref{H2D}).

At sufficiently large energies, the energy spectra of both the bosonic and the supersymmetric Hamiltonians are chaotic, with $r$-ratios in perfect agreement with random matrix theory predictions for the GOE ensemble. At sufficiently high temperatures, these energy levels dominate the real-time dynamics, which in this regime is expected to be similar to the classical chaotic dynamics for both systems. The change between the low-energy, effectively one-dimensional states with non-chaotic level statistics and the high-energy chaotic states appears to be quite sharp (see the middle plot on Fig.~\ref{fig:r_ratio}). It would be interesting to understand what might be the counterpart of this change in the BFSS model, and how it can be interpreted in terms of the holographic dual theory. An intriguing possibility is the transition to the $M$-theory regime that was discussed recently in \cite{Hanada:2110.01312}.

An interesting observation is that the correspondence with the classical dynamics at (moderately) high temperatures can only be established for OTOCs in the supersymmetric model, but not in the bosonic one. This suggests that supersymmetry cancels out some of the quantum corrections to OTOCs at high temperatures, a property that is also likely to hold for supersymmetric Yang-Mills theory and the BFSS model. For the bosonic model, low-temperature OTOC oscillations with amplitude $\sim T^{1/4}$ (labelled as ``Oscill.'' on Fig.~\ref{fig:final_summary_plots}) are continuously transforming into very quick early-time growth, without approaching the classical OTOCs sufficiently closely anywhere. At very high temperatures ($T \gtrsim 20$), this quick early-time growth also makes it impossible to establish the quantum-classical correspondence for the supersymmetric Hamiltonian. The fact that a similar growth occurs also for the free two-dimensional Hamiltonian suggests that it might be a truncation artifact. The invalidity of the quantum-classical correspondence for the four-point OTOCs in simple quantum mechanical systems was also discussed recently in \cite{Hashimoto:1703.09435}.

Our results suggest that the supersymmetric Hamiltonian (\ref{HS}) has more intricate chaotic dynamics than the simple bosonic Hamiltonian $\hat{H}_B = \hat{p}_1^2 + \hat{p}_2^2 + \hat{x}_1^2 \, \hat{x}_2^2$, both at low and at high temperatures. The supersymmetric model definitely reproduces more features of the BFSS model that are expected from the holographic dual description. It would be interesting to work out an analytic description of the effectively one-dimensional eigenstates of this Hamiltonian that saturate the OTOC growth at low temperatures. Such a description might also help to understand the behavior of OTOCs in the BFSS model, which provides one of the most elaborate examples of holographic duality \cite{Susskind:hep-th/9610043,Witten:hep-th/9510135}.

\begin{acknowledgments}
This work used the DiRAC@Durham facility managed by the Institute for Computational Cosmology on behalf of the STFC DiRAC HPC Facility (www.dirac.ac.uk). The equipment was funded by BEIS capital funding via STFC capital grants ST/P002293/1, ST/R002371/1 and ST/S002502/1, Durham University and STFC operations grant ST/R000832/1. DiRAC is part of the National e-Infrastructure.

This work was performed using the DiRAC Data Intensive service at Leicester, operated by the University of Leicester IT Services, which forms part of the STFC DiRAC HPC Facility (www.dirac.ac.uk). The equipment was funded by BEIS capital funding via STFC capital grants ST/K000373/1 and ST/R002363/1 and STFC DiRAC Operations grant ST/R001014/1. DiRAC is part of the National e-Infrastructure.
\end{acknowledgments}


%

\appendix

\section{Symmetries of the supersymmetric and the bosonic Hamiltonians}
\label{apdx:symmetries}

As discussed in the literature \cite{Haller:PRL1984,Akutagawa:2004.04381}, the symmetry group of the bosonic Hamiltonian (\ref{HB}) is a finite non-Abelian group $C_{4v}$ with $8$ elements. This group is generated by the following transformations:
\begin{itemize}
\item Rotation $\hat{R}$ by $\pi/2$: $x_1' = -x_2$, $x_2' = x_1$. Rotations act on wave functions as $\lr{\hat{R} \Psi}\lr{x_1, x_2} = \Psi\lr{x_2, -x_1}$. The group $C_{4v}$ contains $\hat{R}$ as well $\hat{R}^2$ and $\hat{R}^3$.
\item Reflections $\hat{P}_1$ and $\hat{P}_2$ with respect to the horizontal and vertical lines $x_2 = 0$ and $x_1 = 0$, respectively: $\lr{\hat{P}_1 \Psi}\lr{x_1, x_2} = \Psi\lr{-x_1, x_2}$, $\lr{\hat{P}_2 \Psi}\lr{x_1, x_2} = \Psi\lr{x_1, -x_2}$.
\item Reflections $\hat{P}_+$ and $\hat{P}_-$ with respect to the diagonal lines $x_2 = \pm x_1$: $\lr{\hat{P}_+ \Psi}\lr{x_1, x_2} = \Psi\lr{x_2, x_1}$, $\lr{\hat{P}_- \Psi}\lr{x_1, x_2} = \Psi\lr{-x_2, -x_1}$.
\end{itemize}
Together with the identity $\lr{\hat{I} \Psi}\lr{x_1, x_2} = \Psi\lr{x_1, x_2}$, these operators implement the functional representation of $C_{4v}$. It is straightforward to check that all these operators commute with the bosonic Hamiltonian (\ref{HB}).

The group $C_{4v}$ has four Abelian irreps: the trivial irrep $\mathcal{A}_1$ as well as $Z_2$-valued irreps $\mathcal{A}_2$, $\mathcal{B}_1$ and $\mathcal{B}_2$, which can be inferred from the character tables in \cite{GelessusC4v}. There is also one real, two-dimensional, non-Abelian irrep $\mathcal{E}_0$. It coincides with the non-Abelian irrep $\mathcal{E}_0$ of the symmetry group of the supersymmetric Hamiltonian, see the Table (\ref{D16_elements}).

The symmetry group of the supersymmetric Hamiltonian (\ref{HS}) is $D_{4d}$, a finite non-Abelian group with $16$ elements \cite{GelessusD4d}. The transformations that generate this group are closely related to the transformations that leave the bosonic Hamiltonian invariant, with some additional operations that arise due to the fermionic structure of the Hilbert space of two-component wave functions $\Psi\lr{x_1, x_2} = \lrc{\psi\lr{x_1, x_2}, \chi\lr{x_1, x_2}}$:
\begin{enumerate}
  \item Rotations $\hat{\mathcal{R}}$ by $\pi/2$. Rotations act on the two-component wave functions as
      \begin{eqnarray}
      \label{SUSY_rotation}
       \lr{\hat{\mathcal{R}} \Psi}\lr{x_1, x_2}
        =
       \frac{I - \epsilon}{\sqrt{2}} \lr{\hat{R} \Psi}\lr{x_1, x_2}
        = \nonumber \\ =
       \frac{I - \epsilon}{\sqrt{2}} \, \Psi\lr{x_2, -x_1} ,
      \end{eqnarray}
      where $I$ is the $2 \times 2$ identity matrix, and $\epsilon$ is the $2 \times 2$ anti-symmetric matrix with $\epsilon_{12} = 1$. The unitary matrix $\frac{I - \epsilon}{\sqrt{2}}$ multiplies the two-component vector $\Psi\lr{x_1, x_2} = \lrc{\psi\lr{x_1, x_2}, \chi\lr{x_1, x_2}}$. In contrast to the rotation operator $\hat{R}$ acting on the Hilbert space of the bosonic Hamiltonian (\ref{HB}), for the supersymmetric Hamiltonian the operator $\hat{\mathcal{R}}$ is a fermionic representation of the rotation operator, for which the rotation by $2 \pi$ results in a change of sign: $\hat{\mathcal{R}}^4 = - \hat{\mathcal{I}}$, where $\hat{\mathcal{I}}$ is the identity operator. Correspondingly, the group $D_{4d}$ contains all powers of $\hat{\mathcal{R}}$ up to $\hat{\mathcal{R}}^7$, and only the eighth power of $\hat{\mathcal{R}}$ yields the identity operator: $\hat{\mathcal{R}}^8 = \hat{\mathcal{I}}$.
  \item Reflections $\hat{\mathcal{P}}_1$ and $\hat{\mathcal{P}}_2$ with respect to the horizontal and vertical lines $x_2 = 0$ and $x_1 = 0$, respectively:
      \begin{eqnarray}
      \label{SUSY_P12}
       \lr{\hat{\mathcal{P}}_1 \Psi}\lr{x_1, x_2}
       =
       \sigma_3 \lr{\hat{P}_1 \Psi}\lr{x_1, x_2}
       = \\
       \sigma_3 \, \Psi\lr{-x_1,  x_2} ,
       \nonumber \\
       \lr{\hat{\mathcal{P}}_2 \Psi}\lr{x_1, x_2}
       =
       \sigma_1 \lr{\hat{P}_2 \Psi}\lr{x_1, x_2}
       = \nonumber \\ =
       \sigma_1 \, \Psi\lr{ x_1, -x_2} ,
      \end{eqnarray}
      where $\sigma_1$ and $\sigma_3$ are the Pauli matrices that multiply the two-component vector $\Psi\lr{x_1, x_2} = \lrc{\psi\lr{x_1, x_2}, \chi\lr{x_1, x_2}}$.
  \item Reflections $\hat{\mathcal{P}}_{\pm}$ with respect to the diagonal lines $x_2 = \pm x_1$:
      \begin{eqnarray}
      \label{SUSY_Pdiag}
       \lr{\hat{\mathcal{P}}_+ \Psi}\lr{x_1, x_2}
       =
       \frac{\sigma_3 + \sigma_1}{\sqrt{2}} \, \lr{\hat{P}_+ \Psi}\lr{x_1, x_2}
       = \nonumber \\ =
       \frac{\sigma_3 + \sigma_1}{\sqrt{2}} \, \Psi\lr{ x_2,  x_1} ,
       \nonumber \\
       \lr{\hat{\mathcal{P}}_- \Psi}\lr{x_1, x_2}
        =
       \frac{\sigma_3 - \sigma_1}{\sqrt{2}} \, \lr{\hat{P}_- \Psi}\lr{x_1, x_2}
        = \nonumber \\ =
        \frac{\sigma_3 - \sigma_1}{\sqrt{2}} \, \Psi\lr{-x_2, -x_1} .
      \end{eqnarray}
\end{enumerate}
Note that because of the non-commutativity of Pauli matrices that enter $\hat{\mathcal{P}}_{1,2}$ and $\hat{\mathcal{P}}_{\pm}$, these transformations do not commute with each other. In contrast, reflections that act on the Hilbert space of the bosonic Hamiltonian commute with each other: $\hat{P}_1 \hat{P}_2 = \hat{P}_2 \hat{P}_1$, $\hat{P}_+ \hat{P}_- = \hat{P}_- \hat{P}_+$.

The group $D_{4d}$ has four Abelian irreducible representations (irreps): the trivial irrep $\mathcal{A}_1$ as well as $Z_2$-valued irreps $\mathcal{A}_2$, $\mathcal{B}_1$ and $\mathcal{B}_2$, which can be inferred from the character tables in \cite{GelessusD4d}. There are also three real, two-dimensional, non-Abelian irreps $\mathcal{E}_0$, $\mathcal{E}_1$ and $\mathcal{E}_2$. For completeness, in Table~\ref{D16_elements} we list all the elements of the $D_{4d}$ group together with the corresponding matrices of irreps $\mathcal{E}_0$, $\mathcal{E}_1$ and $\mathcal{E}_2$. The format of this Table is the following:
\begin{itemize}
\item In the \textbf{second column}, we give the functional representation of the corresponding element as a direct product of the form $\hat{U}_B \otimes U_F$, where the operator $\hat{U}_B$ acts on each of the two components of the wave function as $\lrc{\lr{\hat{U}_B \phi}\lr{x_1, x_2}, \lr{\hat{U}_B \chi}\lr{x_1, x_2} }$, and the matrix $U_F$ multiplies these two components as a vector, as in equations (\ref{SUSY_rotation}), (\ref{SUSY_P12}) and (\ref{SUSY_Pdiag}). The operators $\hat{U}_B$ generate a functional representation of the group $C_{4v}$.
\item In the \textbf{third column}, we give the symbol that corresponds to this element in the character tables in \cite{GelessusD4d}.
\item We note that first factors in the direct product expressions in the second column are the elements of $C_{4v}$, the symmetry group of the bosonic Hamiltonian (\ref{HB}). In the \textbf{fourth column}, we give the symbol that corresponds to these element of $C_{4v}$ in the character tables in \cite{GelessusC4v}.
\item In the \textbf{fifth column}, we show how the operators in the second column act on the basis state ${\ket{k_1, k_2, \uparrow}} = \lrc{\ket{k_1} \ket{k_2}, 0}$ with the wave function $\lrc{\psi_{k_1}\lr{x_1} \psi_{k_2}\lr{x_2}, 0}$, where $\braket{x_1}{k_1} = \psi_{k_1}\lr{x_1}$ and $\braket{x_2}{k_2} = \psi_{k_2}\lr{x_2}$ are the one-dimensional harmonic oscillator wave functions (\ref{basis_wavefunc_1D}). Note that we can still apply the Wigner's theorem to the finite matrices (\ref{HS_matrix}) and (\ref{HB_matrix}) of the Hamiltonian operators (\ref{HS}) and (\ref{HB}) in the harmonic oscillator basis, because two-dimensional harmonic oscillator eigenstates $\ket{k_1} \otimes \ket{k_2}$ inherit an $O\lr{2}$ symmetry group of the two-dimensional harmonic oscillator which contains both $C_{4v}$ and $D_{4d}$ as sub-groups.
\item In the \textbf{sixth}, \textbf{seventh} and \textbf{eighth} columns we give the matrices of non-Abelian irreps $\mathcal{E}_0$, $\mathcal{E}_1$ and $\mathcal{E}_2$ that correspond to the transformations listed in the first and the second columns. All matrices are given as combinations of $2 \times 2$ identity matrix $I$, the anti-symmetric matrix $\epsilon$ and the Pauli matrices $\sigma_1$ and $\sigma_3$. The matrices in the sixth column also form a non-Abelian irrep $\mathcal{E}_0$ of the group $C_{4v}$ (generated by the operators $\hat{U}_B$, which are the first factors in the direct products in the second column.)
\end{itemize}

\begin{widetext}
\begin{eqnarray}
\label{D16_elements}
\arraycolsep=1.4pt\def\arraystretch{1.5}
\begin{array}{|c||c|c|c|c|c|c|c|c|}
 \hline
 g &  \hat{\mathcal{G}}\lr{g}     &         \hat{U}_B\lr{g} \otimes U_F\lr{g}     & D_{4d} & C_{4v} & \hat{\mathcal{G}}\lr{g} \ket{k_1, k_2, \uparrow} & \mathcal{E}_0 & \mathcal{E}_1 & \mathcal{E}_2 \\
 \hline
 1  & \hat{\mathcal{I}}   & \hat{I}   \otimes I                                   & E     & E   &
 \tcv{\ket{k_1}\ket{k_2}}{0} & I & I & I \\ \hline
 2  & \hat{\mathcal{R}}           & \hat{R} \otimes \frac{I - \epsilon}{\sqrt{2}}   & S_8   & C_4 & \frac{\lr{-1}^{k_1}}{\sqrt{2}}\tcv{\ket{k_2}\ket{k_1}}{\ket{k_2}\ket{k_1}} & -\epsilon & \frac{I - \epsilon}{\sqrt{2}} & \frac{-I + \epsilon}{\sqrt{2}}  \\ \hline
 3  & \hat{\mathcal{R}}^2         & \hat{R}^2 \otimes \lr{-\epsilon}              & C_4   & C_2 & \lr{-1}^{k_1 + k_2}\tcv{0}{\ket{k_1} \ket{k_2}} & -I & -\epsilon & -\epsilon \\ \hline
 4  & \hat{\mathcal{R}}^3         & \hat{R}^3 \otimes \lr{\frac{-I - \epsilon}{\sqrt{2}}} & S_8^3 & C_4 & \frac{\lr{-1}^{k_2}}{\sqrt{2}} \tcv{-\ket{k_2} \ket{k_1}}{\ket{k_2} \ket{k_1}}  & \epsilon & \frac{-I - \epsilon}{\sqrt{2}} & \frac{I + \epsilon}{\sqrt{2}}  \\ \hline
 5  & \hat{\mathcal{R}}^4         & \hat{I}   \otimes \lr{-I}                             & C_2   & E   & \tcv{-\ket{k_1}\ket{k_2}}{0} & I & -I & -I  \\ \hline
 6  & \hat{\mathcal{R}}^5         & \hat{R}   \otimes \frac{-I + \epsilon}{\sqrt{2}}      & S_8^3 & C_4 & \frac{\lr{-1}^{k_1 + 1}}{\sqrt{2}} \tcv{\ket{k_2} \ket{k_1}}{\ket{k_2} \ket{k_1}} & -\epsilon & \frac{-I + \epsilon}{\sqrt{2}} & \frac{I - \epsilon}{\sqrt{2}}  \\ \hline
 7  & \hat{\mathcal{R}}^6         & \hat{R}^2 \otimes \epsilon                            & C_4   & C_2 & \lr{-1}^{k_1 + k_2 + 1} \tcv{0}{\ket{k_1}\ket{k_2}} & -I & \epsilon & \epsilon  \\ \hline
 8  & \hat{\mathcal{R}}^7         & \hat{R}^3 \otimes \lr{\frac{I + \epsilon}{\sqrt{2}}}  & S_8   & C_4 & \frac{\lr{-1}^{k_2}}{\sqrt{2}} \tcv{\ket{k_2}\ket{k_1}}{-\ket{k_2}\ket{k_1}} & \epsilon & \frac{I + \epsilon}{\sqrt{2}} & \frac{-I - \epsilon}{\sqrt{2}}   \\ \hline
 9  & \hat{\mathcal{P}}_1             & \hat{P}_1 \otimes \sigma_3                & C_2'  & \sigma_v & \lr{-1}^{k_1} \tcv{\ket{k_1} \ket{k_2}}{0} & -\sigma_3 & \sigma_3 & \sigma_3  \\ \hline
 10 & \hat{\mathcal{P}}_2         & \hat{P}_2 \otimes \sigma_1 & C_2' & \sigma_v & \lr{-1}^{k_2} \tcv{0}{\ket{k_1} \ket{k_2}} & \sigma_3 & \sigma_1 & \sigma_1  \\ \hline
 11 & \hat{\mathcal{P}}_+         & \hat{P}_+ \otimes \frac{\sigma_3 + \sigma_1}{\sqrt{2}}  & \sigma_d & \sigma_d & \frac{1}{\sqrt{2}} \tcv{\ket{k_2}\ket{k_1}}{\ket{k_2}\ket{k_1}} & \sigma_1 & \frac{\sigma_3 + \sigma_1}{\sqrt{2}} & \frac{-\sigma_3 - \sigma_1}{\sqrt{2}}  \\ \hline
 12 & \hat{\mathcal{P}}_-         & \hat{P}_- \otimes \frac{\sigma_3 -  \sigma_1}{\sqrt{2}} &  \sigma_d & \sigma_d, & \frac{\lr{-1}^{k_1 + k_2}}{\sqrt{2}} \tcv{\ket{k_2}\ket{k_1}}{-\ket{k_2}\ket{k_1}} & -\sigma_1 & \frac{\sigma_3 -  \sigma_1}{\sqrt{2}} & \frac{-\sigma_3 + \sigma_1}{\sqrt{2}}  \\ \hline
 13 & -\hat{\mathcal{P}}_1 & \hat{P}_1 \otimes \lr{-\sigma_3} & C_2' & \sigma_v & \lr{-1}^{k_1 + 1} \tcv{\ket{k_1}\ket{k_2}}{0} & -\sigma_3 & -\sigma_3 & -\sigma_3 \\ \hline
 14 & -\hat{\mathcal{P}}_2 & \hat{P}_2 \otimes \lr{-\sigma_1} & C_2' & \sigma_v & \lr{-1}^{k_2 + 1} \tcv{0}{\ket{k_1}\ket{k_2}} & \sigma_3 & -\sigma_1  & -\sigma_1  \\ \hline
 15 & -\hat{\mathcal{P}}_+ & \hat{P}_+ \otimes \frac{-\sigma_3 - \sigma_1}{\sqrt{2}} & \sigma_d & \sigma_d & -\frac{1}{\sqrt{2}} \tcv{\ket{k_2}\ket{k_1}}{\ket{k_2}\ket{k_1}} & \sigma_1 & \frac{-\sigma_3 - \sigma_1}{\sqrt{2}} & \frac{\sigma_3 + \sigma_1}{\sqrt{2}}  \\ \hline
 16 & -\hat{\mathcal{P}}_- & \hat{P}_- \otimes \frac{-\sigma_3 + \sigma_1}{\sqrt{2}} & \sigma_d & \sigma_d & \frac{\lr{-1}^{k_1 + k_2}}{\sqrt{2}} \tcv{-\ket{k_2} \ket{k_1}}{\ket{k_2} \ket{k_1}} & -\sigma_1 & \frac{-\sigma_3 + \sigma_1}{\sqrt{2}} & \frac{\sigma_3 - \sigma_1}{\sqrt{2}}  \\ \hline
 \end{array}
\end{eqnarray}
\end{widetext}

By virtue of Wigner's theorem, eigenstates of the Hamiltonians (\ref{HS}) or (\ref{HB}) should form multiplets $\ket{\Psi_i^{\alpha}}$ with degenerate energy levels that transform under one of the irreps of the groups $D_{4d}$ (for $\hat{H}_S$) or $C_{4v}$ (for $\hat{H}_B$):
\begin{eqnarray}
\label{multiplet_transform}
 \hat{\mathcal{G}}\lr{g} \ket{\Psi_i^{\alpha}} = \sum\limits_{\beta=1}^{d_R} G^R_{\alpha \beta}\lr{g} \ket{\Psi_i^{\beta}} ,
\end{eqnarray}
where the index $i$ labels distinct energy levels $E_i$, $\alpha = 1 \ldots d_R$ labels all linearly independent eigenstates with $\hat{H} \ket{\Psi_i^{\alpha}} = E_i \, \ket{\Psi_i^{\alpha}}$, $G^R_{\alpha \beta}$ is the $d_R \times d_R$ matrix of the group element $g$ in the irrep $R$. In our case, the operators $\hat{\mathcal{G}}\lr{g}$ are listed in the first column of Table~\ref{D16_elements} and defined in the second column of this Table. For irreps with dimension $d_R = 2$ the matrices $G^R_{\alpha \beta}\lr{g}$ are given in the sixth, seventh and eighth columns. For one-dimensional irreps ($d_R = 1$), $G^R_{\alpha \beta}\lr{g} \equiv G^R\lr{g}$ can be easily reconstructed from character tables in \cite{GelessusD4d,GelessusC4v}.

To classify all eigenvectors of the Hamiltonian matrices (\ref{HS}) and (\ref{HB}) according to irreps of the groups $C_{4v}$ or $D_{4d}$, we can use the identity (\ref{multiplet_transform}) and the Schur orthogonality relations
\begin{eqnarray}
\label{schur_orthogonality}
 \sum\limits_{g} G^R_{\alpha \beta}\lr{g} \bar{G}^{R'}_{\gamma \delta}\lr{g}
 =
 \delta_{RR'} \, \delta_{\alpha\gamma} \, \delta_{\beta \delta} \, \frac{|G|}{d_R} ,
\end{eqnarray}
where $|G|$ is the total number of elements in the symmetry group of the Hamiltonian. To this end, we multiply the equality (\ref{multiplet_transform}) by $\bar{G}^R_{\gamma \delta}\lr{g}$ (that is equal to $G^R_{\gamma \delta}\lr{g}$ for our real irreps) and sum over all group elements $g$:
\begin{eqnarray}
\label{multiplet_classify}
 \sum\limits_{g} G^{R'}_{\gamma \delta}\lr{g} \mathcal{G}\lr{g} \ket{\Psi_i^{\alpha}}
 = \nonumber \\ =
 \sum\limits_{\beta=1}^{d_R} \sum\limits_{g} \bar{G}^{R'}_{\gamma \delta}\lr{g} G^R_{\alpha \beta}\lr{g} \ket{\Psi_i^{\beta}}
 = \nonumber \\ =
 \frac{|G|}{d_R} \, \delta_{RR'} \, \delta_{\alpha\gamma} \, \ket{\Psi_i^{\delta}} .
\end{eqnarray}
Therefore if we apply all transformations within the symmetry group to some eigenvector of the Hamiltonian matrix, and sum up the results with the weights given by the elements of the matrices of some irrep $R'$, we only get nonzero result if $R'$ is the irrep of the multiplet to which this eigenvector belongs. Of course, for one-dimensional irreps, the vector transforms into itself, and group transformations amount to multiplications by a complex phase factor.

The identity (\ref{multiplet_classify}) can also be used to prove that eigenvectors of the supersymmetric Hamiltonian can only transform under the two-dimensional non-Abelian irreps $\mathcal{E}_1$ and $\mathcal{E}_2$. Indeed, the matrices that multiply the two-component wave function as a vector ($U_F$ factors in the second column in Table~(\ref{D16_elements})) form the irrep $\mathcal{E}_1$. The structure of irreps $\mathcal{E}_1$ and $\mathcal{E}_2$ is such that for any group element $g$ with representation matrix $G^R_{\alpha\beta}\lr{g}$ there is another element $g'$ with $G^R_{\alpha\beta}\lr{g'} = - G^R_{\alpha\beta}\lr{g}$. Looking at the Table~(\ref{D16_elements}) as well as on the character tables in \cite{GelessusD4d}, we notice that for all other irreps $R' = \mathcal{A}_{1,2}, \quad \mathcal{B}_{1,2}, \quad \mathcal{E}_0$ we have $G^{R'}_{\alpha\beta}\lr{g'} = + G^{R'}_{\alpha\beta}\lr{g}$, where $g$ and $g'$ is the same pair of group elements. Bosonic transformations $\hat{U}_B$ that transform the coordinate dependence of wave functions are also insensitive to the sign factors of $\mathcal{E}_1$. Therefore, for any $R' = \mathcal{A}_{1,2}, \quad \mathcal{B}_{1,2}, \quad \mathcal{E}_0$ the product $G^{R'}_{\gamma \delta}\lr{g} \hat{U}_B\lr{g} \otimes U_F\lr{g} \ket{\Psi}$ in the first line of Eq.~(\ref{multiplet_classify}) will be equal to $- G^{R'}_{\gamma \delta}\lr{g'} \hat{U}_B\lr{g'} \otimes U_F\lr{g'} \ket{\Psi}$. In the sum over all group elements, summands that correspond to $g$ and $g'$ will cancel each other. Therefore, for any eigenstate of the supersymmetric Hamiltonian the sum in the first line of Eq.~(\ref{multiplet_classify}) will be equal to zero for $R' = \mathcal{A}_{1,2}, \quad \mathcal{B}_{1,2}, \quad \mathcal{E}_0$.

We conclude therefore that eigenstates of the supersymmetric Hamiltonian can only transform under the irreps $\mathcal{E}_1$ or $\mathcal{E}_2$ and therefore should all be doubly degenerate once we impose the IR cutoff (finite $M$ truncation) that makes the continuous spectrum discrete.

\section{Block diagonal representation of Hamiltonian matrices within parity sectors and basis state enumeration}
\label{apdx:harmonic_basis}

In this work we perform numerical diagonalization of the Hamiltonian matrices (\ref{HB_matrix}) and (\ref{HS_matrix}). To achieve faster convergence of diagonalization algorithms, it is advantageous to reduce the size of matrices being diagonalized, for example, by representing them in block diagonal form. The largest reduction of the matrix block size could be achieved by considering combinations of basis states $\ket{k_1} \otimes \ket{k_2}$ that transform under irreps of the groups $C_{4v}$ (for the bosonic Hamiltonian) or $D_{4d}$ (for the supersymmetric Hamiltonian). By virtue of Wigner's theorem, the Hamiltonian matrices (\ref{HB_matrix}) and (\ref{HS_matrix}) would then reduce to a block diagonal form such that each block corresponds to one of the irreps. Explicit construction of such basis states is quite involved, and would result in a significantly more complicated code.

In this work we choose a different but equally efficient strategy and use an eigenbasis of a single element $\hat{\mathcal{P}}_1 = \hat{P}_1 \otimes \sigma_3$ of the symmetry group $D_{4d}$ of the supersymmetric Hamiltonian $\hat{H}_S$. Since $\hat{\mathcal{P}}_1$ commutes with $\hat{H}_S$ as well as with the corresponding Hamiltonian matrix and $\hat{\mathcal{P}}_1^2 = \hat{\mathcal{I}}$, the Hamiltonian matrix should have a block diagonal form with blocks corresponding to $\pm 1$ eigenvalues of $\hat{\mathcal{P}}_1$. Note that since the parity operators $\hat{\mathcal{P}}_{1,2}$ and $\hat{\mathcal{P}}_{\pm}$ do not commute with each other, only one of these operators can be diagonalized simultaneously with the Hamiltonian.

An immediate consequence of the identity (\ref{multiplet_transform}) is that any two states $\ket{\Psi_i^1}$, $\ket{\Psi_i^2}$ that belong to a doublet of states transforming under irreps $\mathcal{E}_1$ or $\mathcal{E}_2$ of the $D_{4d}$ symmetry group should have opposite parities. As one can infer from Table~(\ref{D16_elements}), for both irreps the parity transformation $\hat{\mathcal{P}}_1$ is represented by the Pauli matrix $\sigma_3$. Rewriting (\ref{multiplet_transform}) in explicit form, we obtain
\begin{eqnarray}
\label{doublet_parities}
 \hat{\mathcal{P}}_1 \ket{\Psi_i^1} = \lr{\sigma_3}_{11} \ket{\Psi_i^1} + \lr{\sigma_3}_{12} \ket{\Psi_i^2} = +\ket{\Psi_i^1} ,
 \nonumber \\
 \hat{\mathcal{P}}_1 \ket{\Psi_i^2} = \lr{\sigma_3}_{21} \ket{\Psi_i^1} + \lr{\sigma_3}_{22} \ket{\Psi_i^2} = - \ket{\Psi_i^2} .
\end{eqnarray}
Together with the fact that eigenstates of $\hat{H}_S$ can only transform under irreps $\mathcal{E}_1$ or $\mathcal{E}_2$, this observation implies that both diagonal blocks of the supersymmetric Hamiltonian matrix (\ref{HS_matrix}) have identical energy spectra. Using the fact that the operators $\hat{x}_2$ and $\hat{p}_2$ are invariant under $\hat{\mathcal{P}}_1$, one can also show that the products of matrix elements of $\hat{x}_{2}$ and $\hat{p}_{2}$ that enter the OTOCs $C\lr{t}$ in (\ref{otoc_reg}) are identical for both parity sectors. Therefore the contributions of both diagonal blocks to OTOCs are identical, which allows us to save CPU time by diagonalizing only the positive-parity block of the supersymmetric Hamiltonian matrix. Note that this does not apply to the bosonic Hamiltonian, for which both parity sectors need to be diagonalized in order to get the full OTOCs.

All harmonic oscillator eigenstates (\ref{basis_wavefunc_1D}) have a definite parity under coordinate reflections, therefore the direct product states $\ket{k_1} \otimes \ket{k_2}$ are also eigenstates of the bosonic parity operator $\hat{P}_1$ with $\hat{P}_1 \ket{k_1} \otimes \ket{k_2} = \lr{-1}^{k_1} \ket{k_1} \otimes \ket{k_2}$. Using this fact, it is straightforward to find the two-component eigenstates of the operator $\hat{\mathcal{P}}_1 = \hat{P}_1 \otimes \sigma_3$ with eigenvalues $\pm 1$:
\begin{eqnarray}
\label{SUSY_parity_eigenstates}
 \ket{k_1, k_2, +}
 =
 \left(
 \begin{array}{c}
   \ket{k_1} \otimes \ket{k_2}  \\
   0  \\
 \end{array}
 \right), \quad k_1 = 2 \, m ,
 \nonumber \\
  \ket{k_1, k_2, +}
 =
 \left(
 \begin{array}{c}
    0 \\
   \ket{k_1} \otimes \ket{k_2}  \\
 \end{array}
 \right), \quad k_1 = 2 \, m + 1 ,
 \nonumber \\
  \ket{k_1, k_2, -}
 =
 \left(
 \begin{array}{c}
   \ket{k_1} \otimes \ket{k_2}  \\
   0                            \\
 \end{array}
 \right), \quad k_1 = 2 \, m + 1 ,
 \nonumber \\
  \ket{k_1, k_2, -}
 =
 \left(
 \begin{array}{c}
   0                           \\
   \ket{k_1} \otimes \ket{k_2} \\
 \end{array}
 \right), \quad k_1 = 2 \, m ,
\end{eqnarray}
where $m = 0, 1, 2, \ldots$. Matrix elements $\bra{k_1, k_2, +} \hat{H}_S \ket{k_1, k_2, -}$ are equal to zero because $\hat{H}_S$ commutes with $\hat{\mathcal{P}}_1$.

\begin{figure}
  \centering
  \includegraphics[width=0.47\textwidth]{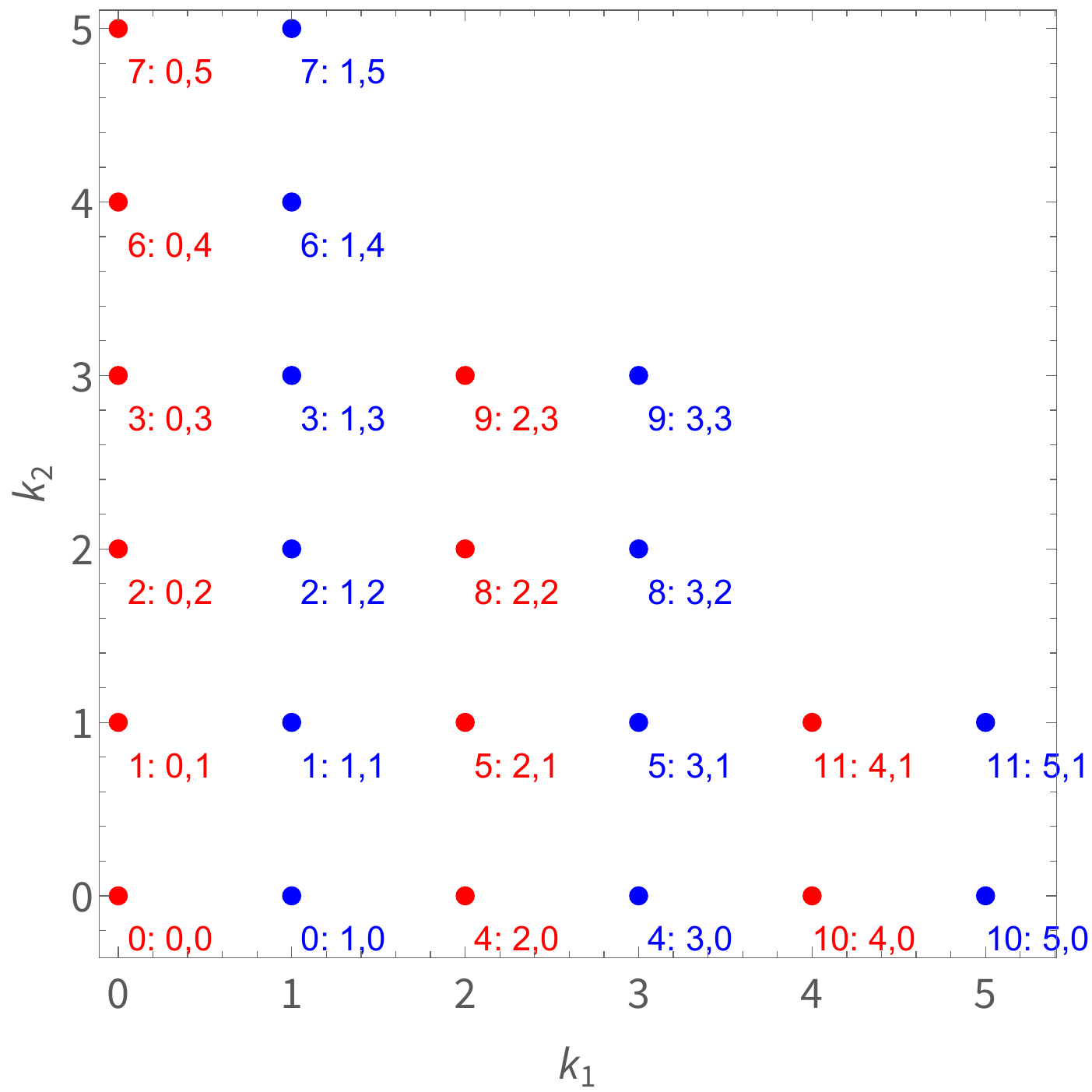}
  \caption{Enumeration of two-dimensional harmonic oscillator basis states $\ket{k_1} \otimes \ket{k_2}$. States that correspond to red and blue points are even and odd under reflections of $x_1$, respectively. The label ``$i: k_1, k_2$'' near each point shows the serial number $i$ of the state and the corresponding values of $k_1$ and $k_2$. States with even and odd $x_1$-parity are enumerated independently. For this plot, the index $i$ changes between $0$ and $M^2 + M - 1$ with $M  = 3$.}
  \label{fig:enumeration_example}
\end{figure} 

To work with the basis states (\ref{SUSY_parity_eigenstates}), we need a way to efficiently enumerate the states $\ket{k_1} \otimes \ket{k_2}$ with $k_1 + k_2 \leq M$ and a definite parity of $k_1$. We use the enumeration shown on Fig.~\ref{fig:enumeration_example}. Red and blue points have positive and negative parity $\hat{P}_1$ (even and odd $k_1$). The labels of points on the plot have the format $i: k_1, k_2$, where $i$ is the serial number of the state. We enumerate states separately in each of the two parity sectors. If we limit the values of $i$ to lie in the range $i = 0 \ldots M^2 + M - 1$, the set of basis states is symmetric with respect to the interchange $x_1 \leftrightarrow x_2$. Since this is one of the basic symmetries of our system, we always use ranges of this form.

The transformation from $k_1$, $k_2$ to the one-dimensional index $i$ can be written as
\begin{eqnarray}
\label{cart2idx}
 i = 2 \lr{\mathrm{div}\lr{k_1, 2} + \mathrm{div}\lr{r + r^2, 2}}
 + \nonumber \\ +
 \mathrm{mod}\lr{k_2, 2} ,
 \quad
 r = \mathrm{div}\lr{k_1, 2} + \mathrm{div}\lr{k_2, 2} ,
\end{eqnarray}
where $\mathrm{div}\lr{a, b}$ and $\mathrm{mod}\lr{a, b}$ are the integer division of $a$ over $b$ and the integer modulo of $a$ over $b$. The inverse transformation is
\begin{eqnarray}
\label{idx2cart}
    k_1 = 2 \lr{\mathrm{div}\lr{i, 2} - \mathrm{div}\lr{r + r^2, 2}} + p_1 ,
    \nonumber \\
	k_2 = 2 \lr{r + \mathrm{div}\lr{r + r^2, 2} - \mathrm{div}\lr{i, 2}}
   + \nonumber \\ + \mathrm{mod}\lr{i, 2} ,
    \quad
    r = \left\lfloor \sqrt{2 \mathrm{div}\lr{i, 2} + \frac{1}{4}} - \frac{1}{2} \right\rfloor ,
\end{eqnarray}
where $\left\lfloor \ldots \right\rfloor$ is the floor function and $\lr{-1}^{p_1}$ is the parity sign: $p_1 = 0$, $p_1 = 1$ for the positive and negative parity sectors, correspondingly.

Note that for the bosonic Hamiltonian (\ref{HB}), we could have used the two commuting operators $\hat{P}_1$ and $\hat{P}_2$ (or $\hat{P}_+$ and $\hat{P}_-$) to define four parity sectors and thus reduce the size of matrix blocks even further. However, here we only used $\hat{P}_1$ to reduce the diagonal block size of the bosonic Hamiltonian matrix (\ref{HB_matrix}). The reason is that the primary focus of this paper is the supersymmetric Hamiltonian (\ref{HS}), and for debugging purposes we wanted to maintain state enumeration that would be similar for both Hamiltonians. Even with this choice, the bosonic Hamiltonian is much easier to treat numerically than the supersymmetric one, and using only $\hat{P}_1$ parity is enough to obtain good-quality numerical data for the bosonic model.

\section{Calculating the classical analogue of thermal OTOCs}
\label{apdx:lyapunov_classical}

To check whether in the high temperature limit the quantum dynamics of the bosonic and supersymmetric Hamiltonians (\ref{HB}) and (\ref{HS}) agrees with the classical dynamics of the bosonic Hamiltonian, we calculate the classical analogue of OTOCs. It is given by the thermal expectation value of the square of the Poisson brackets $\lrc{x_2\lr{t}, p_2\lr{0}} = \frac{\partial x_2\lr{t}}{\partial x_2\lr{0}}$:
\begin{widetext}
\begin{eqnarray}
\label{class_Lyapunov_extended}
 C_{cl}\lr{t} = \frac{1}{Z_{cl}}\int dx_1\lr{0} \, dx_2\lr{0} \, dp_1\lr{0} \, dp_2\lr{0} \,
  \exp\lr{ -\frac{p_1^2\lr{0} + p_2^2\lr{0} + x_1^2\lr{0} \, x_2^2\lr{0}}{T} }
 \,
 \lr{\frac{\partial x_2\lr{t}}{\partial x_2\lr{0}} }^2  ,
\end{eqnarray}
\end{widetext}
where $x_1\lr{0}$, $x_2\lr{0}$, $p_1\lr{0}$, $p_2\lr{0}$ specify the initial conditions for the time-dependent coordinates and momenta $x_1\lr{t}$, $x_2\lr{t}$, $p_1\lr{t}$, $p_2\lr{t}$. These in turn satisfy the classical equations of motion
\begin{eqnarray}
\label{class_eom}
 \frac{d}{dt} x_1\lr{t} = 2 \, p_1\lr{t}, \quad
 \frac{d}{dt} x_2\lr{t} = 2 \, p_2\lr{t},
 \nonumber \\
 \frac{d}{dt} p_1\lr{t} = - 2 \, x_1\lr{t} \, x_2^2\lr{t},
 \nonumber \\
 \frac{d}{dt} p_2\lr{t} = - 2 \, x_2\lr{t} \, x_1^2\lr{t} .
\end{eqnarray}
The classical partition function $Z_{cl}$ is
\begin{eqnarray}
\label{classical_partition}
 \mathcal{Z}_{cl}
 =
 \int d p_1 \, d p_2 \exp\lr{-\frac{p_1^2 + p_2^2}{T}}
 \times \nonumber \\ \times
 \int d x_1 \, d x_2 \exp\lr{-\frac{x_1^2 x_2^2}{T} } .
\end{eqnarray}
To obtain the classical Poisson bracket $\lrc{x_2\lr{t}, p_2\lr{0}} = \frac{\partial x_2\lr{t}}{\partial x_2\lr{0}}$, we can differentiate the classical equations of motion with respect to $x_2\lr{0}$, thereby obtaining a system of differential equations that govern the time evolution of $\frac{\partial x_1\lr{t}}{\partial x_2\lr{0}}$, $\frac{\partial x_2\lr{t}}{\partial x_2\lr{0}}$, $\frac{\partial p_1\lr{t}}{\partial x_2\lr{0}}$ and $\frac{\partial p_2\lr{t}}{\partial x_2\lr{0}}$:
\begin{widetext}
\begin{eqnarray}
\label{class_lyapunov_eom}
 \frac{d}{dt} \frac{\partial x_1\lr{t}}{\partial x_2\lr{0}} = 2 \frac{\partial p_1\lr{t}}{\partial x_2\lr{0}}, \quad
 \frac{d}{dt} \frac{\partial x_2\lr{t}}{\partial x_2\lr{0}} = 2 \frac{\partial p_2\lr{t}}{\partial x_2\lr{0}},
 \nonumber \\
 \frac{d}{dt} \frac{\partial p_1\lr{t}}{\partial x_2\lr{0}} = - 2 \frac{\partial x_1\lr{t}}{\partial x_2\lr{0}} \, x_2^2\lr{t} - 4 \, x_1\lr{t} x_2\lr{t} \frac{\partial x_2\lr{t}}{\partial x_2\lr{0}},
 \nonumber \\
 \frac{d}{dt} \frac{\partial p_2\lr{t}}{\partial x_2\lr{0}} = - 2 \frac{\partial x_2\lr{t}}{\partial x_2\lr{0}} \, x_1^2\lr{t} - 4 \, x_1\lr{t} x_2\lr{t} \frac{\partial x_1\lr{t}}{\partial x_2\lr{0}} .
\end{eqnarray}
\end{widetext}
These equations have to be solved simultaneously with the classical equations of motion (\ref{class_eom}) with the initial conditions $\frac{\partial x_1\lr{0}}{\partial x_2\lr{0}} = 0$, $\frac{\partial x_2\lr{0}}{\partial x_2\lr{0}} = 1$, $\frac{\partial p_1\lr{0}}{\partial x_2\lr{0}} = 0$, $\frac{\partial p_2\lr{0}}{\partial x_2\lr{0}} = 0$.

To calculate the ``classical OTOC'' (\ref{class_Lyapunov_extended}), we carry out a small-scale Monte-Carlo simulation, generating $O\lr{10^5}$ random initial conditions $x_1\lr{0}$, $x_2\lr{0}$, $p_1\lr{0}$, $p_2\lr{0}$ with the probability distribution that is proportional to $\exp\lr{ -\frac{p_1^2\lr{0} + p_2^2\lr{0} + x_1^2\lr{0} \, x_2^2\lr{0}}{T} }$. The evolution equations (\ref{class_eom}) and (\ref{class_lyapunov_eom}) are then solved numerically, and $\lr{\frac{\partial x_2\lr{t}}{\partial x_2\lr{0}}}^2$ is averaged over sufficiently many random initial conditions. A technical difficulty is that both the classical partition function (\ref{classical_partition}) and the probability distribution of initial coordinate values $x_1\lr{0}$, $x_2\lr{0}$ contain a non-normalizable weight function $\exp\lr{-\frac{x_1^2 x_2^2}{T}}$. To regularize the diverging integrals $\int dx_1 dx_2 \exp\lr{-\frac{x_1^2 x_2^2}{T}}$ and to generate $x_1\lr{0}$ and $x_2\lr{0}$ with the required probability distribution, we express $x_1$ and $x_2$ in terms of ``hyperbolic'' coordinates $-\infty < r < +\infty$ and $-\infty < \phi < +\infty$ as $x_1 = r \, e^{\phi}$, $x_2 = r \, e^{-\phi}$. In terms of the new coordinates, the above integrals can be written as
\begin{eqnarray}
\label{classical_integral_regularization}
\int dx_1 dx_2 \exp\lr{-\frac{x_1^2 x_2^2}{T}} = 4 \int\limits_{-\infty}^{+\infty} d\phi \int\limits_0^{+\infty} d r^2 \, e^{-\frac{r^4}{T}} .
\end{eqnarray}
We see that the integral divergence is related to the infinite integration limits for the ``hyperbolic angle'' variable $\phi$, and the probability distribution of the ``radial'' coordinate $r$ is perfectly normalizable. To regularize the integrals over $\phi$ and make the probability distribution in (\ref{class_Lyapunov_extended}), we introduce a cutoff $-\phi_{max} < \phi < \phi_{max}$ on the $\phi$ variable. While the integrals in the classical partition function $Z_{cl}$ and in the Lyapunov distance definition (\ref{class_Lyapunov_extended}) are divergent, this divergence cancels out in the ratio of the two integrals. As a result, $C_{cl}\lr{t}$ is practically independent of the cutoff $\phi_{max}$ once $\phi_{max}$ is large enough. In practice, we used $\phi_{max}$ in the range $2 < \phi_{max} < 5$ and observed that the results are independent of $\phi_{max}$ within statistical errors of Monte-Carlo expectation values.

The expression (\ref{classical_Lyapunov_scaling}) is obtained by fitting $\log\lr{C_{cl}\lr{t}}$ at sufficiently late times with a linear function. The corresponding slope is twice the classical Lyapunov exponent in (\ref{classical_Lyapunov_scaling}).

\end{document}